\documentclass[letterpaper,11pt]{article}
\pdfoutput=1

\usepackage{jheppub}
\usepackage{amsmath}
\usepackage{graphicx}
\usepackage{xspace}
\usepackage{slashed}

\newcommand{\refcite}[1]{ref.~\cite{#1}}
\newcommand{\refscite}[1]{refs.~\cite{#1}}
\newcommand{\eq}[1]{eq.~\eqref{eq:#1}}
\newcommand{\eqs}[2]{eqs.~\eqref{eq:#1} and \eqref{eq:#2}}
\newcommand{\Equations}[2]{Equations \eqref{eq:#1} and \eqref{eq:#2}}
\renewcommand{\sec}[1]{sec.~\ref{sec:#1}}
\newcommand{\secs}[2]{secs.~\ref{sec:#1} and \ref{sec:#2}}
\newcommand{\fig}[1]{fig.~\ref{fig:#1}}
\newcommand{\figs}[2]{figs.~\ref{fig:#1} and \ref{fig:#2}}
\newcommand{\app}[1]{appendix~\ref{app:#1}}

\newcommand{\ord}[1]{\mathcal{O}(#1)}
\newcommand{\altM}[1]{\overline{\Msquared}^{(#1)}}
\newcommand{\altMp}[1]{\overline{\Msquared}^{(#1)\,\prime}}
\newcommand{\altMpp}[1]{\overline{\Msquared}^{(#1)\,\prime\prime}}
\newcommand{\df}{\mathrm{d}}
\newcommand{\img}{\mathrm{i}}
\newcommand{\eps}{\epsilon}

\newcommand{\qt}{{\vec q}_T}

\newcommand{\kt}{{\vec k}_T}
\newcommand{\bt}{{\vec b}_T}

\newcommand{\cO}{{\mathcal O}}

\newcommand{\cI}{{\mathcal I}}

\newcommand{\cL}{{\mathcal L}}
\newcommand{\cM}{{\mathcal M}}
\newcommand{\cN}{{\mathcal N}}

\newcommand{\cP}{{\mathcal P}}

\newcommand{\bn}{{\bar{n}}}
\newcommand{\bq}{{\bar{q}}}
\newcommand{\GeV}{\,\mathrm{GeV}}

\newcommand{\nn}{\nonumber}
\newcommand{\Ecm}{E_\mathrm{cm}}
\newcommand{\LO}{\mathrm{LO}}

\newcommand{\Msquared}{A}
\newcommand{\as}{\alpha_s}
\newcommand{\muMS}{\mu_{\rm MS}}
\newcommand{\MS}{\overline{\mathrm{MS}}}
\newcommand{\LQCD}{\Lambda_{\rm QCD}}
\newcommand{\sceti}{SCET$_{\rm I}$}
\newcommand{\scetii}{SCET$_{\rm II}$}
\newcommand{\softregulator}{R(k,\eta)}
\newcommand{\nregulator}{R(k,\eta)}
\newcommand{\bnregulator}{R(k,\eta)}

\newcommand{\Isoftgen}[1]{I_s^{(#1)}(R)}
\newcommand{\Ingen}[1]{I_n^{(#1)}(R)}
\newcommand{\Ibngen}[1]{I_\bn^{(#1)}(R)}
\newcommand{\Isoftz}[1]{I_s^{(#1)}(R_z)}
\newcommand{\Inz}[1]{I_n^{(#1)}(R_z)}
\newcommand{\Ibnz}[1]{I_\bn^{(#1)}(R_z)}
\newcommand{\IsoftY}[1]{I_s^{(#1)}(R_Y)}
\newcommand{\InY}[1]{I_n^{(#1)}(R_Y)}
\newcommand{\IbnY}[1]{I_\bn^{(#1)}(R_Y)}
\newcommand{\bigupsilon}{\upsilon}
\newcommand{\smallupsilon}{\eta}

\allowdisplaybreaks[4]

\begin{document} 


\preprint{\vbox{
\hbox{MIT-CTP 5087}
\hbox{DESY 18-207}
}}

\title{\boldmath Subleading Power Rapidity Divergences and\\ Power Corrections for $q_T$}

\author[1]{Markus A.~Ebert,}
\emailAdd{ebert@mit.edu}

\author[2,3]{Ian Moult,}
\emailAdd{ianmoult@lbl.gov}

\author[1]{Iain W.~Stewart,}
\emailAdd{iains@mit.edu}

\author[4]{Frank J.~Tackmann,}
\emailAdd{frank.tackmann@desy.de}

\author[1]{Gherardo Vita,}
\emailAdd{vita@mit.edu}

\author[5]{and Hua Xing Zhu}
\emailAdd{zhuhx@zju.edu.cn}

\affiliation[1]{Center for Theoretical Physics, Massachusetts Institute of Technology, Cambridge, MA 02139, USA}
\affiliation[2]{Berkeley Center for Theoretical Physics, University of California, Berkeley, CA 94720, USA}
\affiliation[3]{Theoretical Physics Group, Lawrence Berkeley National Laboratory, Berkeley, CA 94720, USA}
\affiliation[4]{Theory Group, Deutsches Elektronen-Synchrotron (DESY), D-22607 Hamburg, Germany\vspace{0.48ex}}
\affiliation[5]{Department of Physics, Zhejiang University, Hangzhou, Zhejiang 310027, China\vspace{0.48ex}}

\abstract{A number of important observables exhibit logarithms in their perturbative
description that are induced by emissions at widely separated rapidities.
These include transverse-momentum ($q_T$) logarithms, logarithms involving heavy-quark or electroweak gauge boson masses, and small-$x$ logarithms. In this paper, we initiate the study of rapidity logarithms, and the associated rapidity divergences, at subleading order in the power expansion.
This is accomplished using the soft collinear effective theory (SCET).
We discuss the structure of subleading-power rapidity divergences and
how to consistently regulate them.
We introduce a new pure rapidity regulator and a corresponding $\overline{\rm MS}$-like scheme, which handles rapidity divergences while maintaining the homogeneity of the power expansion.
We find that power-law rapidity divergences appear at subleading power, which give rise
to derivatives of parton distribution functions.
As a concrete example, we consider the $q_T$ spectrum for color-singlet production,
for which we compute the complete $q_T^2/Q^2$ suppressed power corrections at $\mathcal{O}(\alpha_s)$, including both logarithmic and nonlogarithmic terms.
Our results also represent an important first step towards carrying out a resummation
of subleading-power rapidity logarithms.
}

\maketitle

\section{Introduction}\label{sec:intro}

Observables in quantum field theory that are sensitive to soft and collinear emissions
suffer from potentially large logarithms in their perturbative predictions.
The structure of these logarithms depends on the observable in question.
For a large class of phenomenologically relevant observables,
these logarithms arise from emissions that are widely separated in rapidity, as opposed to, or in addition to, the more standard case of logarithms from a hierarchy of virtualities.
At leading order in the associated power expansion, these rapidity logarithms can be resummed to all orders in $\alpha_s$ using rapidity evolution equations. Historically these include the well-known massive Sudakov form factor \cite{Collins:1989bt},  Collins-Soper \cite{Collins:1981uk,Collins:1981va,Collins:1984kg}, BFKL  \cite{Kuraev:1977fs,Balitsky:1978ic,Lipatov:1985uk},  and rapidity renormalization group \cite{Chiu:2011qc,Chiu:2012ir} equations.

The resummation of such rapidity logarithms is necessary for a number of applications, including the $q_T$ spectrum for small $q_T$ in color-singlet processes (see e.g.~\refscite{Ji:2004wu, Bozzi:2005wk, Becher:2010tm, GarciaEchevarria:2011rb, Chiu:2012ir, Wang:2012xs, Neill:2015roa, Ebert:2016gcn, Bizon:2017rah, Chen:2018pzu, Bizon:2018foh}), double parton scattering (see e.g.~\refscite{Diehl:2011yj, Manohar:2012jr, Buffing:2017mqm}), jet-veto resummation (see e.g.~\refscite{Banfi:2012jm, Becher:2013xia, Stewart:2013faa}), recoil sensitive event-shape observables (see e.g.~\refscite{Dokshitzer:1998kz, Becher:2012qc, Larkoski:2014uqa, Moult:2018jzp}), multi-differential observables (see e.g.~\refscite{Laenen:2000ij,Procura:2014cba, Marzani:2015oyb, Lustermans:2016nvk, Muselli:2017bad, Hornig:2017pud, Kang:2018agv, Michel:2018hui}), processes involving massive quarks or gauge bosons (see e.g.~\refscite{Ciafaloni:1998xg, Fadin:1999bq, Kuhn:1999nn, Chiu:2007yn,Chiu:2009mg, Gritschacher:2013pha, Hoang:2015vua, Pietrulewicz:2017gxc}), and small-$x$ resummations that go beyond the simplest applications of BFKL (see e.g.~\refscite{Catani:1990eg,Balitsky:1995ub,Kovchegov:1999yj,JalilianMarian:1996xn,JalilianMarian:1997gr,Iancu:2001ad}).
In all these cases, the resummation was performed at leading power (LP), and at present very little is known about the structure of rapidity logarithms and their associated evolution equations at subleading power.

There has been significant interest and progress in studying power corrections \cite{Manohar:2002fd, Beneke:2002ph, Pirjol:2002km, Beneke:2002ni, Bauer:2003mga,Laenen:2008gt,Laenen:2010uz,Larkoski:2014bxa} both in the context of $B$-physics (see e.g.~\refscite{Mantry:2003uz,Hill:2004if, Mannel:2004as, Lee:2004ja, Bosch:2004cb, Beneke:2004in, Tackmann:2005ub, Trott:2005vw,Paz:2009ut,Benzke:2010js}) and for collider-physics
cross sections (see e.g.~\refscite{Dokshitzer:2005bf,Laenen:2008ux,Freedman:2013vya,Freedman:2014uta,Bonocore:2014wua,Bonocore:2015esa,Kolodrubetz:2016uim,Bonocore:2016awd,Moult:2016fqy,Boughezal:2016zws,DelDuca:2017twk,Balitsky:2017flc,Moult:2017jsg,Goerke:2017lei,Balitsky:2017gis,Beneke:2017ztn,Feige:2017zci,Moult:2017rpl,Chang:2017atu,Boughezal:2018mvf,Ebert:2018lzn,Bahjat-Abbas:2018hpv}). Recently, progress has been made also in understanding the behaviour of matrix elements in the subleading soft and collinear limit \cite{Bhattacharya:2018vph} in the presence of multiple collinear directions using spinor-helicity formalism.
In \refcite{Moult:2018jjd} the first all-order resummation at subleading power for collider observables was achieved for a class of power-suppressed kinematic logarithms in thrust including both soft and collinear radiation. More recently in~\refcite{Beneke:2018gvs} subleading power logarithms for a class of corrections in the threshold limit have also been resummed. In both cases the subleading power logarithms arise from widely separated virtuality scales, and their resummation make use of effective field theory techniques.
Given the importance of observables involving nontrivial rapidity scales, it is essential to extend these recent subleading-power results to such observables, and more generally, to understand the structure of rapidity logarithms and their evolution equations at subleading power.

In this paper, we initiate the study of rapidity logarithms at subleading power, focusing on their structure in fixed-order perturbation theory. We show how to consistently regularize subleading-power rapidity divergences, and highlight several interesting features regarding their structure. In particular, power-law divergences appear at subleading power, which give nontrivial contributions and must be handled properly. We introduce a new ``pure rapidity'' regulator and an associated ``pure rapidity'' $\overline{\rm MS}$-like renormalization scheme. This procedure is homogeneous in the power expansion, meaning that it does not mix different orders in the power expansion, which significantly simplifies the analysis of subleading power corrections. We envision that it will benefit many applications.

As an application of our formalism, we compute the complete $\cO(\alpha_s)$ power-suppressed contributions for $q_T$ for color-singlet production, which provides a strong check on our regularization procedure.
We find the interesting feature that the appearing power-law rapidity divergences yield derivatives of PDFs in the final cross section. Our results provide an important ingredient for improving the understanding of $q_T$ distributions at next-to-leading power (NLP). They also have immediate practical applications for understanding and improving the performance of fixed-order subtraction schemes based on the $q_T$ observable \cite{Catani:2007vq}.

To systematically organize the power expansion, we use the soft collinear effective theory (SCET) \cite{Bauer:2000ew, Bauer:2000yr, Bauer:2001ct, Bauer:2001yt}, which provides operator and Lagrangian based techniques for studying the power expansion in the soft and collinear limits. The appropriate effective field theory for observables with rapidity divergences is \scetii~\cite{Bauer:2002aj}. In this theory, rapidity logarithms can be systematically resummed using the rapidity renormalization group (RRG) \cite{Chiu:2011qc,Chiu:2012ir} in a similar manner to virtuality logarithms. The results derived here extend the rapidity renormalization procedure to subleading power, and we anticipate that they will enable the resummation of rapidity logarithms at subleading power.

The outline of this paper is as follows. In \sec{rapidity_general}, we give a general discussion of the structure and regularization of rapidity divergences at subleading power.
We highlight the issues appearing for rapidity regulators that are not homogeneous in the power-counting parameter, focusing on the $\eta$ regulator as an explicit example. We then introduce and discuss the pure rapidity regulator, which is homogeneous.
In \sec{qT}, we derive a master formula for the power corrections to the color-singlet $q_T$ spectrum at $\cO(\as)$, highlighting several interesting features of the calculation.
We also give explicit results for Higgs and Drell-Yan production, and
perform a numerical cross check to validate our results. We conclude in \sec{conc}.

\section{Rapidity Divergences and Regularization at Subleading Power}
\label{sec:rapidity_general}

Rapidity divergences naturally arise in the calculation of observables sensitive to the transverse momentum of soft emissions. In a situation where we have a hard interaction scale $Q$ and the relevant transverse momentum $k_T$ of the fields is small compared to that scale, $\lambda \sim k_T / Q \ll 1$, the appropriate effective field theory (EFT) is \scetii~\cite{Bauer:2002aj}, which contains modes with the following momentum scalings
\begin{align} \label{eq:modes_intro}
 &&&n{-}\text{collinear}: \quad k_n \sim Q \, (\lambda^2, 1, \lambda) &&\implies &&k^-/Q \sim 1
\,,\\ \nn
&&&\bn{-}\text{collinear}: \quad k_{\bn} \sim Q \, (1, \lambda^2, \lambda) &&\implies &&k^-/Q \sim \lambda^2
\,,\\\nn
&&&\text{soft}: \hspace{1.8cm} k_s \sim Q \, (\lambda, \lambda, \lambda) &&\implies &&k^-/Q \sim \lambda\,.
\end{align}
Here we have used lightcone coordinates $(n\cdot k, \bn \cdot k, k_\perp) \equiv (k^+, k^-, k_\perp)$, defined with respect to two lightlike reference vectors $n^\mu$ and $\bn^\mu$. For concreteness, we take them to be $n^\mu =(1,0,0,1)$ and $\bn^\mu=(1,0,0,-1)$.
Unlike \sceti\ where the modes are separated in virtuality, in \scetii~the modes in the EFT have the same virtuality, but are distinguished by their longitudinal momentum ($k^+$ or $k^-$), or equivalently, their rapidity $e^{2y_k} = k^-/k^+$. This separation into modes at hierarchical rapidities introduces divergences, which arise when $k^+/k^-\to \infty$ or $k^+/k^-\to 0$  \cite{Collins:1992tv,Manohar:2006nz,Collins:2008ht,Chiu:2012ir,Vladimirov:2017ksc}. These so-called rapidity divergences are not regulated by dimensional regularization, which is boost invariant and therefore cannot distinguish modes that are only separated in rapidity.

Rapidity divergences can be regulated by introducing a rapidity regulator that breaks boost invariance, allowing the modes to be distinguished, and logarithms associated with the different rapidity scales to be resummed. The rapidity divergences cancel between the different sectors of the effective theory, since they are not present in the full theory. They should not be thought of as UV, or IR, but as arising from the factorization in the EFT. By demanding invariance with respect to the regulator, one can derive renormalization group evolution equations (RGEs) in rapidity.
In SCET, a generic approach to rapidity evolution was introduced in \refscite{Chiu:2011qc, Chiu:2012ir}. These rapidity RGEs allow for the resummation of large logarithms associated with hierarchical rapidity scales.

At leading power in the EFT expansion, the structure of rapidity divergences and the associated rapidity renormalization group are well understood by now, and they have been studied to high perturbative orders (see e.g.\ \refcite{Li:2016ctv} at three-loop order). Indeed, in certain specific physical situations involving two lightlike directions, rapidity divergences can be conformally mapped to UV divergences \cite{Hatta:2008st,Caron-Huot:2015bja,Caron-Huot:2016tzz,Vladimirov:2016dll,Vladimirov:2017ksc}, giving a relation between rapidity anomalous dimensions and standard UV anomalous dimensions. However, little is known about the structure of rapidity divergences or their renormalization beyond the leading power.\footnote{For some interesting recent progress for the particular case of the subleading power Regge behavior for massive scattering amplitudes in $\cN=4$ super Yang-Mills theory, see \refcite{Bruser:2018jnc}.}

In this section, we discuss several interesting features of rapidity divergences at subleading power, focusing on the perturbative behavior at next-to-leading order (NLO). At subleading power there are no purely virtual corrections at NLO, and so we will focus on the case of the rapidity regularization of a single real emission, which allow us to identify and resolve a number of subtleties.
After a brief review of the structure of rapidity-divergent integrals at leading power in \sec{rapDivLP}, we discuss additional issues that arise at subleading power in \sec{rapDivNLP}. We discuss in detail the behavior of the $\eta$ regulator at subleading power, highlighting effects that are caused by the fact that it is not homogeneous in the power expansion. In \sec{upsilonreg}, we introduce the pure rapidity regularization, which regulates rapidity instead of longitudinal momentum and which we find to significantly simplify the calculation at subleading power.
Finally, in \sec{distribution}, we discuss the distributional treatment of power-law divergences,
which arise at subleading power.

\subsection{Review of Rapidity Divergences at Leading Power}
\label{sec:rapDivLP}

We begin by reviewing the structure of rapidity divergent integrals at leading power.  As mentioned above, we restrict ourselves to the case of a single on-shell real emission, which suffices at NLO. Defining $\delta_+(k^2)=\theta(k^0) \delta(k^2)$,
its contribution to a cross section sensitive to the transverse momentum $\kt$ of the emission is schematically given by
\begin{align} \label{eq:sigma_schematic}
\df\sigma(\kt) &\sim \frac{2}{k_T^2} \int\df k^0 \df k^z \, \delta_+(k^2)\, g(k)
\nn\\
&= \frac{1}{k_T^2} \int_0^\infty \frac{\df k^-}{k^-}\, g(k)\Big|_{k^+ = k_T^2/k^-}
 = \frac{1}{k_T^2} \int_0^\infty \frac{\df k^+}{k^+}\, g(k)\Big|_{k^- = k_T^2/k^+}
\,.\end{align}
Here, we have extracted the overall $1/k_T^2$ behaviour, and $g(k)$ is an observable and process dependent function, containing the remaining phase-space factors and amplitudes.
The precise form of $g(k)$ is unimportant, except for the fact that it includes
kinematic constraints on the integration range of $k^\pm$,
\begin{equation} \label{eq:kin_constr_toy}
 g(k) \sim \theta(k^\pm - k_\text{min}^\pm)\, \theta( k_\text{max}^\pm - k^\pm)
\,.\end{equation}
For our discussion we take $k_T > 0$ such that we can work in $d=4$ dimensions.
In the full theory, \eq{sigma_schematic} is finite, with the apparent singularities for $k^\pm \to 0$ or $k^\pm \to \infty$ being cut off by the kinematic constraints in \eq{kin_constr_toy}.
In the effective theory, one expands \eq{sigma_schematic} in the soft and collinear limits specified in \eq{modes_intro}. This expansion also removes the kinematic constraints,
\begin{equation}
	\underbrace{\,k_\text{min}^\pm \to 0\,}_{\text{soft and collinear limits}}\,,\qquad \underbrace{\,k_\text{max}^\pm \to +\infty\,}_{\text{soft limit}}\,,
\end{equation}
such that individual soft and collinear contributions acquire explicit divergences as $k^\pm \to 0$ or $k^\pm \to \infty$. This is actually advantageous, since the associated logarithms can now be tracked by these divergences. To regulate them, we introduce a regulator $R(k,\eta)$, where $\eta$ is a parameter such that $\lim_{\eta\to0}R(k,\eta)=1$. By construction, inserting $R(k,\eta)$ under the integral in \eq{sigma_schematic} does not affect the value of $\df\sigma(\kt)$ when taking $\eta\to0$ in the full calculation.
To describe the limit $k_T \ll Q$, we expand \eq{sigma_schematic} in the soft and collinear limits described by the modes in \eq{modes_intro}.
To be specific, the soft limit of \eq{sigma_schematic} is obtained by evaluating the integrand together with the regulator $R(k,\eta)$ using the soft scaling $k_s$ of \eq{modes_intro}, and expanding in $\lambda$,
\begin{align} \label{eq:sigma_schematic_soft}
\df\sigma_s(\kt)
&\sim \frac{1}{k_T^2} \int_0^\infty\! \frac{\df k_s^-}{k_s^-}\, g(k_s)\Big|_{k_s^+ = k_T^2/k_s^-}\, R(k_s,\eta)
\nn \\
&= \frac{1}{k_T^2} \int_0^\infty\! \frac{\df k^-}{k^-}\, g_s(0)\, R(k, \eta)
\times \bigl[ 1 + \cO(\lambda) \bigr]
\,.\end{align}
Since the leading-power result must scale like $1/k_T^2$, the
LP soft limit $g_s(k^\mu=0)$ must be a pure constant, which implies that the kinematic constraints in \eq{kin_constr_toy} are removed.
This introduces the aforementioned divergences as $k^-\to0$ or $k^-\to\infty$, which are now regulated by $R(k,\eta)$.

The analogous expansion in the collinear sectors is obtained by inserting the $k_n$ or $k_\bn$ scalings of \eq{modes_intro} into \eq{sigma_schematic}, and expanding in $\lambda$,
\begin{align} \label{eq:sigma_schematic_collinear}
\df\sigma_n(\kt)
&\sim \frac{1}{k_T^2} \int_0^\infty\! \frac{\df k_n^-}{k_n^-}\, g(k_n)\Big|_{k_n^+ = k_T^2/k_n^-}\, R(k_n,\eta)
\nn\\
&= \frac{1}{k_T^2} \int_0^Q\! \frac{\df k^-}{k^-}\, g_n\biggl(\frac{k^-}{Q}\biggr)\, R(k,\eta)
\times \bigl[ 1 + \cO(\lambda) \bigr]
\,,\nn\\
\df\sigma_\bn(\kt)
&\sim \frac{1}{k_T^2} \int_0^\infty\! \frac{\df k_\bn^+}{k_\bn^+}\, g(k_\bn)\Big|_{k_\bn^- = k_T^2/k_\bn^+}\, R(k_\bn,\eta)
\nn\\
&= \frac{1}{k_T^2} \int_0^Q\! \frac{\df k^+}{k^+}\, g_\bn\biggl(\frac{k^+}{Q}\biggr)\, R(k,\eta)
\times \bigl[ 1 + \cO(\lambda) \bigr]
\,.\end{align}
In this case, only the lower bound on $k^\pm$ is removed by the power expansion, while the upper limit is given by the relevant hard scale $Q$.
The expansion of $g(k_n)$ in the collinear limit can still depend on the momentum $k^-/Q \sim \cO(\lambda^0)$, as indicated by the functional form of $g_n(k^-/Q)$, and likewise for the $\bn$-collinear limit.

Without the rapidity regulator, the integrals in \eqs{sigma_schematic_soft}{sigma_schematic_collinear} exhibit a logarithmic divergence as $k^\pm \to 0$ or $k^\pm \to \infty$, which is not regulated by dimensional regularization or any other invariant-mass regulator. Since $k^+ k^- = k_T^2$ is fixed by the measurement, this corresponds to a divergence as the rapidity $y_k = (1/2) \ln(k^-/k^+) \to \pm\infty$.
The rapidity regulator $R(k,\eta)$ regulates these divergence by distinguishing the soft and collinear modes. To ensure a cancellation of rapidity divergences in the effective theory, it should be defined  as a function valid on a full-theory momentum $k$, which can then be expanded in the soft or collinear limits. Since there are no divergences in the full theory, this guarantees the cancellation of divergences in the EFT expansion.

At leading power a variety of regulators have been proposed. Since the divergences are only logarithmic, and the focus has not been on higher orders in the power expansion, there are not many constraints from maintaining the power counting of the EFT. Therefore, a variety of regulators have been used, including hard cutoffs \cite{Balitsky:1995ub,JalilianMarian:1997gr,Kovchegov:1999yj,Manohar:2006nz}, tilting Wilson lines off the lightcone \cite{Collins:1350496},  the delta regulator \cite{Chiu:2009yx}, the $\eta$ regulator \cite{Chiu:2011qc,Chiu:2012ir}, the analytic regulator \cite{Beneke:2003pa,Chiu:2007yn,Becher:2011dz}, and the exponential regulator \cite{Li:2016axz}.

At subleading power, we will discuss in more detail the application of the $\eta$ regulator, which can be formulated at the operator level by modifying the Wilson lines appearing in the SCET fields as \cite{Chiu:2011qc,Chiu:2012ir}
\begin{align} \label{eq:reg_soft}
S_n(x)&= \sum\limits_{\text{perms}} \exp \biggl[ -\frac{g}{n\cdot \cP} \frac{w\, |2\,\cP^z|^{-\eta/2}}{\nu^{-\eta/2}}\, n \cdot A_s \biggr] \,, \\
\label{eq:reg_coll}
W_{n}(x) &= \sum\limits_{\text{perms}} \exp \biggl[  -\frac{g}{ \bar n\cdot \cP} \frac{w^2 \,|2\,\cP^z|^{-\eta}}{\nu^{-\eta}}\,  \bar n \cdot A_{n} \biggr] \,,
\end{align}
where $S_n$ and $W_n$ are soft and collinear Wilson lines.
The operator $\cP$ picks out the large (label) momentum flowing into the Wilson line, $\nu$ is a rapidity regularization scale, $\eta$ a parameter exposing the rapidity divergences as $1/\eta$ poles, and $w$ a bookkeeping parameter obeying
\begin{align}\label{eq:omegadef}
\nu\frac{\partial w(\nu)}{\partial\nu} = -\frac{\eta}{2} w(\nu)\,,\qquad\lim_{\eta \to 0} w(\nu)=1
\,.\end{align}
Note that at leading power, one can replace $|2\cP^z| \to |\bn \cdot \cP|$ in \eq{reg_coll},
as employed in \refscite{Chiu:2011qc,Chiu:2012ir},
while at subleading power we will show that this distinction is actually important.
The $\eta$ regulator was extended in \refcite{Rothstein:2016bsq} to also regulate Glauber exchanges in forward scattering, where regulating Wilson lines alone does not suffice.

\subsection{Rapidity Regularization at Subleading Power}
\label{sec:rapDivNLP}

We now extend our discussion to subleading power, where we will find several new features.
First, while at leading power, rapidity divergences arise only from gluons,  at subleading power rapidity divergences can arise also from soft quarks.  Soft quarks have also been rapidity-regulated to derive the quark Regge trajectory \cite{Moult:2017xpp}.
Here, since we consider only the case of a single real emission crossing the cut, this simply means that we must regulate both quarks and gluons. More generally, one would have to apply a rapidity regulator to all operators in the EFT, as has been done for the case of forward scattering in \refcite{Rothstein:2016bsq}. It would be interesting to understand if these subleading rapidity divergences can also be conformally mapped to UV divergences of matrix elements, as was done for the rapidity divergences in the leading power $q_T$ soft function in \refscite{Vladimirov:2016dll,Vladimirov:2017ksc}.

Second, the structure of rapidity divergences becomes much richer at subleading power, placing additional constraints on the form of the rapidity regulator to maintain a simple power expansion. This more interesting divergence structure follows directly from power counting.
For example, the subleading corrections to the soft limit can be obtained by expanding the integrand in \eq{sigma_schematic_soft} to higher orders in $\lambda$.
The power counting for soft modes in \eq{modes_intro} implies that the first $\cO(\lambda)$ power suppression can only be given by additional factors of $k^-/Q$ or $k^+/Q$ in \eq{sigma_schematic_soft}.
At the next order, $\cO(\lambda^2)$, one can encounter additional factors $(k^+/Q)^2, (k^-/Q)^2$.
The possible structure of rapidity-divergent integrals in the soft limit up to $\cO(\lambda^2)$ is thus given by%
\footnote{We can also have integrals with an additional factor of $k_T/Q$ or $k^2_T/Q^2$, which however do not change the structure of the integrand and can thus be treated with the same techniques as at leading power.}
\begin{align}\label{eq:sublintegrals}
 \cO(\lambda^0): \qquad
 & \int_0^\infty \frac{\df k^-}{k^-} \softregulator
\,,\\\nn
 \cO(\lambda^1): \qquad
 & \int_0^\infty \frac{\df k^-}{k^-} \biggl(\frac{k^-}{Q}\biggr) \softregulator
 \,,\quad
 \int_0^\infty \frac{\df k^-}{k^-} \biggl(\frac{k^+}{Q}\biggr) \softregulator
\,,\\\nn
 \cO(\lambda^2): \qquad
 & \int_0^\infty \frac{\df k^-}{k^-} \biggl(\frac{k^-}{Q}\biggr)^2 \softregulator
 \,, \quad
 \int_0^\infty \frac{\df k^-}{k^-} \biggl(\frac{k^+}{Q}\biggr)^2 \softregulator
\,,\end{align}
where it is understood that $k^+ = k_T^2 / k^-$.
We can see that the $\cO(\lambda^0)$ limit only produces logarithmic divergences,
while the power-suppressed corrections give rise to power-law divergences.
The prototypical rapidity-divergent integral encountered in the soft limit is thus given by
\begin{align}\label{eq:toy_integrals_soft}
\Isoftgen{\alpha}
&= \int_0^\infty \frac{\df k^-}{k^-} \biggl(\frac{k^-}{Q}\biggr)^\alpha \softregulator
\,,\end{align}
where $\alpha$ counts the additional powers of $k^-$.

A similar situation occurs in the collinear sectors.
In the $n$-collinear limit, $k \sim Q(\lambda^2, 1, \lambda)$, the large momentum $k^-$ is not suppressed with respect to $Q$, such that the power suppression can only arise from explicit factors of $k_T^2$.
(Of course, $k^+ \sim \cO(\lambda^2)$ can also give a suppression, but it can always be reduced back to $k^+ = k_T^2 / k^-$.)
Similarly, in the $\bn$-collinear limit $k^+$ is unsuppressed, and power suppressions only arise from $k_T^2$.
However, the structure of the collinear expansion of $g(k)$ is richer than in the soft case, because there is always a nontrivial dependence on the respective unsuppressed ratio $k^\mp/Q$.
To understand this intuitively, consider the splitting of a $n$-collinear particle into two on-shell $n$-collinear particles with momenta
\begin{align}
 p_1^\mu&=(Q-k^-)\frac{n^\mu}{2}+k_\perp^\mu+\frac{k_T^2}{Q-k^-} \frac{\bar n^\mu}{2}
\,,\qquad
 p_2^\mu=k^-\frac{n^\mu}{2}-k_\perp^\mu+\frac{k_T^2}{k^-} \frac{\bar n^\mu}{2}
\,.\end{align}
The associated Lorentz-invariant kinematic variable is given by
\begin{align}
s_{12} = (p_1 + p_2)^2 = \frac{k_T^2 Q^2}{k^-(Q-k^-)}\,.
\end{align}
Expanding any function of $s_{12}$ in $k_T$ thus gives rise to additional factors of the large momentum $k^-$.
Thus, in general, expanding $g(k_n)$ in the collinear limit can give rise to both positive and negative powers of $k^-$ that accompany the power-suppression in $k_T^2$.
These factors are of course not completely independent, as the sum of all soft and collinear contributions must be rapidity finite, i.e., any rapidity divergences induced by these additional powers of $k^-$ must in the end cancel against corresponding divergences in the soft and/or other collinear contributions.
In summary, the generic form of integrals in the collinear expansion is given by
\begin{align}\label{eq:toy_integrals_collinear}
\Ingen{\alpha}
&= \int_0^Q \frac{\df k^-}{k^-} \biggl(\frac{k^-}{Q}\biggr)^\alpha g_n \biggl(\frac{k^-}{Q}\biggr) \nregulator
\,, \\
\Ibngen{\alpha}
&=\int_0^Q \frac{\df k^+}{k^+} \biggl(\frac{k^+}{Q}\biggr)^\alpha g_\bn \biggl(\frac{k^+}{Q} \biggr)\bnregulator
\,.\end{align}
Here, $g_n(x)$ and $g_\bn(x)$ are regular functions as $x \to 0$.
At LP, only $\alpha=0$ contributes, which gives rise to logarithmic divergences,
while at subleading power for $\alpha\neq 0$ we again encounter power-law divergences.
As we will see in \sec{distribution}, these power-law divergences have a nontrivial effect,
namely they lead to derivatives of PDFs in the perturbative expansion for hadron collider processes.

The presence of power-law divergences at subleading power also implies that more care must be taken to ensure that the regulator does not unnecessarily complicate the power counting of the EFT. For example, with the exponential regulator \cite{Li:2016axz}, or with a hard cutoff, power-law divergences lead to the appearance of powers of the regulator scale, and hence break the homogeneity of the power expansion of the theory.

Furthermore, at leading power one also has the freedom to introduce and then drop subleading terms to simplify any stage of the calculation. While this may seem a general feature and not appear very related to the regularization of rapidity divergences, we will see in a moment that this freedom, explicitly or not, is actually used in most of the rapidity regulators in the literature.

In summary, having a convenient-to-use regulator at subleading power imposes stronger constraints than at leading power.
In particular, we find that the regulator
\begin{itemize}
 \item must be able to regulate not only Wilson lines, but all operators, including those generating soft quark emissions,
 \item must be able to deal not only with logarithmic divergences, but also with power-law divergences without violating the power counting of the EFT by inducing power-law mixing,
 \item and should be homogeneous in the power-counting parameter $\lambda$ to minimize mixing between different powers.
\end{itemize}
The first requirement means one cannot use regulators acting only on Wilson lines, such as taking Wilson lines off the light-cone as in~\refcite{Collins:1350496}, the $\delta$ regulator as used in \refscite{Chiu:2009yx,GarciaEchevarria:2011rb}, and the $\eta$ regulator as used in \refscite{Chiu:2011qc,Chiu:2012ir}, while the $\eta$ regulator as modified and employed in \refscite{Rothstein:2016bsq,Moult:2017xpp} and the analytic regulator of \refcite{Becher:2011dz} can be used.
The second requirement is satisfied by all dimensional regularization type regulators, such as the $\eta$ regulator or analytic regulator, but not by those that are more like a hard cutoff, including the exponential regulator \cite{Li:2016axz}.
To highlight the last point, in the following we discuss in more detail the properties of the $\eta$ regulator at subleading power.

\subsubsection[The \texorpdfstring{$\eta$}{eta} Regulator at Subleading Power]
{\boldmath The $\eta$ Regulator at Subleading Power}
\label{sec:etareg}

In the $\eta$ regulator, one regulates the $k^z$ momentum of emissions through the regulator function
(see \eq{reg_soft})
\begin{equation} \label{eq:Rz}
R_z(k,\eta) = w^2 \biggl|\frac{2 k^z}{\nu}\biggr|^{-\eta} = w^2 \nu^\eta |k^- - k^+|^{-\eta}
\,.\end{equation}
For a single massless emission this corresponds to regulating its phase-space integral as
\begin{equation}
 \int \df^d k\, \delta_+(k^2) \quad\to\quad
 \int\df^d k\, \delta_+(k^2)\, R_z(k,\eta)
 = w^2\nu^\eta \int\df^d k \,\delta_+(k^2)\, |k^- - k^+|^{-\eta}
\,.\end{equation}
In the soft limit $k^+ \sim k^- \sim \lambda Q$, the regulator is homogeneous in $\lambda$ and therefore does not need to be expanded. The prototypical soft integral in \eq{toy_integrals_soft} evaluates to
\begin{align}\label{eq:soft_integrals_a}
\Isoftz{\alpha}
&= w^2 \nu^\eta \int_0^\infty \frac{\df k^-}{k^-} \biggl(\frac{k^-}{Q}\biggr)^\alpha \biggl|k^- - \frac{k_T^2}{k^-}\biggr|^{-\eta}
 \nn\\&
 = w^2 \biggl(\frac{\nu}{k_T}\biggr)^{\eta} \biggl(\frac{k_T}{Q}\biggr)^\alpha \cos\biggl(\frac{\alpha \pi}{2}\biggr)
   \, \sin\biggl(\frac{\eta \pi}{2}\biggr) \,
   \frac{1}{\pi} \Gamma(1-\eta) \Gamma\Bigl(\frac{\eta}{2} - \frac{\alpha}{2}\Bigr)
   \Gamma\Bigl(\frac{\eta}{2} + \frac{\alpha}{2}\Bigr)
\,.\end{align}
Symmetry  under $\alpha \leftrightarrow -\alpha$ implies that
\begin{align}
\Isoftz{-\alpha} =  \biggl(\frac{k_T^2}{Q^2}\biggr)^{-\alpha} \Isoftz{\alpha}
\,.\end{align}
This reflects the symmetry under exchanging $k^- \leftrightarrow k^+$,
which is not broken by the $\eta$ regulator.
One can easily deduce the behavior as $\eta\to0$ from \eq{soft_integrals_a}.
Since $\sin(\eta) \sim \eta$, a pole in $\eta$ can only arise if both $\Gamma$ functions have poles,
which requires $\alpha = 0$.
A finite result is obtained if exactly one $\Gamma$ function yields a pole, which requires $\alpha$ to be even.
For odd $\alpha$, the expression vanishes at $\eta = 0$.
Hence, the exact behavior for $\eta\to0$ is given by
\begin{alignat}{2} \label{eq:soft_integrals}
\Isoftz{0} &= \frac{2}{\eta} + \ln\frac{\nu^2}{k_T^2} + \cO(\eta)
\,, \nn\\
\Isoftz{\alpha} &=0 \qquad &&(\alpha~\mathrm{odd})
\,, \nn\\
\Isoftz{\alpha} &= \frac{2}{|\alpha|}\biggl(\frac{k_T}{Q}\biggr)^\alpha + \cO(\eta)
\qquad &&(\alpha~\mathrm{even})
\,.\end{alignat}
In particular, since the $\eta$ regulator behaves like dimensional regularization,
it is well-behaved for power-law divergences and the soft integrals only give rise
to poles from the logarithmic divergences.

In the collinear sector, the behavior is more complicated at subleading power,
because the regulator factor $2k^z = k^- - k^+$ is not homogeneous in $\lambda$.
At leading power~\cite{Chiu:2011qc,Chiu:2012ir,Rothstein:2016bsq}, one takes advantage of the fact that $2 k^z \to k^-$ in the $n$-collinear limit and $2 k^z \to k^+$ in the $\bn$-collinear limit, so that the expanded result correctly regulates the collinear cases, and makes it symmetric under the exchange $n \leftrightarrow \bn$. A fact that will be important for our analysis is that this power expansion induces higher order terms.  These terms have never been considered in the literature since they are not important at leading power.
However, at subleading power one can no longer neglect the subleading component of the regulator.
Implementing the $\eta$ regulator at subleading power in the collinear limits
thus requires to expand the regulator \eq{Rz} itself,
\begin{align} \label{eq:Rz_expanded}
R_z(k_n,\eta) &= w^2 \nu^{\eta} \biggl| k_n^- - \frac{k_T^2}{k_n^-} \biggl|^{-{\eta}}
 = w^2 \, \biggl|\frac{k_n^-}{\nu}\biggl|^{-\eta} \, \biggl[1 + {\eta}\, \frac{k_T^2}{(k_n^-)^2} +\cO(\lambda^4) \biggr]
\,, \nn\\
 R_z(k_\bn,\eta) &= w^2 \biggl| \frac{k_\bn^+}{\nu} \biggr|^{-\eta} \biggl[1 + {\eta}\, \frac{k_T^2}{(k_\bn^+)^2} +\cO(\lambda^4) \biggr]
\,.\end{align}
Applying this to the general LP integral in the $n$-collinear sector, \eq{toy_integrals_collinear} with $\alpha=0$, we obtain
\begin{align}\label{eq:218}
\Inz{0}
&= w^2 \int_0^Q \frac{\df k^-}{k^-} \biggl|\frac{k^-}{\nu}\biggl|^{-\eta} g_n \biggl(\frac{k^-}{Q}\biggr)
\nn\\ & \quad
+ \eta\, w^2\, \frac{k_T^2}{Q^2} \int_0^Q \frac{\df k^-}{k^-}  \biggl|\frac{k^-}{\nu}\biggl|^{-\eta} \biggl(\frac{k^-}{Q}\biggr)^{-2} g_n \biggl(\frac{k^-}{Q}\biggr)
+ \cO(\lambda^4)
\,,\end{align}
and analogously for $\Ibnz{0}$.
Here, the first line is the standard LP integral, while the second line arises from expanding the regulator and is suppressed by $k_T^2/Q^2 \sim \lambda^2$.
While it is also proportional to $\eta$, the remaining integral can produce a $1/\eta$ rapidity divergence to yield an overall finite contribution.

In \sec{qT}, we will see explicitly that these terms from expanding the regulator are crucial to obtain the correct final result at subleading power.
However, in practice they are cumbersome to track in the calculation and yield complicated structures.
To establish an all-orders factorization theorem, the mixing of different orders in the power expansion due to the regulator becomes a serious complication. Hence, it is desirable to employ a rapidity regulator that is homogeneous in $\lambda$. We will present such a regulator in the following \sec{upsilonreg}.

\subsection{Pure Rapidity Regularization}
\label{sec:upsilonreg}

We wish to establish a rapidity regulator that is homogeneous at leading power such that it does not mix LP and NLP integrals, as observed in \sec{etareg} for the $\eta$ regulator.
This can be achieved by implementing the regulator similar to the $\eta$ regulator of \refscite{Chiu:2011qc,Chiu:2012ir,Rothstein:2016bsq}, but instead of regulating the momentum $k^z$ with factors of $w |2k^z/\nu|^{-\eta/2}$, one regulates the rapidity $y_k$ of the momentum $k^\mu$, where
\begin{align}
y_k \equiv \frac{1}{2}\ln\frac{\bn\cdot k}{n\cdot k }
\,.\end{align}
To implement a regulator involving rapidity we use\footnote{
	Note that we can implement the pure rapidity regulator in terms of label and residual momentum operators for example as
	\begin{equation}
	w^2\, \upsilon^{\eta}\,
	\biggl|\frac{\bn\cdot(\cP + \partial)}{n\cdot(\cP + \partial)}\biggr|^{-\eta/2}
	\,.\end{equation}
	where the label momentum operator $\cP$ picks out the large $\cO(\lambda^0)$ momentum component
	of the operator it acts on, while $\partial$ picks out the $\cO(\lambda)$ or $\cO(\lambda^2)$ components. 
	In this case, the operator
	\begin{align}
	\hat Y = \frac{1}{2}\ln\frac{\bn\cdot(\cP + \partial)}{n\cdot(\cP + \partial)}
	\end{align}
	picks out the rapidity of the operator it acts on.
}
factors of
\begin{equation}\label{eq:def_of_upsilon}
w^2\upsilon^{\eta}\left|\frac{\bn\cdot k}{n\cdot k}\right|^{-\eta/2} 
 = w^2 \upsilon^{\eta} e^{-y_k\eta}
\,.\end{equation}
Here we have defined a rapidity scale $\upsilon$ (\verb|\upsilon|)
which is the analog of the scale $\nu$ (\verb|\nu|) in the $\eta$ regulator.
Although $\upsilon$ is dimensionless, in contrast to the dimensionful $\nu$, it still shares the same properties as pure dimensional regularization. In particular, it will give rise to poles in $\eta$ that can be absorbed in $\MS$-like rapidity counterterms.
To ensure $\upsilon$ independence of \eq{def_of_upsilon}, we introduced a bookkeeping parameter $w=w(\upsilon)$ in analogy to the bookkeeping parameter $w(\nu)$ in the $\eta$ regulator, see \eq{omegadef} and \refcite{Chiu:2012ir}.
Also note that this regulator does not affect UV renormalization, which in \scetii~arises from transverse momenta going to infinity and thus is orthogonal to regulating rapidity.

We call \eq{def_of_upsilon} the \emph{pure rapidity regulator}, and pure rapidity regularization the procedure of regulating rapidity divergences using \eq{def_of_upsilon}. When only the $1/\eta$ poles are subtracted we then refer to the renormalized result as being in the \emph{pure rapidity renormalization scheme}.

If we want to make the rapidity scale $\upsilon$ into a true rapidity scale $\Upsilon$, then we can change variables as
\begin{align}
\upsilon \equiv e^\Upsilon  \,.
\end{align}
With this definition \eq{def_of_upsilon} becomes
\begin{align}  \label{eq:def_of_CapUpsilon}
w^2 \upsilon^{\eta} e^{-y_k\eta}
\equiv w^2 e^{\eta(\Upsilon - y_k)} 
 \,,
\end{align}
and the factor regulating divergences depends on a rapidity difference between the scale parameter $\Upsilon$ and $y_k$.

It is interesting to consider the behavior of amplitudes regulated with \eq{def_of_CapUpsilon} under a reparameterization transformation known as RPI-III~\cite{Manohar:2002fd}, which takes $n^\mu \to e^{-\beta} n^\mu$ and $\bn^\mu \to e^{\beta} \bn^\mu$ for some, not necessarily infinitesimal, constant $\beta$.
For a single collinear sector, this can be interpreted as a boost transformation.
Since RPI transformations can be applied independently for each set of collinear basis vectors $\{n_i,\bar n_i\}$ they in general constitute a broader class of symmetry transformations in SCET. 
Prior to including a regulator for rapidity divergences all complete SCET amplitudes are invariant under such transformations. 
All previous rapidity regulators violate this symmetry.
For the pure rapidity regulator in \eq{def_of_CapUpsilon} we have $y_k\to y_k+\beta$, so the transformation is quite simple.%
\footnote{
	Any operators that are defined such that they transform under RPI-III, will do so by a factor $e^{k\beta}$, where $k$ is their RPI-III charge.
	The pure rapidity regulator therefore has an RPI-III charge of $-\eta$. 
	This leads to rapidity-renormalized collinear and soft functions in SCET which carry this charge.
	When considering any observable like a cross section, the combined charge of the renormalized functions describing this observable is zero. 
} 
It can be compensated by defining the rapidity scale to transform like a rapidity,
$\Upsilon \to \Upsilon+\beta$.
Therefore, the $\upsilon^\eta$ factor in the regulator does for RPI-III what the usual
$\mu^\epsilon$ factor does for the mass-dimensionality in dimensional regularization.

As an example of the application of this new regulator, we consider again a real emission with momentum $k^\mu$. The regulator function $R(k,\eta)$ that follows from \eq{def_of_upsilon} is given by
\begin{align} \label{eq:vita_regulatorrapidity}
R_Y(k,\eta) &= w^2\, \upsilon^{\eta}\, \biggl|\frac{k^-}{k^+} \biggr|^{-\eta/2}
 = w^2\, \upsilon^{\eta}\, e^{-\eta\, y_k}
\,.\end{align}
The real-emission phase space is then regulated as
\begin{align}
\int \df^d k\, \delta_+(k^2) &\quad\to\quad  \int \df^d k \,\delta_+(k^2)\,  R_Y(k,\eta)
 =  \int \df^d k\, \delta_+(k^2) \, w^2\, \upsilon^{\eta}\, e^{-\eta\, y_k}
\,.\end{align}
A peculiar feature of the pure rapidity regulator is that it renders the prototypical soft integrals scaleless such that they vanish. That is, using \eq{vita_regulatorrapidity} in
\eq{toy_integrals_soft}, we obtain
\begin{align}\label{eq:vanishingsoft}
\IsoftY{\alpha}
&= \int_0^\infty \frac{\df k^-}{k^-} \biggl(\frac{k^-}{Q}\biggr)^\alpha R_Y(k,\eta)
= w^2\, \upsilon^\eta\, k_T^\eta\, Q^{-\alpha} \int_0^\infty \df k^-\, (k^-)^{\alpha - \eta - 1} = 0
\,.\end{align}
The final integrals are scaleless and vanish for all integer values of $\alpha$, just like scaleless integrals vanish in dimensional regularization.\footnote{Technically one can find terms of the form $1/\eta - 1/\eta$, which can be set to zero via analytic continuation in the standard manner.}

Considering the collinear sectors, the prototypical collinear integrals in \eq{toy_integrals_collinear} with $R_Y(k,\eta)$ become
\begin{align} \label{eq:RYcollinear}
\InY{\alpha}
&= w^2\, \upsilon^\eta\, k_T^{+\eta}\,Q^{-\alpha}
\int_0^Q \df k^-\, (k^-)^{\alpha-\eta-1} \, g_n \biggl(\frac{k^-}{Q}\biggr)
\,,\nn\\
\IbnY{\alpha}
&= w^2\, \upsilon^\eta\, k_T^{-\eta}\,Q^{-\alpha}
\int_0^Q \df k^+\, (k^+)^{\alpha+\eta-1} \, g_\bn \biggl(\frac{k^+}{Q} \biggr)
\,.\end{align}
Although the regulator does not act symmetrically in the $n$-collinear and $\bn$-collinear sectors, the asymmetry is easy to track by taking $\eta \leftrightarrow -\eta$ and $\upsilon\leftrightarrow 1/\upsilon$ when swapping $n\leftrightarrow \bn$ and $k^+\leftrightarrow k^-$.
Since $R_Y(k,\eta)$ is homogeneous in $\lambda$, it does not generate any subleading power terms, in contrast to \eq{218} for the $\eta$ regulator. In particular, the LP integral becomes
\begin{align}
\InY{0}
&= w^2\, \upsilon^\eta\, k_T^{+\eta}
\int_0^Q \frac{\df k^-}{(k^-)^{1+\eta}} \, g_n \biggl(\frac{k^-}{Q}\biggr)
\nn \\*
&= w^2\, \Bigl(\upsilon \frac{k_T}{Q}\Bigr)^\eta
\int_0^Q \df k^- \biggl[-\frac{1}{\eta}\, \delta(k^-) + \frac{1}{Q} \cL_0\biggl(\frac{k^-}{Q}\biggr)\,
 + \ord{\eta} \biggr]g_n \biggl(\frac{k^-}{Q}\biggr)
\,,\end{align}
where we used the standard distributional identity $1/x^{1+\eta} = -\delta(x)/\eta + \cL_0(x) + \ord{\eta}$ to extract the $1/\eta$ divergence. (See \sec{distribution} below for a more general
discussion.)
Taking $\eta \to -\eta$, the analogous $1/\eta$ pole in the $\bn$-collinear sector
has the opposite sign, such that the $1/\eta$ poles cancel when adding the $n$-collinear and $\bn$-collinear contributions.
This is a general feature in all cases where the soft contribution vanishes as in \eq{vanishingsoft}.

Some comments about the features of the pure rapidity regulator are in order:
\begin{itemize}
  \item It involves the rapidity
  \begin{equation}
   e^{2y_k} \equiv \frac{\bn\cdot k}{n\cdot k}\,,
   \end{equation}
  and therefore breaks boost invariance as required to regulate rapidity divergences. The boost invariance is restored by the dimensionless $\upsilon$ rapidity scale, analogous to how the dimensionful mass scale $\mu$ in dimensional regularization restores the dimensionality.
  \item Rapidity divergences appear as $1/\eta$ poles, allowing the definition of the pure rapidity renormalization scheme as a dimensional regularization-like scheme.
  \item At each order in perturbation theory, the poles in $\eta$ and the $\upsilon$-dependent pieces cancel when combining the results for the $n$-collinear, $\bn$-collinear, and soft sectors.
  \item The pure rapidity regulator is homogeneous\footnote{In cases where it is possible to combine label and residual momenta in the phase space integral that needs to be rapidity regulated.} in the SCET power counting parameter $\lambda$. Therefore it does not need to be power expanded, and hence does not mix contributions at different orders in the power expansion.
  \item For the case of a single real emission considered here:
  \begin{itemize}
     \item Soft integrals and zero-bin \cite{Manohar:2006nz} integrals are scaleless and vanish.
     \item It follows that the $\eta$ poles and the $\upsilon$ dependent pieces cancel between the $n$-collinear and $\bn$-collinear sectors.
     \item The results for the $n$-collinear and $\bn$-collinear sectors are not identical but are trivially related by taking $\eta \leftrightarrow -\eta$ and $\upsilon\leftrightarrow 1/\upsilon$ when swapping $n\leftrightarrow \bn$.
  \end{itemize}
\end{itemize}
The introduction of this new pure rapidity regulator allows us to regulate rapidity divergences at any order in the EFT power expansion, while maintaining the power counting of the EFT independently at each order.

Although in this paper we will only use pure rapidity regularization for a single real emission at fixed order, we note that one can derive a rapidity renormalization group for the pure rapidity regulator by imposing that the cross section must be independent of $\upsilon$.
Similar to the $\eta$ regulator, this regulator is not analytical and can also be used to properly regulate virtual and massive loops. This will be discussed in detail elsewhere.

To conclude this section we note that the pure rapidity regulator can be seen as a particular case of a broader class of homogeneous rapidity regulators given by
\begin{align} \label{eq:vita_regulator}
R_c(k,\eta)
= w^2\, \upsilon^{(1-c)\eta/2}\,
\biggl|\frac{k^-}{\nu}\biggr|^{-\eta/2} \biggl|\frac{k^+}{\nu}\biggr|^{-c \eta/2}
\,,\end{align}
where $c\ne1$ is an arbitrary parameter governing the antisymmetry between the $n$-collinear and $\bn$-collinear sectors.
As for the pure rapidity regulator, this regulator is homogeneous in $\lambda$ and renders the same class of soft integrals scaleless.
However, it requires an explicit dimensionful scale $\nu$ to have the correct mass dimension.
Note that for $c=1$, \eq{vita_regulator} only depends on the boost invariant product $k^+k^-$ and therefore does not regulate rapidity divergences.
For $c=-1$, it recovers the pure rapidity regulator and the dependence on $\nu$ cancels.
Lastly, for $c=0$ and massless real emissions, \eq{vita_regulator} essentially reduces to the regulator of \refcite{Becher:2011dz}.
We choose $c=-1$ because it yields the same finite terms in the $n$-collinear and $\bn$-collinear functions, and thus has enhanced symmetry.
Choosing a different value of $c$ shifts terms between the two sectors, see also \app{master_formula_c}. The combined result is always independent of $c$.

\subsection{Distributional Treatment of Power Law Divergences}
\label{sec:distribution}

To complete our treatment of rapidity divergences at subleading power, we show how their distributional structure can be consistently treated when expanded against a general test function.
In particular, we will see that the power-law rapidity divergences lead to derivatives of PDFs. 

In the collinear limit at NLP, we obtain divergent integrals of the form
\begin{equation} \label{eq:toy_dist}
 \int_0^Q \frac{\df k^-}{Q} \frac{g_n(k^-/Q)}{(k^-/Q)^{a + \eta}}
\,,\end{equation}
which appear for both the $\eta$ regulator (with $a=1-\alpha=1,2,3$ at NLO) and the pure rapidity regulator (with $a=1-\alpha=1,2$ at NLO).

The function $g_n(k^-/Q)$ is defined to be regular for $k^-/Q\to 0$.
If it is known analytically, we can in principle evaluate the integral in \eq{toy_dist} analytically and expand the result for $\eta \to 0$ to obtain the regularized expression.
However, $g_n(k^-/Q)$ is typically not given in analytic form.
In particular, for $pp$ collisions it contains the parton distribution functions (PDFs) $f(x)$.
Therefore, to extract the rapidity divergence, we need to expand $1/{(k^-)^{a + \eta}}$ in $\eta$ in a distributional sense.
To do so, we first change the integration variable from $k^-$ to the dimensionless variable $z$ defined through $k^- = Q (1-z)$,
such that \eq{toy_dist} becomes
\begin{align} \label{eq:toy_dist2}
\int_0^1 \df z\,\frac{\tilde g(z)}{(1-z)^{a+\eta}} \,,\qquad \tilde g(z) = g_n(1-z)
\,.\end{align}
In \eq{toy_dist2}, the rapidity divergence arises as $z\to1$.
For $a=1$, it can be extracted using the standard distributional identity
\begin{align} \label{eq:plus_dist_1}
 \frac{1}{(1-z)^{1+\eta}} = - \frac{\delta(1-z)}{\eta} + \cL_0(1-z) + \cO(\eta)
\,,\end{align}
where $\cL_0(y) = [\theta(y)/y]_+$ is the standard plus distribution and we remind the reader that its convolution against a test function $\tilde g(z)$ is given by
\begin{align}
 \int_x^1 \df z\, \tilde g(z) \cL_0(1-z) = \int_x^1 \df z \, \frac{\tilde g(z) - \tilde g(1)}{1-z} + \tilde g(1) \underbrace{\int_x^1 \df z \, \cL_0(1-z)}_{\ln(1-x)}
 \,,\quad x \in [0,1]
\,.\end{align}
For $a>1$, these distributions need to be generalized to higher-order plus distributions subtracting higher derivatives as well.
For example, for $a=2$ one obtains
\begin{align} \label{eq:double_plus}
  \frac{1}{(1-z)^{2+\eta}} &
 = \frac{\delta'(1-z)}{\eta} - \delta(1-z) + \cL_0^{++}(1-z) + \cO(\eta)
\,,\end{align}
where the second-order plus function $\cL_0^{++}(1-z)$ regulates the quadratic divergence $1/(1-z)^2$.
Its action on a test function $\tilde g(z)$ is given by a double subtraction,
\begin{align}
 \int_x^1 \df z \, \tilde g(z) \cL_0^{++}(1-z) &
 = \int_x^1 \df z \, \frac{\tilde g(z) - [\tilde g(1) + \tilde g'(1)(z-1)]}{(1-z)^2}
 \nn\\*&\quad
 + \tilde g(1) \underbrace{\int_x^1 \df z\, \cL_0^{++}(1-z)}_{-x/(1-x)}
 +\, \tilde g'(1) \underbrace{\int_x^1 \df z\, (z-1)  \cL_0^{++}(1-z)}_{-\ln(1-x)}
\,.\end{align}
In \app{plus_distr}, we give more details on these distributions, generalizing to arbitrary $a \ge 1$.
Note that the second-order plus function has also appeared for example in \refcite{Mateu:2012nk}.

Eq.~\eqref{eq:double_plus} implies the appearance of derivatives of delta functions, $\delta'(1-z)$, which will induce derivatives of the PDFs that are contained in $\tilde g(z)$. The appearance of such derivatives in subleading power calculations was first shown in \refcite{Moult:2016fqy} in the context of \sceti-like observables. However, in such cases they arose simply from a Taylor expansion of the momentum being extracted from the PDF.
Here, they also arise from power-law divergences, a new mechanism to induce derivatives of PDFs.
Recently, power-law divergences inducing derivatives of PDFs have appeared also in the study of \sceti-like observables involving multiple collinear directions at subleading power~\cite{Bhattacharya:2018vph}. We believe they are a general feature of calculations beyond leading power.

In practice, the higher-order distributions can be cumbersome to work with.
Instead, we find it more convenient to use integration-by-parts relations to reduce the divergence in \eq{toy_dist2} to the linear divergence $1/(1-z)$, which yields explicit derivatives of the test function.
For the cases $a=2$ and $a=3$ we encounter in \sec{qT}, this gives
\begin{align} \label{eq:plus_dist_2}
 \int_x^1 \df z \frac{\tilde g(z)}{(1-z)^{2+\eta}} &
 = \tilde g'(1) \biggl(\frac{1}{\eta} - 1\biggr) - \frac{\tilde g(x)}{1-x}
   - \int_x^1 \df z \, \tilde g'(z) \cL_0(1-z) + \cO(\eta)
\,,\\ \label{eq:plus_dist_3}
 \int_x^1 \df z \frac{\tilde g(z)}{(1-z)^{3+\eta}} &
 = \tilde g''(1) \biggl(- \frac{1}{2\eta} + \frac34\biggr) - \frac{\tilde g(x)+(x-1) \tilde g'(x)}{2(1-x)^{2}}
 \nn\\*&\quad + \frac{1}{2} \int_x^1 \df z \, \tilde g''(z) \cL_0(1-z) + \cO(\eta)
\,.\end{align}
\Equations{plus_dist_2}{plus_dist_3} can be used to write the kernels fully in terms of a standard $\cL_0$, but they must be applied within the integral to directly yield derivatives of the test function $\tilde g(z)$.

In our application in \sec{qT}, $\tilde g(z)$ will always involve the PDF $f(x/z)$ and vanish at $z=x$.
We can thus also write \eqs{plus_dist_2}{plus_dist_3} as operator equations,
\begin{align} \label{eq:plus_dist_2_operator}
 \frac{1}{(1-z)^{2+\eta}} &\quad\to\quad
 \biggl[ \biggl(\frac{1}{\eta}-1\biggr) \delta(1-z) - \cL_0(1-z)\biggr] \frac{\df}{\df z}
\,, \\  \label{eq:plus_dist_3_operator}
 \frac{1}{(1-z)^{3+\eta}} &\quad\to\quad
 \biggl[ \delta(1-z) \biggl(- \frac{1}{2 \eta} + \frac{3}{4} \biggr) + \frac{1}{2} \cL_0(1-z) \biggr] \frac{\df^2}{\df^2 z}
 + \frac{\tilde g'(x)}{2(1-x)} \delta(1-z)
\,.\end{align}
Note that the second relation is quite peculiar, as we have to add the boundary term proportional to $g'(x)$,
and thus cannot be interpreted as a distributional relation.
In our calculation in \sec{qT}, this term will not contribute due to an overall suppression by $\eta$, such that only the divergent term in \eq{plus_dist_3_operator} needs to be kept.

\section{Power Corrections for Color-Singlet \texorpdfstring{\boldmath $q_T$}{qT} Spectra}
\label{sec:qT}

In this section we use our understanding of rapidity regularization at subleading power
to compute the perturbative power corrections to the transverse momentum $q_T$ in color-singlet production
at invariant mass $Q$, which is one of the most well studied observables in QCD.
Schematically, the cross section differential in $q_T$ can be expanded as
\begin{align}
 \frac{\df\sigma}{\df q_T^2} = \frac{\df\sigma^{(0)}}{\df q_T^2} + \frac{\df\sigma^{(2)}}{\df q_T^2} + \cdots
\,,\end{align}
where $\sigma^{(0)}$ is the leading-power cross section and $\sigma^{(2n)}$ the N$^n$LP cross section.
In general, in this section we will denote power suppression in $\cO(\lambda^n)$ with $\lambda \sim q_T/Q$ relative to the leading-power result through superscripts $^{(n)}$.
The $\sigma^{(2n)}$ terms scale like
\begin{align}
 \frac{\df\sigma^{(2n)}}{\df q_T^2} \sim \frac{1}{q_T^2} \biggl(\frac{q_T^2}{Q^2}\biggl)^{n}
\,,\end{align}
and hence only the LP cross section is singular as $q_T \to 0$.
In particular, $\sigma^{(0)}$ contains Sudakov double logarithms $\log^2(Q/q_T)$.

The factorization of $\sigma^{(0)}$ in terms of transverse-momentum dependent PDFs (TMDPDFs) was first shown by Collins, Soper, and Sterman in \refscite{Collins:1981uk,Collins:1981va,Collins:1984kg} and later elaborated on by Collins in \refcite{Collins:1350496}.
Its structure was also studied in \refscite{Catani:2000vq,deFlorian:2001zd,Catani:2010pd}.
The factorization was also studied in the framework of SCET by various groups, see e.g.\ \refscite{Becher:2010tm, GarciaEchevarria:2011rb, Chiu:2012ir}.
Using the notation of \refcite{Chiu:2012ir}, the factorized LP cross section for the production of a color-singlet final state $L$ with invariant mass $Q$ and total rapidity $Y$ in a proton-proton collision can be written as%
\footnote{We suppress that for gluon-gluon fusion, $H$ and $B$ carry polarization indices.}
\begin{align} \label{eq:sigma0}
 \frac{\df \sigma^{(0)}}{\df Q^2 \df Y \df^2 \qt} &
 = \sigma_0 \!\sum_{i,j} H_{ij}(Q, \mu)
   \!\int\!\! \df^2\bt \, e^{\img \qt \cdot \bt}
   \tilde B_i\bigl(x_a, \bt, \mu, \nu\bigr) \tilde B_j\bigl(x_b, \bt, \mu, \nu\bigr)
   \tilde S(b_T, \mu, \nu)
\,,\!\end{align}
where $x_{a,b} = Q e^{\pm Y}/\Ecm$ are the momentum fractions carried by the incoming partons.
In \eq{sigma0}, $H_{ij}$ is the hard function describing virtual corrections to the underlying hard process $ij \to L$,
the $\tilde B_i$ are TMD beam functions in Fourier space and $\tilde S$ is the TMD soft function in Fourier space.
While $H_{ij}$ only depends on the $\MS$ renormalization scale $\mu$, the beam and soft functions also depend on the rapidity renormalization scale $\nu$.

For nonperturbative $q_T \sim b_T^{-1} \sim \LQCD$, the $\tilde B_i$ become genuinely nonperturbative functions,
while for perturbative $q_T \sim b_T^{-1} \gg \LQCD$ they can be matched perturbatively onto PDFs,
\begin{align}\label{eq:q_match}
 \tilde B_i(x,\bt,\mu,\nu) &= \sum_j \int_x^1 \frac{\df z}{z} \, \tilde \cI_{ij}\Bigl(\frac{x}{z},\bt,\mu,\nu\Bigr) f_j(z,\mu)
\,.\end{align}
The perturbative kernels $\cI_{ij}$ are known to two loops \cite{Catani:2011kr,Catani:2012qa,Gehrmann:2012ze,Gehrmann:2014yya},
and the soft function $\tilde S$ is known to three loops \cite{Echevarria:2015byo,Luebbert:2016itl,Li:2016ctv}.
This has allowed resummation to next-to-next-to-next-to leading logarithmic accuracy \cite{Bizon:2017rah,Chen:2018pzu,Bizon:2018foh}.

Recently, there has been some progress towards a nonperturbative factorization of the NLP cross section $\df\sigma^{(2)}/\df q_T^2$, which involves higher twist PDFs \cite{Balitsky:2017gis,Balitsky:2017flc}.
Here, we are interested in studying the perturbative power corrections to the NLP terms,
where one can perform an OPE to match onto standard PDFs.
At subleading power, the perturbative kernels also involve (higher) derivatives of distributions,
which can always be reduced to standard distributions acting on derivatives of PDFs.
The NLP cross section at $\cO(\as)$ thus takes the form
\begin{align} \label{eq:sigma_NLP_intro}
 &\frac{\df\sigma^{(2,1)}}{\df Q^2 \df Y \df q_T^2}
= \hat\sigma^\LO(Q,Y) \, \frac{\as}{4\pi}
    \int_{x_a}^1 \frac{\df z_a}{z_a} \int_{x_b}^1 \frac{\df z_b}{z_b}
\nn\\&\quad\times \biggl[
   f_i\biggl(\frac{x_a}{z_a}\biggr) f_j\biggl(\frac{x_b}{z_b}\biggr)
   C_{f_i f_j}^{(2,1)}(z_a, z_b, q_T)
  + \frac{x_a}{z_a} f'_i\biggl(\frac{x_a}{z_a}\biggr) \frac{x_b}{z_b} f'_j\biggl(\frac{x_b}{z_b}\biggr)
  C_{f'_i f_j'}^{(2,1)}(z_a, z_b, q_T)
  \nn\\&\qquad
   + \frac{x_a}{z_a} f'_i\biggl(\frac{x_a}{z_a}\biggr) f_j\biggl(\frac{x_b}{z_b}\biggr)
   C_{f_i' f_j}^{(2,1)}(z_a, z_b, q_T)
  + f_i\biggl(\frac{x_a}{z_a}\biggr) \frac{x_b}{z_b} f'_j\biggl(\frac{x_b}{z_b}\biggr)
  C_{f_i f_j'}^{(2,1)}(z_a, z_b, q_T)
\biggr]
\,,\end{align}
where $\hat \sigma^\LO$ is the LO partonic cross section which serves as an overall normalization.
The $C^{(2,1)}_{a b}$ are perturbative coefficients, expressed in terms of distributions, and we suppress the explicit $Q$ and $Y$ dependence in the kernels $C^{(2,1)}_{a b}$.
In general, at order $\as^n$ their logarithmic structure is
\begin{align}
 C^{(2,n)}_{ab}(z_a, z_b, q_T) = \sum_{m=0}^{2n-1} C^{(2,n)}_{ab,m}(z_a, z_b) \ln^m\frac{Q^2}{q_T^2}
\,.\end{align}
More explicitly, at NLO they have the form
\begin{align}
 C^{(2,1)}_{ab}(z_a, z_b, q_T) = C^{(2,1)}_{ab,1}(z_a, z_b) \ln\frac{Q^2}{q_T^2} + C^{(2,1)}_{ab,0}(z_a, z_b)
\,,\end{align}
i.e.\ they only contain a single logarithm $\ln(Q^2/q_T^2)$ and a $q_T$-independent piece.
(Note that due to the dependence on $z_{a,b}$, it will yield a $Q^2$ and $Y$ dependence.)
We emphasize that in the form given here, all logarithms have been extracted,
and the $q_T$ distribution is directly expressed in terms of PDFs and their derivatives.

In the following, we will derive a master formula to obtain the NLO NLP kernels $C^{(2,1)}_{ab}$
for arbitrary color-singlet processes, as well as the explicit results for Higgs and Drell-Yan production.
The study of higher perturbative orders, and the derivation of a factorization and resummation is left to future work.
However, we do wish to comment on one complication which occurs for $q_T$ at higher orders, that we have not addressed. Unlike for beam thrust, at NNLO and beyond, one can have power-suppressed contributions at small $q_T$ from two hard partons in the final state that are nearly back-to-back such that their transverse momenta balance to give a small total $q_T$. At NNLO, this is at most a constant power correction, since it is not logarithmically enhanced. but at higher orders it can have a logarithmic contribution. These power corrections are of a different nature than those discussed here, and are not captured as an expansion in the soft and collinear limits about the Born process.

The remainder of this section is organized as follows. In \sec{master_formula}, we derive the
master formula for the NLP corrections using the $\eta$ regulator, showing in particular that the terms from expanding the regulator contribute. In \sec{master_formula_2}, we rederive this master formula in  pure rapidity regularization, which will be simpler due to the fact that one does not have additional terms from the expansion of the regulator, and due to the fact that the soft sector is scaleless.
In \sec{results_singlet}, we then apply the master formula to derive explicit results for Drell-Yan and gluon-fusion Higgs production. In \sec{discuss}, we discuss our results and compare them with
the known NLP results for beam thrust. Finally in \sec{numerics}, we provide a numerical
validation of our results.

\subsection{Master Formula for Power Corrections to Next-to-Leading Power}
\label{sec:master_formula}

We consider the production of a color-singlet final state $L$ at fixed invariant mass $Q$ and rapidity $Y$,
measuring the magnitude of its transverse momentum $q_T^2 = |\qt|^2$.
The underlying partonic process is
\begin{align}
 a(p_a) + b(p_b) \to L(p_1, \cdots) + X(k_1, \cdots)
\,,\end{align}
where $a,b$ are the incoming partons and $X$ denotes additional QCD radiation.
Following the notation of \refcite{Ebert:2018lzn}, we express the cross section as
\begin{align} \label{eq:sigma1}
 \frac{\df\sigma}{\df Q^2 \df Y \df q_T^2} &
 = \int_{0}^{1}\!\! \df \zeta_a \df \zeta_b\, \frac{f_a(\zeta_a)\, f_b(\zeta_b)}{2 \zeta_a \zeta_b \Ecm^2}
   \int\!\biggl(\prod_i \frac{\df^d k_i}{(2\pi)^d} (2\pi) \delta_+(k_i^2) \biggr)
   \int\!\!\frac{\df^d q}{(2\pi)^d} \, |\cM(p_a, p_b; \{k_i\}, q)|^2
   \nn\\* &\quad\times
   (2\pi)^d \delta^{(d)}(p_a + p_b - k - q) \, \delta(Q^2 - q^2)
   \, \delta\biggl(Y - \frac{1}{2}\ln\frac{q^-}{q^+}\biggr)
   \, \delta\bigl(q_T^2 - |\kt|^2\bigr)
\,.\end{align}
Here, the incoming momenta are given by
\begin{align} \label{eq:p_ab}
 p_a^\mu = \zeta_a \Ecm \frac{n^\mu}{2}
\,,\qquad
 p_b^\mu = \zeta_b \Ecm\frac{\bn^\mu}{2}
\,,\end{align}
$k = \sum_i k_i$ is the total outgoing hadronic momentum, and $q$ is the total leptonic momentum.
In particular, $\kt = \sum_i \vec k_{i,T}$ is the vectorial sum of the transverse momenta of all emissions.
Since the measurements are not affected by the details of the leptonic final state,
the leptonic phase-space integral has been absorbed into the matrix element,
\begin{align} \label{eq:M2_1}
 |\cM(p_a, p_b; \{k_i\}, q)|^2 &= \int\df\Phi_L(q) \, |\cM(p_a, p_b; \{k_i\}, \{p_j\})|^2
\,,\nn\\
  \df\Phi_L(q) &= \prod_j \frac{\df^d p_{j}}{(2\pi)^d} (2\pi) \delta_+(p_{j}^2 - m_j^2)
\, (2\pi)^d \delta^{(d)}\Bigl(q - \sum_j p_j\Bigr)
\,.\end{align}
The matrix element $\cM$ also contains the renormalization scale $\mu^{2\eps}$,
as usual associated with the renormalized coupling $\alpha_s(\mu)$, and may also contain virtual corrections.

There is an important subtlety when measuring the transverse momentum $q_T$ using dimensional regularization,
as the individual transverse momenta $\vec k_{i,T}$ are continued to $2-2\eps$ dimensions.
The measurement function $\delta(q_T^2 - |\kt|^2)$ in \eq{sigma1} can thus be interpreted
either as measuring the magnitude in $2-2\eps$ dimensions or the projection onto $2$ dimensions.
This scheme dependence cancels in the final result, but can lead to different intermediate results.
At the order we are working, both choices give identical results, so for simplicity of the following manipulations
we specify to measuring the magnitude in $2-2\eps$ dimension.
For detailed discussions, see e.g.\ \refscite{Jain:2011iu,Luebbert:2016itl}.

The $\delta$ functions measuring the invariant mass $Q$ and rapidity $Y$ fix the incoming momenta to be
\begin{align} \label{eq:zeta_ab}
 \zeta_a(k) &= \frac{1}{\Ecm} \Bigl(k^- +  e^{+Y} \sqrt{Q^2 + k_T^2} \Bigr)
\,,\quad
 \zeta_b(k) = \frac{1}{\Ecm} \Bigl(k^+ +  e^{-Y} \sqrt{Q^2 + k_T^2} \Bigr)
\,.\end{align}
Equation~(\ref{eq:sigma1}) can now be simplified to
\begin{align} \label{eq:sigma2}
 \frac{\df\sigma}{\df Q^2 \df Y \df q_T^2} &
 = \int\!\biggl(\prod_i \frac{\df^d k_i}{(2\pi)^d} (2\pi) \delta_+(k_i^2) \biggr)
   \frac{f_a(\zeta_a)\, f_b(\zeta_b)}{2 \zeta_a \zeta_b \Ecm^4}
   \Msquared(Q, Y; \{k_i\}) \, \delta\bigl(q_T^2 - |\kt|^2\bigr)
\,,\end{align}
where we introduced the abbreviation
\begin{align} \label{eq:Msquared}
 \Msquared(Q, Y; \{k_i\}) \equiv |\cM(p_a, p_b, \{k_i\}, q=p_a+p_b - k)|^2
\,.\end{align}
This emphasizes that the squared matrix element depends only on the Born measurements $Q$ and $Y$,
which fix the incoming momenta through \eqs{p_ab}{zeta_ab}, and the emission momenta $k_i$.
The restriction that $\zeta_{a,b} \in [0,1]$ is kept implicit in \eq{sigma2} through the support of the proton PDFs.

\subsubsection{General Setup at NLO}
\label{sec:generalsetup}

For reference, we start with the LO cross section following from \eq{sigma2},
\begin{align} \label{eq:sigmaLO}
 \frac{\df\sigma^\LO}{\df Q^2 \df Y \df q_T^2} &
 = \frac{f_a(x_a)\, f_b(x_b)}{2 x_a x_b \Ecm^4}
   \Msquared^\LO(Q, Y)\, \delta\bigl(q_T^2\bigr)
\,,\end{align}
where
\begin{align} \label{eq:xab}
 x_a = \frac{Q e^Y}{\Ecm} \,,\quad x_b = \frac{Q e^{-Y}}{\Ecm}
\,,\end{align}
and $\Msquared^\LO$ is the squared matrix element in the Born kinematics, see \eq{Msquared}.
For future reference, we also define the LO partonic cross section, $\hat \sigma^{\rm LO}(Q,Y)$, by
\begin{align} \label{eq:sigmaLO_2}
  \frac{\df\sigma^\LO}{\df Q^2 \df Y \df q_T^2} &
 = \hat\sigma^\LO(Q,Y) \, f_a(x_a)\, f_b(x_b) \, \delta(q_T^2)
\,,\qquad
  \hat\sigma^\LO(Q,Y) = \frac{\Msquared^\LO(Q, Y)}{2 x_a x_b \Ecm^4}
\,.\end{align}
At NLO, the virtual correction only contributes at leading power and is proportional to $\delta(q_T^2)$.
At subleading power, it suffices to consider the real correction, given from \eq{sigma2} by
\begin{align} \label{eq:sigmaNLO_1}
 \frac{\df\sigma}{\df Q^2 \df Y \df q_T^2} &
 = \int\! \frac{\df^d k}{(2\pi)^d}\, (2\pi) \delta_+(k^2)\,
   \frac{f_a(\zeta_a)\, f_b(\zeta_b)}{2 \zeta_a \zeta_b \Ecm^4}\,
   \Msquared(Q, Y; \{k\}) \, \delta\bigl(q_T^2 - |\kt|^2\bigr)
\nn\\&
 = \frac{q_T^{-2\eps}}{(4\pi)^{2-\eps} \Gamma(1-\eps)} \int_0^\infty \frac{\df k^-}{k^-}
   \frac{f_a(\zeta_a)\, f_b(\zeta_b)}{2 \zeta_a \zeta_b \Ecm^4}\,
   \Msquared(Q, Y; \{k\})\bigg|_{k^+ = q_T^2/k^-}
\,.\end{align}
In the following, we will mostly keep the symbol $k^+$ often leaving the use of the relation $k^+ = k_T^2/k^- = q_T^2/k^-$ to the end,
since this makes the symmetry under $k^+ \leftrightarrow k^-$ manifest.
The integral in \eq{sigmaNLO_1} is finite as the physical support of the PDFs, $0 \le \zeta_{a,b} \le 1$, cuts off the integral in $k^-$.
As discussed in \sec{rapDivLP}, these constraints will be expanded for small $q_T \ll Q$, after which the integral becomes rapidity divergent.
To regulate the integral, we use the $\eta$ regulator where one inserts a factor of $w^2 |2 k^z/\nu|^{-\eta}$ into the integral,
\begin{align} \label{eq:sigmaNLO_2}
 \frac{\df\sigma}{\df Q^2 \df Y \df q_T^2} &
 = \frac{q_T^{-2\eps}}{(4\pi)^{2-\eps} \Gamma(1-\eps)} \int_0^\infty \frac{\df k^-}{k^-}
   w^2 \nu^\eta \biggl|k^- - \frac{q_T^2}{k^-}\biggr|^{-\eta}
   \frac{f_a(\zeta_a)\, f_b(\zeta_b)}{2 \zeta_a \zeta_b \Ecm^4}\,
   \Msquared(Q, Y; \{k\})
\,.\end{align}
We now wish to expand \eq{sigmaNLO_2} in the limit of small $\lambda \sim q_T / Q \ll 1$.
Using the knowledge from the EFT, this can be systematically achieved by employing the scaling of \eq{modes_intro},
\begin{align} \label{eq:modes}
 &n{-}\text{collinear}: \quad k_n \sim Q \, (\lambda^2, 1, \lambda)
\,,\\
 \nn&\bn{-}\text{collinear}: \quad k_{\bn} \sim Q \, (1, \lambda^2, \lambda)
\,,\\
 \nn&\text{soft}: \hspace{1.8cm} k_s \sim Q \, (\lambda, \lambda, \lambda)
\,,\end{align}
for the momentum $k$.
By inserting each of these scalings into \eq{sigmaNLO_2} and expanding the resulting expression
to first order in $\lambda$, one precisely obtains the soft and beam functions as defined
in the $\eta$ regulator. This illustrative exercise is shown explicitly in \app{LP}.
Here, we are interested in the first nonvanishing power correction, which occurs at $\cO(\lambda^2) \sim \cO(q_T^2/Q^2)$. We will explicitly show that the $\cO(\lambda)$ linear power correction vanishes.
To compute the $\cO(\lambda^2)$ result, we will consider the soft and collinear cases separately, deriving master formulas for all scalings
applicable to any color-singlet production.
The power-suppressed operators and Lagrangian insertions required to calculate these directly will be presented in \refcite{Chang:2019xxxxx}.

\subsubsection[Soft Master Formula for \texorpdfstring{$q_T$}{qT}]
{\boldmath Soft Master Formula for $q_T$}
\label{sec:soft_master}

We first consider the case of a soft emission $k \sim Q(\lambda,\lambda,\lambda)$.
In this limit, the incoming momenta from \eq{zeta_ab} are expanded as
\begin{align}
 \zeta_a(k) &
 = x_a \biggl[ 1 + \frac{k^- e^{-Y}}{Q} + \frac{k_T^2}{2 Q^2} + \cO(\lambda^3) \biggr]
 \equiv x_a \biggl[1 + \Delta_a^{(1)} + \Delta_a^{(2)} + \cO(\lambda^3) \biggr]
\,,\nn\\
 \zeta_b(k) &
 = x_b \biggl[ 1 + \frac{k^+ e^{+Y}}{Q} + \frac{k_T^2}{2 Q^2} + \cO(\lambda^3) \biggr]
 \equiv x_b \biggl[1 + \Delta_b^{(1)} + \Delta_b^{(2)} + \cO(\lambda^3) \biggr]
\,,\end{align}
where as usual $k^+ = k_T^2/k^- = q_T^2/k^-$, $x_{a,b} = Q e^{\pm Y}/\Ecm$ as in \eq{xab},
and the terms in square brackets correspond to $\cO(\lambda^0)$,
$\cO(\lambda^1)$, and $\cO(\lambda^2)$, respectively.
It follows that the PDFs and flux factor are expanded as
\begin{align} \label{eq:phi_soft}
 \Phi \equiv \frac{f_a(\zeta_a) f_b(\zeta_b)}{\zeta_a \zeta_b} &
 = \frac{f_a(x_a) f_b(x_b)}{x_a x_b}
 + \frac{1}{x_a x_b} \biggl\{ \frac{k^- e^{-Y}}{Q} \bigl[ x_a f'_a(x_a) \, f_b(x_b) - f_a(x_a) f_b(x_b) \bigr]
   + (\rm{sym.}) \biggr\}
 \nn\\&\quad
 + \frac{1}{x_a x_b} \biggl\{
   \frac{(k^- e^{-Y})^2}{Q^2} \Bigl[ f_a(x_a) f_b(x_b) - x_a f'_a(x_a) \, f_b(x_b) + \frac{1}{2} x_a^2 f''_a(x_a) \, f_b(x_b) \Bigr]
\nn\\&\qquad\qquad\quad
   + \frac{k_T^2}{2Q^2} \bigl[ x_a f'_a(x_a) \, x_b f'_b(x_b) - x_a f'_a(x_a) \, f_b(x_b) \bigr]
   + (\rm{sym.})
 \biggr\} + \cO(\lambda^3)
\nn\\&
 \equiv \frac{1}{x_a x_b} \bigl[ \Phi^{(0)} + \Phi^{(1)} + \Phi^{(2)} \bigr] + \cO(\lambda^3)
\,.\end{align}
Here, $(\rm{sym.})$ denotes simultaneously flipping  $a \leftrightarrow b$ and letting $k^- \to k^+, Y \to -Y$.
For brevity, we introduced the abbreviation $\Phi^{(n)}$ for the $\cO(\lambda^n)$ pieces.
Note that we expanded to the second order in $\lambda$, as the $\cO(\lambda^1)$ piece will vanish
and the first nonvanishing correction in fact arises at $\cO(\lambda^2)$.

The expansion of the matrix element is process dependent, and we define the expansion in the soft limit through
\begin{align} \label{eq:M_soft}
 \Msquared_s(Q, Y; \{k\}) = \Msquared_s^{(0)}(Q, Y; \{k\}) + \Msquared_s^{(1)}(Q, Y; \{k\}) + \Msquared_s^{(2)}(Q, Y; \{k\}) + \cO(\lambda)
\,.\end{align}
The LP matrix element scales as $\Msquared_s^{(0)} \sim \lambda^{-2}$,
such that $\int \df k_T^2 \, \Msquared_s^{(0)} \sim \lambda^0$.
The next two matrix elements are each suppressed by an additional order in $\lambda$ relative to the one before.

Plugging the expansions \eqs{phi_soft}{M_soft} back into \eq{sigmaNLO_2} and collecting terms in $\lambda$,
the soft limit through $\cO(\lambda^2)$ is obtained as
\begin{align} \label{eq:sigma_soft_2}
 \frac{\df\sigma_s}{\df Q^2 \df Y \df q_T^2} &
 = \frac{q_T^{-2\eps}}{(4\pi)^{2-\eps} \Gamma(1-\eps)} \frac{1}{2 x_a x_b \Ecm^4}
   \int_0^\infty \frac{\df k^-}{k^-} w^2 \nu^{\eta} \biggl|k^- - \frac{q_T^2}{k^-}\biggr|^{-\eta}
   \\*\nn&\quad
   \times \biggl\{
      \Phi^{(0)} \Msquared_s^{(0)}(Q, Y; \{k\})
    + \Bigl[ \Phi^{(0)} \Msquared_s^{(1)}(Q, Y; \{k\}) + \Phi^{(1)} \Msquared_s^{(0)}(Q, Y; \{k\}) \Bigr]
   \\*\nn&\qquad
    + \Bigl[ \Phi^{(0)} \Msquared_s^{(2)}(Q, Y; \{k\}) + \Phi^{(1)} \Msquared_s^{(1)}(Q, Y; \{k\}) + \Phi^{(2)} \Msquared_s^{(0)}(Q, Y; \{k\}) \Bigr]
   \biggr\}
\,.\end{align}
The first term in curly brackets is the leading-power result, the second term the $\cO(\lambda)$ contribution, and the last line contains the $\cO(\lambda^2)$ contribution.
Since each of these terms has a homogeneous scaling in $\lambda$,
they can only contribute integer powers of $k^-$, yielding integrals of
the form $\Isoftz{\alpha}$ given in \eq{soft_integrals_a}.

\paragraph{\boldmath Leading Power  [$\cO(\lambda^0)$]}

The leading soft limit of the squared amplitude $\Msquared$ is universal and given by
\begin{align} \label{eq:Msquared_soft_LP}
 \Msquared_s^{(0)}(Q,Y; \{k\}) &
 = \frac{16 \pi \as \muMS^{2\eps} \mathbf{C}}{k_T^2} \times \Msquared^\LO(Q,Y)
\,,\end{align}
where $\muMS$ is the renormalization scale in the MS scheme and $\mathbf{C} = C_F, C_A$ is the Casimir constant for the $q\bar{q}$ and $gg$ channel,
and the limit vanishes for any other channel.
The cross section at LP thus becomes
\begin{align} \label{eq:sigma_soft_LP}
 \frac{\df\sigma^{(0)}_s}{\df Q^2 \df Y \df q_T^2} &
 = \Isoftz{0}\, \frac{q_T^{-2\eps}}{(4\pi)^{2-\eps} \Gamma(1-\eps)}
   \frac{\Phi^{(0)} \Msquared_s^{(0)}(Q, Y; \{k\})}{2 x_a x_b \Ecm^4}
\,.\end{align}
In \sec{LP_soft}, we use this to compute the known bare LP soft function at NLO as a cross check.

\paragraph{\boldmath $\cO(\lambda)$}

Here, we show that power corrections at $\cO(\lambda) \sim \cO(q_T/Q)$ vanish at NLO.
At this order, we can let $\eps\to0$ to obtain the cross section from \eq{sigma_soft_2} as
\begin{align} \label{eq:sigma_soft_lambda1}
 \frac{\df\sigma_s^{(1)}}{\df Q^2 \df Y \df q_T^2} &
 = \frac{1}{2 (4\pi)^2 x_a x_b \Ecm^4}
   \int_0^\infty \frac{\df k^-}{k^-} w^2 \nu^{\eta} \biggl|k^- - \frac{q_T^2}{k^-}\biggr|^{-\eta}
   \\*\nn&\quad\times
   \Bigl[ \Phi^{(0)} \Msquared_s^{(1)}(Q, Y; \{k\}) + \Phi^{(1)} \Msquared_s^{(0)}(Q, Y; \{k\}) \Bigr]
\,.\end{align}
From \eq{phi_soft}, the expansion of the phase space is given by
\begin{align}
 \Phi^{(1)} &
 = \frac{k^- e^{-Y}}{Q} \bigl[ x_a f'_a(x_a) \, f_b(x_b) - f_a(x_a) f_b(x_b) \bigr]
 + (\rm{sym.})
\,.\end{align}
From \eq{Msquared_soft_LP}, we know that $\Msquared^{(0)}_s \sim \Msquared^\LO / k_T^2$,
so $\Phi^{(1)} \Msquared^{(0)} \sim k^-, k_T^2/k^-$.
Hence, this contribution to \eq{sigma_soft_lambda1} is proportional to $\Isoftz{\pm1} = 0$, see \eq{soft_integrals} for odd $\alpha$, and therefore vanishes.
The NLP expansion $\Msquared_s^{(1)}$ of the matrix element is suppressed by $\cO(\lambda)$ relative to $\Msquared^\LO$,
which from power counting can only be given by either $k^-$ or $k^+ = k_T^2/k^-$.
Hence, the $\Phi^{(0)} \Msquared^{(1)}$ term is also proportional to $\Isoftz{\pm1} = 0$ and vanishes as well.

More generally, power counting combined with the behavior of the integrals in \eq{soft_integrals} shows that at NLO, the power expansion is in $q_T^2/Q^2$. It would be interesting to extend this proof to higher perturbative orders.
We also remark that the collinear limit will not have a $\cO(\lambda)$ expansion at all,
and thus the consistency condition that rapidity divergences cancel between soft and collinear sectors
already implies that the soft NLP result cannot contribute to the leading logarithm.

\paragraph{\boldmath Next-to-Leading Power [$\cO(\lambda^2)$]}

The first nonvanishing power correction thus arises at $\cO(\lambda^2)\sim \cO(q_T^2/Q^2)$.
To derive a general master formula at this order, we decompose the expansion of the matrix element
according to the possible dependence on $k^\pm$, which follows from power counting and mass dimension,
\begin{align} \label{eq:altM}
 \Msquared_s^{(0)}(Q,Y; \{k\}) &= \frac{1}{k_T^2} \,\altM{0}(Q,Y)
\,,\nn\\
 \Msquared_s^{(1)}(Q,Y; \{k\}) &= \frac{1}{k_T^2} \biggl[ \frac{k^+}{Q} \,\altM{1}_+(Q,Y) + \frac{k^-}{Q} \,\altM{1}_-(Q,Y) \biggr]
\,,\nn\\
 \Msquared_s^{(2)}(Q,Y; \{k\}) &= \frac{1}{k_T^2} \biggl[ \frac{(k^+)^2}{Q^2} \,\altM{2}_{++}(Q,Y) + \frac{k_T^2}{Q^2} \,\altM{2}_{00} + \frac{(k^-)^2}{Q^2}  \,\altM{2}_{--}(Q,Y) \biggr]
\,.\end{align}
The expansion is defined such that all $\altM{i}$ have the same mass dimension.
We now only need to plug \eq{altM} back into \eq{sigma_soft_2}, collect the powers of $k^-$
(using that $k^+ = k_T^2 / k^-$) and apply \eq{soft_integrals_a}.
Only terms proportional to $\Isoftz{0}$ will yield a divergence in $\eta$,
and thus constitute the LL correction at NLP, while all other terms contribute at NLL.
We find
\begin{align} \label{eq:sigma_soft_NLP_LL}
 \frac{\df\sigma^{(2),\text{LL}}_s}{\df Q^2 \df Y \df q_T^2} &
 = \frac{1}{2 (4\pi)^2 x_a x_b \Ecm^4} \frac{1}{Q^2}
   w^2 \biggl(\frac{2}{\eta} + \ln\frac{\nu^2}{q_T^2} \biggr) \times \biggl\{
   \nn\\*&\hspace{1.cm}
   f_a(x_a) f_b(x_b) \biggl[ \altM{2}_{00}
      - e^{-Y} \altM{1}_+(Q,Y) - e^{Y} \altM{1}_-(Q,Y)
   \biggr]
   \nn\\*&\hspace{1.cm}
   + x_a f'_a(x_a) \, f_b(x_b) \biggl[ e^{-Y} \altM{1}_+(Q,Y) - \frac{1}{2} \altM{0}(Q,Y) \biggr]
   \nn\\*&\hspace{1.cm}
   + f_a(x_a) \, x_b f'_b(x_b) \biggl[ e^{+Y} \altM{1}_-(Q,Y) - \frac{1}{2} \altM{0}(Q,Y)  \biggr]
   \nn\\*&\hspace{1.cm}
   + x_a f'_a(x_a) \, x_b f'_b(x_b)\, \altM{0}(Q,Y)
   \biggr\}
\,,\end{align}
and
\begin{align} \label{eq:sigma_soft_NLP_NLL}
 \frac{\df\sigma^{(2),\text{NLL}}_s}{\df Q^2 \df Y \df q_T^2} &
 = \frac{1}{2 (4\pi)^2 x_a x_b \Ecm^4} \frac{1}{Q^2} \times \biggl\{
   \nn\\*&\hspace{1.cm}
   f_a(x_a) f_b(x_b) \biggl[
      \altM{0}(Q,Y) \Bigl( e^{-2Y} + e^{+2Y} \Bigr)
      + \altM{2}_{++}(Q,Y) + \altM{2}_{--}(Q,Y)
      \nn\\*&\hspace{3.cm}
      - e^{-Y} \altM{1}_-(Q,Y) - e^{Y} \altM{1}_+(Q,Y)
   \biggr]
   \nn\\*&\hspace{1.cm}
   + x_a f'_a(x_a) \, f_b(x_b) \biggl[ e^{-Y} \altM{1}_-(Q,Y) - e^{-2Y} \altM{0}(Q,Y) \biggr]
   \nn\\*&\hspace{1.cm}
   + f_a(x_a) \, x_b f'_b(x_b) \biggl[ e^{+Y} \altM{1}_+(Q,Y) - e^{+2Y} \altM{0}(Q,Y) \biggr]
   \nn\\*&\hspace{1.cm}
   + x_a^2 f''_a(x_a) \, f_b(x_b) \, \frac{e^{-2Y}}{2} \altM{0}(Q,Y)
   \nn\\*&\hspace{1.cm}
   + f_a(x_a) \, x_b^2 f''_b(x_b) \, \frac{e^{+2Y}}{2} \altM{0}(Q,Y)
   \biggr\}
\,.\end{align}
An interesting feature of \eq{sigma_soft_NLP_NLL} is the appearance of double derivatives of the PDFs,
arising from the expansion of $f[\zeta(k)]$ through $\cO(\lambda^2)$.
Most terms in \eqs{sigma_soft_NLP_LL}{sigma_soft_NLP_NLL} also exhibit an explicit
rapidity dependence, which is surprising for the boost-invariant observable $q_T$.
In fact, we will see explicitly that the full soft expansion exactly cancels against
rapidity-dependent terms in the collinear expansions, yielding a rapidity-independent final result. This behavior is expected since the rapidity dependence arises from the rapidity-dependent regulator, and therefore we expect that they should cancel in the final regulator independent result.

\subsubsection[Collinear Master Formula for \texorpdfstring{$q_T$}{qT}]
{\boldmath Collinear Master Formula for $q_T$}
\label{sec:collinear_master}

We next consider the case of a $n$-collinear emission $k \sim Q (\lambda^2,1,\lambda)$,
from which one can easily obtain the $\bn$-collinear case from symmetry.
Here, it is important to consistently expand the rapidity regulator in \eq{sigmaNLO_2} in the $n$-collinear limit,
\begin{equation} \label{eq:regulator_NLP}
  w^2 \nu^\eta \biggl|k^- - \frac{q_T^2}{k^-} \biggr|^{-\eta}
= w^2 \biggl|\frac{k^-}{\nu} \biggr|^{-\eta} \biggl[1 + \eta\,\frac{q_T^2}{(k^-)^2} + \cO(\lambda^4) \biggr]
\,.\end{equation}
Applying this to \eq{sigmaNLO_2} yields
\begin{align} \label{eq:sigma_coll_1}
 \frac{\df\sigma}{\df Q^2 \df Y \df q_T^2} &
 = \frac{q_T^{-2\eps}}{(4\pi)^{2-\eps} \Gamma(1-\eps)} \int_0^\infty \frac{\df k^-}{k^-}
   w^2 \biggl|\frac{k^-}{\nu} \biggr|^{-\eta} \biggl(1 + \eta\,\frac{q_T^2}{(k^-)^2} \biggr)
   \frac{f_a(\zeta_a)\, f_b(\zeta_b)}{2 \zeta_a \zeta_b \Ecm^4}\,
   \Msquared(Q, Y; \{k\})
\,.\end{align}
We now expand all pieces in $\lambda$.
The incoming momenta from \eq{zeta_ab} are expanded as
 \begin{align}
 \zeta_a(k) &
 = x_a \biggl[ \biggl(1 + \frac{k^- e^{-Y}}{Q}\biggr) + \frac{q_T^2}{2 Q^2} \biggr] + \cO(\lambda^4)
 \equiv x_a \biggl[ \frac{1}{z_a} + \Delta_a^{(2)} \biggr]  + \cO(\lambda^4)
\,,\nn\\
 \zeta_b(k) &
 = x_b \biggl[ 1 + \biggl(\frac{k^+ e^{+Y}}{Q} + \frac{q_T^2}{2 Q^2} \biggr) \biggr] + \cO(\lambda^4)
 \equiv x_b \biggl[1 + \Delta_b^{(2)} \biggr] + \cO(\lambda^4)
\,,\end{align}
where we grouped the terms of common scaling together
and defined $k^- = Q e^Y (1 - z_a)/z_a$.
(Recall that the superscript $^{(2)}$ denotes the suppression by $\lambda^2$.)
Expanding the PDFs and flux factors in $\lambda$, we obtain
\begin{align} \label{eq:flux_ncoll}
 \frac{f_a(\zeta_a)\, f_b(\zeta_b)}{\zeta_a \zeta_b} &
 = \frac{z_a}{x_a x_b} f_a\Bigl(\frac{x_a}{z_a}\Bigr) f_b(x_b)
 \nn\\* & \quad
 + \frac{z_a}{x_a x_b} \frac{q_T^2}{2Q^2} \biggl[
    \frac{(1-z_a)^2 - 2}{1-z_a} f_a\Bigl(\frac{x_a}{z_a}\Bigr) f_b(x_b)
   + x_a f'_a\Bigl(\frac{x_a}{z_a}\Bigr) f_b(x_b)
   \nn\\*&\qquad\qquad\qquad\quad
   + \frac{1+z_a}{1-z_a} f_a\Bigl(\frac{x_a}{z_a}\Bigr) \, x_b f'_b(x_b)
 \biggr] + \cO(\lambda^4)
\,.\end{align}
The expansion of the matrix element is process dependent, and we define it by
\begin{align} \label{eq:M_coll}
 \Msquared(Q, Y; \{k\}) = \Msquared_n^{(0)}(Q, Y; \{k\}) + \Msquared_n^{(2)}(Q, Y; \{k\}) + \cO(\lambda^4)
\,.\end{align}
Note that in contrast to the soft limit, there is no $\cO(\lambda)$ suppressed term here.

Next, we switch the integration variable in \eq{sigma_coll_1} via
\begin{align} \label{eq:km_to_za}
 k^- = Q e^Y \frac{1 - z_a}{z_a}
\,,\qquad
 \int_0^\infty \frac{\df k^-}{k^-} = \int_{x_a}^1 \frac{\df z_a}{z_a (1-z_a)}
\,,\end{align}
where the lower bound on the $z_a$ integral follows from the physical support of the PDF $f_a(x_a/z_a)$.
Inserting eqs.\ \eqref{eq:flux_ncoll} -- \eqref{eq:km_to_za} into \eq{sigma_coll_1}
and collecting the $\cO(\lambda^0)$ and $\cO(\lambda^2)$ pieces,
we obtain the leading $n$-collinear limit as
\begin{align} \label{eq:sigma_coll_LP}
 \frac{\df\sigma_n^{(0)}}{\df Q^2 \df Y \df q_T^2} &
 = \frac{q_T^{-2\eps}}{(4\pi)^{2-\eps} \Gamma(1-\eps)}
   w^2 \biggl|\frac{Q e^Y}{\nu}\biggr|^{-\eta}
   \int_{x_a}^1 \frac{\df z_a}{z_a}  \frac{z_a^{1+\eta}}{(1-z_a)^{1+\eta}}
   \frac{f_a(x_a/z_a) f_b(x_b)}{2 x_a x_b \Ecm^4}
   \nn\\*&\hspace{6.5cm}\times
   \Msquared_n^{(0)}(Q, Y; \{k\})
\,,\end{align}
which we evaluate in \app{LP} to obtain the known LP beam function.
For the NLP correction, we can let $\eps\to0$ to obtain
\begin{align} \label{eq:sigma_coll_NLP}
 \frac{\df\sigma^{(2)}_n}{\df Q^2 \df Y \df q_T^2} &
 = \frac{w^2}{(4\pi)^2} \biggl|\frac{Q e^Y}{\nu}\biggr|^{-\eta}
   \int_{x_a}^1 \frac{\df z_a}{z_a}  \frac{z_a^{1+\eta}}{(1-z_a)^{1+\eta}}
   \frac{1}{2 x_a x_b \Ecm^4} \, \biggl\{
    f_a\Bigl(\frac{x_a}{z_a}\Bigr) f_b(x_b) \Msquared_n^{(2)}(Q, Y; \{k\})
   \nn\\*&\quad
   + \frac{q_T^2}{2 Q^2} \Msquared_n^{(0)}(Q, Y; \{k\}) \biggl[ \frac{(1-z_a)^2 - 2}{1-z_a} f_a\Bigl(\frac{x_a}{z_a}\Bigr) f_b(x_b)
   + x_a f'_a\Bigl(\frac{x_a}{z_a}\Bigr) \, f_b(x_b)
   \nn\\*&\qquad
   + \frac{1+z_a}{1-z_a} \,f_a\Bigl(\frac{x_a}{z_a}\Bigr) \, x_b f'_b(x_b)
   + \frac{2 \eta}{e^{2Y}} \frac{z_a^2}{(1-z_a)^2} \, f_a\Bigl(\frac{x_a}{z_a}\Bigr) f_b(x_b)
   \biggr]
   \biggr\}
\,.\end{align}
The corresponding result in the $\bn$-collinear case reads
\begin{align} \label{eq:sigma_bncoll_NLP}
 \frac{\df\sigma^{(2)}_{\bn}}{\df Q^2 \df Y \df q_T^2} &
 = \frac{w^2}{(4\pi)^2} \biggl|\frac{Q e^{-Y}}{\nu}\biggr|^{-\eta}
   \int_{x_b}^1 \frac{\df z_b}{z_b}  \frac{z_b^{1+\eta}}{(1-z_b)^{1+\eta}}
   \frac{1}{2 x_a x_b \Ecm^4} \, \biggl\{
    f_a(x_a) f_b\Bigl(\frac{x_b}{z_b}\Bigr) \Msquared_\bn^{(2)}(Q, Y; \{k\})
   \nn\\*&\quad
   + \frac{q_T^2}{2 Q^2} \Msquared_\bn^{(0)}(Q, Y; \{k\}) \biggl[ \frac{(1-z_b)^2 - 2}{1-z_b} f_a(x_a) f_b\Bigl(\frac{x_b}{z_b}\Bigr)
   + f_a(x_a) \, x_b f'_b\Bigl(\frac{x_b}{z_b}\Bigr)
   \nn\\*&\qquad
   + \frac{1+z_b}{1-z_b} \,x_a f'_a(x_a) \, f_b\Bigl(\frac{x_b}{z_b}\Bigr)
   + \frac{2 \eta}{e^{-2Y}} \frac{z_b^2}{(1-z_b)^2} \, f_a(x_a) f_b\Bigl(\frac{x_b}{z_b}\Bigr)
   \biggr]
   \biggr\}
\,.\end{align}
As discussed in \sec{distribution}, a striking feature of \eqs{sigma_coll_NLP}{sigma_bncoll_NLP} is the appearance of power divergences $1/(1-z)^{2+\eta}$ and even $1/(1-z)^{3+\eta}$, which can be regulated using higher-order plus distributions, see also \app{plus_distr}.
Here, we find it more convenient to employ the integration-by-parts relations in \eqs{plus_dist_2}{plus_dist_3} to write the kernels fully in terms of standard plus distributions, at the cost of inducing explicit derivatives of the PDFs.
In order to apply these relations, we need to identify all divergences in $1/(1-z)^2$ and $1/(1-z)^3$.
To do so, first note that the LP matrix element scales as
\begin{equation}
 \Msquared_n^{(0)} \sim \frac{k^-}{k_T^2} P(z,\eps) \sim (1-z) P(z,\eps)
\,,\end{equation}
where $P$ is the appropriate splitting function in $d=4-2\eps$ dimensions, which itself scales like $P(z,\epsilon)\sim 1/(1-z)$.
Due to the overall prefactor of $k^- \sim (1-z)$, the LP matrix element is finite as $z\to1$. Power counting implies that the subleading matrix element can at most yield one additional pole $1/(1-z)$.
Motivated by these two observations, we write the expanded squared amplitude as
\begin{align} \label{eq:altM_coll}
 \Msquared_n^{(0)}(Q,Y;\{k\}) &= \altM{0}_n(z_a)
\,,\nn\\
 \Msquared_n^{(2)}(Q, Y; \{k\}) &= \frac{k_T^2}{2Q^2} \frac{\altM{2}_n(z_a)}{1-z_a}
\,,\end{align}
and likewise for $\Msquared_\bn$ in the $\bn$-collinear limit.
The power suppression of $\Msquared_n^{(2)}$ is made manifest by extracting the factor $k_T^2/Q^2$.
For brevity, we suppress any dependence of $\altM{0}_n$ and $\altM{2}_n$ on $Q$ and $Y$.

Inserting \eq{altM_coll} into \eq{sigma_coll_NLP}, collecting powers of $(1-z_a)$,
and applying the distribution identities eqs.\ \eqref{eq:plus_dist_1}, \eqref{eq:plus_dist_2} and \eqref{eq:plus_dist_3},
the LL contribution at NLP is obtained as
\begin{align} \label{eq:sigma_coll_NLP_LL}
 \frac{\df\sigma^{(2),\text{LL}}_n}{\df Q^2 \df Y \df\cO} &
 = \frac{1}{(4\pi)^2} \frac{q_T^2}{2Q^2}
   \frac{1}{2 x_a x_b \Ecm^4}\,w^2 \biggl(\frac{1}{\eta} - \ln\frac{Q e^Y}{\nu} \biggr)
   \nn\\&\quad \times  \biggl\{
   f_a(x_a) f_b(x_b) \Bigl[ \altMp{2}_n(1) - 2 \altMp{0}_n(1) \Bigr]
   + f_a(x_a) \, x_b f'_b(x_b) \Bigl[ \altM{0}_n(1) + 2 {\altMp{0}_n}(1) \Bigr]
   \nn\\*&\qquad
   + x_a f'_a(x_a) f_b(x_b) \Bigl[ \altM{0}_n(1) - \altM{2}_n(1) \Bigr]
   - 2 x_a f'_a(x_a) \, x_b f'_b(x_b)  \altM{0}_n(1)
   \biggr\}
\,.\end{align}
Here, we used that the LL result is proportional to $\delta(1-z_a)$ to cancel the $z_a$ integral in \eq{sigma_coll_LP},
and the $\altMp{i}_n(1)$ are the derivative of $\altM{i}_n(z_a)$ at $z_a=1$.
Similarly, we obtain the NLL contribution as
\begin{align} \label{eq:sigma_coll_NLP_NLL}
 \frac{\df\sigma^{(2),\text{NLL}}_n}{\df Q^2 \df Y \df q_T} &
 = \frac{1}{(4\pi)^2} \frac{q_T^2}{2Q^2}
   \frac{1}{2 x_a x_b \Ecm^4}
   \int_{x_a}^1 \frac{\df z_a}{z_a}
   \nn\\*&\quad \times  \biggl\{
   f_a\Bigl(\frac{x_a}{z_a}\Bigr) f_b(x_b) \Bigl\{
      \delta(1-z_a) \Bigl[ \altM{2}_n(1) - \altMp{2}_n(1) - 2 \altM{0}_n(1) + 2 \altMp{0}_n(1) \Bigr]
      \nn\\*&\hspace{3.5cm}
      {- e^{-2Y} \delta(1-z_a) \Bigl[ 2 \altM{0}_n(1) + 4 \altMp{0}_n(1) + \altMpp{0}_n(1) \Bigr] }
      \nn\\*&\hspace{3.5cm}
      + z_a \cL_0(1-z_a) \Bigl[ 2 \altMp{0}_n(z_a) - \altMp{2}_n(z_a)\Bigr]
      + z_a \altM{0}_n(z_a)
   \Bigr\}
   \nn\\*&\qquad
   + \frac{x_a}{z_a} f'_a\Bigl(\frac{x_a}{z_a}\Bigr) f_b(x_b) \biggl\{
      \delta(1-z_a) \Bigl[ \altM{2}_n(1) - 2 \altM{0}_n(1) \Bigr]
      \nn\\*&\hspace{4cm}
      {+ 2e^{-2Y} \delta(1-z_a) \Bigl[ \altM{0}_n(1) + \altMp{0}_n(1) \Bigr] }
      \nn\\*&\hspace{4cm}
      + \cL_0(1-z_a) \Bigl[ \altM{2}_n(z_a) + (z_a^2-2)\altM{0}_n(z_a) \Bigr]
   \biggr\}
   \nn\\*&\qquad
   + f_a\Bigl(\frac{x_a}{z_a}\Bigr) \, x_b f'_b(x_b) \biggl\{
      \delta(1-z_a) \Bigl[ \altM{0}_n(1) - 2 \altMp{0}_n(1) \Bigr]
      \nn\\*&\hspace{4cm}
     - z_a \cL_0(1-z_a) \Bigl[ \altM{0}_n(z_a) + (1+z_a) \altMp{0}_n(z_a) \Bigr]
   \biggr\}
   \nn\\*&\qquad
   + \frac{x_a}{z_a} f'_a\Bigl(\frac{x_a}{z_a}\Bigr) \, x_b f'_b(x_b) \Bigl[
      2 \delta(1-z_a) \altM{0}_n(1) + (1+z_a) \altM{0}_n(z_a) \cL_0(1-z_a)
   \Bigr]
   \nn\\*&\qquad
   {
   - \Bigl(\frac{x_a}{z_a}\Bigr)^2 f''_a\Bigl(\frac{x_a}{z_a}\Bigr) f_b(x_b)
     \, \delta (1-z_a) e^{-2Y} \altM{0}_n(1)
   }
   \biggr\}
\,.\end{align}
Here, all terms with an explicit rapidity dependence arise from the expansion of the regulator itself,
see \eq{regulator_NLP}.
In practice, they will exactly cancel against the soft NLL result \eq{sigma_soft_NLP_NLL}.

\subsection{Derivation of the Master Formula in Pure Rapidity Regularization}
\label{sec:master_formula_2}

In \sec{master_formula}, we used the $\eta$ regulator of the form $|2 k^z / \nu|^{-\eta}$ to derive
the master formula.
In this section, we repeat the derivation of the master formula using the pure rapidity regulator introduced in \sec{upsilonreg}.
As discussed there, this regulator has the advantage that it is homogeneous in the power expansion, which reduces the number of terms at subleading power.
Furthermore, it renders the soft sector scaleless. 
The result using the generalization of the pure rapidity regulator, \eq{vita_regulator}, is shown in \app{master_formula_c} for completeness.

The derivation of the $n$-collinear expansion proceeds similar to the calculation shown in \sec{collinear_master}.
In \eq{sigma_coll_NLP}, one has to replace the regulator factor by
\begin{equation}
 \biggl|\frac{k^-}{\nu}\biggr|^{-\eta} = \biggl|\frac{Q e^Y}{\nu}\biggr|^{-\eta} \biggl|\frac{1-z_a}{z_a}\biggr|^{-\eta}
 ~\to~
 \upsilon^{\eta}\biggl|\frac{k^-}{k^+}\biggr|^{-\eta/2} = \upsilon^{\eta} \biggl|\frac{k^-}{q_T}\biggr|^{-\eta} = \upsilon^{\eta} \biggl|\frac{Q e^Y}{q_T}\biggr|^{-\eta}\biggl|\frac{1-z_a}{z_a}\biggr|^{-\eta}
\end{equation}
and drop the terms in $\smallupsilon/e^{2Y}$, as they are fully induced by the expansion of the regulator.
The NLP LL result is then easily obtained from \eq{sigma_coll_NLP_LL} by replacing $\nu \to q_T \upsilon$,
\begin{align} \label{eq:sigma_ncoll_NLP_LL_vita}
 \frac{\df\sigma^{(2),\text{LL}}_n}{\df Q^2 \df Y \df q_T^2} &
 = \frac{1}{(4\pi)^2} \frac{q_T^2}{2Q^2}
   \frac{1}{2 x_a x_b \Ecm^4}\,w^2\biggl(\frac{1}{\smallupsilon} - \ln\frac{Q e^Y}{q_T} + \ln(\upsilon) \biggr)
   \nn\\*&\quad \times  \biggl\{
   f_a(x_a) f_b(x_b) \Bigl[ \altMp{2}_n(1) - 2 \altMp{0}_n(1) \Bigr]
   + f_a(x_a) \, x_b f'_b(x_b) \Bigl[ \altM{0}_n(1) + 2 \altMp{0}_n(1) \Bigr]
   \nn\\*&\qquad
   + x_a f'_a(x_a) f_b(x_b) \Bigl[ \altM{0}_n(1) - \altM{2}_n(1) \Bigr]
   - 2 x_a f'_a(x_a) \, x_b f'_b(x_b)  \altM{0}_n(1)
   \biggr\}
\,.\end{align}
In the $\bn$-collinear limit, one has to replace the regulator factor
\begin{equation}
 \biggl|\frac{k^+}{\nu}\biggr|^{-\eta} = \biggl|\frac{Q e^{-Y}}{\nu}\biggr|^{-\eta} \biggl|\frac{1-z_b}{z_b}\biggr|^{-\eta}
 ~\to~
 \upsilon^{\eta}\biggl|\frac{k^-}{k^+}\biggr|^{-\eta/2} = \upsilon^{\eta} \biggl|\frac{k^+}{q_T}\biggr|^{\eta} = \upsilon^{\eta} \biggl|\frac{Q e^{-Y}}{q_T}\biggr|^{\eta} \biggl|\frac{1-z_b}{z_b}\biggr|^{\eta}
\end{equation}
and drop terms in $\eta/e^{-2Y}$ in \eq{sigma_bncoll_NLP}.
The NLP LL result is then obtained from \eq{sigma_coll_NLP_LL} by replacing $\eta \to -\eta, \nu \to q_T / \upsilon$ and exchanging $a\leftrightarrow b$ as
\begin{align} \label{eq:sigma_nbarcoll_NLP_LL_vita}
 \frac{\df\sigma^{(2),\text{LL}}_\bn}{\df Q^2 \df Y \df q_T^2} &
 = \frac{1}{(4\pi)^2} \frac{q_T^2}{2Q^2}
   \frac{1}{2 x_a x_b \Ecm^4}\,w^2\biggl(-\frac{1}{\smallupsilon} - \ln\frac{Q e^{-Y}}{q_T} - \ln(\upsilon) \biggr)
   \\&\quad \times  \biggl\{
   f_a(x_a) f_b(x_b) \Bigl[ \altMp{2}_\bn(1) - 2 \altMp{0}_\bn(1) \Bigr]
   + f_a(x_a) \, x_b f'_b(x_b) \Bigl[ \altM{0}_\bn(1) - \altM{2}_\bn(1) \Bigr]
   \nn\\*&\qquad
   + x_a f'_a(x_a) f_b(x_b)  \Bigl[ \altM{0}_\bn(1) + 2 {\altMp{0}_\bn}(1) \Bigr]
   - 2 x_a f'_a(x_a) \, x_b f'_b(x_b)  \altM{0}_\bn(1)
   \biggr\}\nn
\,.\end{align}
Summing \eqs{sigma_ncoll_NLP_LL_vita}{sigma_nbarcoll_NLP_LL_vita}, the poles in $\smallupsilon$ precisely cancel,
and the dependence on $e^Y$ and $\bigupsilon$ cancel as well to yield a pure logarithm in $\ln(Q/q_T)$.
This cancellation has to occur between the two collinear sectors,
as there are no contributions from the soft sector.

The NLP NLL result for the pure rapidity regulator is identical to that in \eq{sigma_coll_NLP_NLL} upon dropping all rapidity-dependent pieces,
which we have explicitly verified by repeating the derivation in \sec{collinear_master} using the pure rapidity regulator.
This provides a highly nontrivial check of our regularization procedure, and our understanding of subleading-power rapidity divergences.

\subsection{Next-to-leading Power Corrections at NLO}
\label{sec:results_singlet}

In this section, we give explicit results for the full NLP correction at NLO
for gluon-fusion Higgs and Drell-Yan production in all partonic channels.
Since both are $s$-channel processes, their power corrections
are always proportional to their Born cross sections, and we express the NLP result at $\cO(\as)$ as
\begin{align} \label{eq:sigma_NLP}
 &\frac{\df\sigma^{(2,1)}}{\df Q^2 \df Y \df q_T^2}
= \hat\sigma^\LO(Q) \, \frac{\as}{4\pi}
    \int_{x_a}^1 \frac{\df z_a}{z_a} \int_{x_b}^1 \frac{\df z_b}{z_b}
\nn\\*&\quad\times \biggl[
   f_i\biggl(\frac{x_a}{z_a}\biggr) f_j\biggl(\frac{x_b}{z_b}\biggr)
   C_{f_i f_j}^{(2,1)}(z_a, z_b, q_T)
  + \frac{x_a}{z_a} f'_i\biggl(\frac{x_a}{z_a}\biggr) \frac{x_b}{z_b} f'_j\biggl(\frac{x_b}{z_b}\biggr)
  C_{f'_i f_j'}^{(2,1)}(z_a, z_b, q_T)
\nn\\*&\qquad
   + \frac{x_a}{z_a} f'_i\biggl(\frac{x_a}{z_a}\biggr) f_j\biggl(\frac{x_b}{z_b}\biggr)
   C_{f_i' f_j}^{(2,1)}(z_a, z_b, q_T)
  + f_i\biggl(\frac{x_a}{z_a}\biggr) \frac{x_b}{z_b} f'_j\biggl(\frac{x_b}{z_b}\biggr)
  C_{f_i f_j'}^{(2,1)}(z_a, z_b, q_T)
\biggr]
\,.\end{align}
Here, we suppress the explicit $Q$ and $Y$ dependence in the kernels $C^{(2,1)}_{a b}$.

The required $H+j$ and $Z+j$ amplitudes are conveniently expressed in terms of the Mandelstam variables
\begin{align} \label{eq:mandelstam}
 s_{ab} &= 2 p_a \cdot p_b = Q^2 + 2 q_T^2 + \Bigl( k^+ e^Y + k^- e^{-Y}\Bigr) \sqrt{Q^2 + q_T^2}
\,,\nn \\
 s_{ak} &= -2 p_a \cdot k = - q_T^2 - k^+ e^{+Y} \sqrt{Q^2 + q_T^2}
\,,\nn \\
 s_{bk} &= -2 p_b \cdot k = - q_T^2 - k^- e^{-Y} \sqrt{Q^2 + q_T^2}
\,,\end{align}
which allows us to straightforwardly obtain the LP and NLP expansions in both the soft and collinear limits,
as required by the collinear and soft master formulas.
In the following, we only give the final results after combining soft,
$n$-collinear, and $\bn$-collinear power corrections. The results were computed
separately using both regulators, which provides a highly nontrivial check of our calculation.

\subsubsection{Gluon-Fusion Higgs Production}
\label{sec:ggH_results}

We first consider on-shell Higgs production in gluon fusion in the $m_t \to \infty$ limit,
for which the LO partonic cross section is given by
\begin{align}
 \hat\sigma^\LO(Q)
 = \frac{\Msquared^\LO(Q,Y)}{2 x_a x_b \Ecm^4}
 = 2\pi \delta(Q^2 - m_H^2) \frac{|\cM_{gg\to H}^\LO(Q)|^2}{2 Q^2 \Ecm^2}
\,.\end{align}
The LO matrix element in $d=4-2\epsilon$ dimensions is given by
\cite{Dawson:1990zj,Djouadi:1991tka}
\begin{align}\label{eq:M2_gg_H}
 |\cM^\LO_{gg\to H}(Q)|^2 &= \frac{\as^2 Q^4}{576 \pi^2 v^2} \biggl(\frac{4\pi\muMS^2}{m_t^2}\biggr)^{2\eps} \frac{\Gamma^2(1+\eps)}{1-\eps}
\,.\end{align}
At NLO, there are three distinct partonic channels, $gg\to Hg$, $q\bar q \to H g$, and $gq \to H q$,
which we consider separately.
Here, we calculate the full LL and NLL kernels for all channels.
The LL results will be summarized in \sec{discuss}.

\paragraph{\boldmath $gg \to H g$}

The spin- and color-averaged squared amplitude for $g(p_a) + g(p_b) \to H(q) + g(k)$ is given by \cite{Dawson:1990zj}
\begin{align} \label{eq:M2_ggHg}
 \Msquared_{gg \to H g}(Q, Y, \{k\}) &
 = \Msquared^\LO_{gg\to H}(Q) \times \frac{8 \pi \as C_A \muMS^{2\eps}}{ Q^4 (1-\eps)}
\nn\\* & \quad\times
   \biggl[ (1-2\eps) \frac{Q^8 + s_{ab}^4 + s_{ak}^4 + s_{bk}^4}{s_{ab} s_{ak} s_{bk}}
   + \frac{\eps}{2} \frac{(Q^4 + s_{ab}^2 + s_{ak}^2 + s_{bk}^2)^2}{s_{ab} s_{ak} s_{bk}} \biggr]
\,.\end{align}
The full result from combining the soft, $n$-collinear, and $\bn$-collinear contributions is given by
\begin{align} \label{eq:C_ggHg}
  C_{f_g f_g}^{(2,1)}(z_a, z_b, q_T) &
  = 2 C_A \frac{1}{Q^2} \biggl\{
      \biggl[8 \ln\frac{Q^2}{q_T^2} + 12 \biggr] \delta(1-z_a) \delta(1-z_b)
      \nn\\*&\hspace{2cm}
      + \delta(1-z_a)\biggl[ -8 + \frac{3}{z_b} + z_b - 12 z_b^2 + 9 z_b^3 + 8 \cL_0(1-z_b) \biggr]
      \nn\\*&\hspace{2cm}
      + \biggl[ -8 + \frac{3}{z_a} + z_a - 12 z_a^2 + 9 z_a^3 + 8 \cL_0(1-z_a) \biggr] \delta(1-z_b)
    \biggr\}
\,,\nn\\
  C_{f'_g f_g}^{(2,1)}(z_a, z_b, q_T) &
  = 2 C_A \frac{1}{Q^2} \biggl\{
    \biggl[- \ln\frac{Q^2}{q_T^2} - 1 \biggr] \delta(1-z_a) \delta(1-z_b)
      \nn\\*&\hspace{2cm}
      + \delta(1-z_a)\biggl[ 2 + \frac{2}{z_b^2} + \frac{1}{z_b} + z_b +  3 z_b^2 - \cL_0(1-z_b) \biggr]
      \nn\\*&\hspace{2cm}
      + \biggl[ 4 + \frac{2}{z_a} - 2 z_a + 5 z_a^2 - 3 z_a^3 - \cL_0(1-z_a) \biggr] \delta(1-z_b)
    \biggr\}
\,,\nn\\
  C_{f_g f'_g}^{(2,1)}(z_a, z_b, q_T) &
  = 2 C_A \frac{1}{Q^2} \biggl\{
      \biggl[- \ln\frac{Q^2}{q_T^2} - 1 \biggr] \delta(1-z_a) \delta(1-z_b)
      \nn\\*&\hspace{2cm}
      + \delta(1-z_a)\biggl[ 4 + \frac{2}{z_b} - 2 z_b + 5 z_b^2 - 3 z_b^3 - \cL_0(1-z_b)\biggr]
      \nn\\*&\hspace{2cm}
      + \biggl[ 2 + \frac{2}{z_a^2} + \frac{1}{z_a} + z_a + 3 z_a^2 - \cL_0(1-z_a) \biggr] \delta(1-z_b)
    \biggr\}
\,,\nn\\
  C_{f'_g f'_g}^{(2,1)}(z_a, z_b, q_T) &
  = 2 C_A \frac{1}{Q^2} \biggl\{
       \biggl[2\ln\frac{Q^2}{q_T^2} + 4 \biggr] \delta(1-z_a) \delta(1-z_b)
      \nn\\*&\hspace{2cm}
      + \delta(1-z_a)\biggl[ -1 + \frac{1}{z_b^2} - z_b^2 + 2 \cL_0(1-z_b)\biggr]
      \nn\\*&\hspace{2cm}
      + \biggl[ -1 + \frac{1}{z_a^2} - z_a^2 + 2 \cL_0(1-z_a)\biggr] \delta(1-z_b)
    \biggr\}
\,.\end{align}
Substituting these results into \eq{sigma_NLP} yields the NLP cross section for $gg\to Hg$ at NLO.

\paragraph{\boldmath $gq \to Hq$}

The $gq\to Hq$ channel has power corrections at both LL and NLL.
The spin- and color-averaged squared amplitude for $g(p_a) + q(p_b) \to H(q) + q(k)$ is given by \cite{Dawson:1990zj}
\begin{align} \label{eq:M2_gqHq}
 \Msquared_{gq \to Hq}(Q, Y, \{k\})
 = - \Msquared^\LO_{gg\to H}(Q) \times 8 \pi \as C_F \muMS^{2\eps} \frac{1}{Q^4 s_{bk}}
  \Bigl[ s_{ab}^2 + s_{ak}^2 - \eps (s_{ab}+s_{ak})^2 \Bigr]
\,.\end{align}
The full result from combining the soft, $n$-collinear, and $\bn$-collinear contributions is given by
\begin{align} \label{eq:C_gqHq}
  C_{f_g f_q}^{(2,1)}(z_a, z_b, q_T) &
  = 2 C_F \frac{1}{Q^2} \biggl\{
      \biggl[ \ln\frac{Q^2}{q_T^2} + 3 \biggr] \delta(1-z_a) \delta(1-z_b)
      \nn\\*&\hspace{2cm}
      + \delta(1-z_a) \biggl[ -1 + \frac{3}{z_b} + 2 z_b - 2 z_b^2 + \cL_0(1-z_b) \biggr]
      \nn\\*&\hspace{2cm}
      + \biggl[ \frac{1}{z_a} + \cL_0(1-z_a) \biggr] \delta(1-z_b)
    \biggr\}
\,,\nn\\
  C_{f'_g f_q}^{(2,1)}(z_a, z_b, q_T) &
  = 2 C_F \frac{1}{Q^2} \biggl\{
      \biggl[ \ln\frac{Q^2}{q_T^2} + 2 \biggr] \delta(1-z_a) \delta(1-z_b)
      \nn\\*&\hspace{2cm}
      + \delta(1-z_a) \biggl[ \frac{2+z_b - z_b^3}{z_b^2} + \cL_0(1-z_b) \biggr]
      \nn\\*&\hspace{2cm}
      + \biggl[ \frac{1}{z_a} + \cL_0(1-z_a) \biggr] \delta(1-z_b)
    \biggr\}
\,,\nn\\
  C_{f_g f'_q}^{(2,1)}(z_a, z_b, q_T) &
  = 2 C_F \frac{1}{Q^2} \delta(1-z_a) \biggl(\frac{2}{z_b} - \frac{3}{2} z_b + z_b^2 \biggr)
\,,\nn\\
  C_{f'_g f'_q}^{(2,1)}(z_a, z_b, q_T) &
  = 2 C_F \frac{1}{Q^2} \delta(1-z_a) \biggl(-\frac12 + \frac{1}{z_b^2} + \frac{z_b}{2} \biggr)
\,.\end{align}
Substituting these results into \eq{sigma_NLP} yields the NLP cross section for $gq\to Hq$ at NLO.

\paragraph{\boldmath $qg \to Hq$}

The result for $qg \to Hq$ can be obtained from \eq{C_gqHq}
by exchanging $f_q \leftrightarrow f_g$ and $a \leftrightarrow b$,
\begin{align} \label{eq:C_qgHq}
  C_{f_q f_g}^{(2,1)}(z_a, z_b, q_T) &
  = 2 C_F \frac{1}{Q^2} \biggl\{
      \biggl[ \ln\frac{Q^2}{q_T^2} + 3 \biggr] \delta(1-z_a) \delta(1-z_b)
      \nn\\*&\hspace{2cm}
      + \biggl[ -1 + \frac{3}{z_a} + 2 z_a - 2 z_a^2 + \cL_0(1-z_a) \biggr] \delta(1-z_b)
      \nn\\*&\hspace{2cm}
      + \delta(1-z_a) \biggl[ \frac{1}{z_b} + \cL_0(1-z_b) \biggr]
    \biggr\}
\,,\nn\\
  C_{f'_q f_g}^{(2,1)}(z_a, z_b, q_T) &
  = 2 C_F \frac{1}{Q^2} \biggl(\frac{2}{z_a} - \frac{3}{2} z_a + z_a^2 \biggr) \delta(1-z_b)
\,,\nn\\
  C_{f_q f'_g}^{(2,1)}(z_a, z_b, q_T) &
  = 2 C_F \frac{1}{Q^2} \biggl\{
      \biggl[ \ln\frac{Q^2}{q_T^2} + 2 \biggr] \delta(1-z_a) \delta(1-z_b)
      \nn\\*&\hspace{2cm}
      + \biggl[ \frac{2+z_a - z_a^3}{z_a^2} + \cL_0(1-z_a) \biggr] \delta(1-z_b)
      \nn\\*&\hspace{2cm}
      + \delta(1-z_a) \biggl[ \frac{1}{z_b} + \cL_0(1-z_b) \biggr]
    \biggr\}
\,,\nn\\
  C_{f'_q f'_g}^{(2,1)}(z_a, z_b, q_T) &
  = 2 C_F \frac{1}{Q^2} \biggl(-\frac12 + \frac{1}{z_a^2} + \frac{z_a}{2} \biggr) \delta(1-z_b)
\,.\end{align}
Substituting these results into \eq{sigma_NLP} yields the NLP cross section for $qg\to Hq$ at NLO.

\paragraph{\boldmath $q\bar{q} \to Hg$}

The $q\bar q \to Hg$ channel has no leading logarithms and thus only contributes at NLL.
The spin- and color-averaged squared amplitude is given by \cite{Dawson:1990zj}
\begin{align}
 \Msquared_{q\bar q \to Hg}(Q, Y, \{k\}) &
 = \Msquared^\LO_{gg\to H}(Q) \times \frac{64\pi}{3} \as C_F \muMS^{2\eps} \frac{1-\eps}{Q^4 s_{ab}}
   \bigl[ s_{ak}^2 + s_{bk}^2 - \eps (s_{ak}+s_{bk})^2 \bigr]
\,.\end{align}
The results for the kernels are given by
\begin{align} \label{eq:C_qqHg}
  C_{f_q f_\bq}^{(2,1)}(z_a, z_b, q_T) &
  = \frac{16 C_F}{3} \frac{1}{Q^2} \biggl[ \delta(1-z_a) \biggl(1 + \frac{1}{z_b} - 2 z_b\biggr)
    + \biggl(1 + \frac{1}{z_a} - 2 z_a\biggr) \delta(1-z_b) \biggr]\,,
\nn\\
  C_{f'_q f_\bq}^{(2,1)}(z_a, z_b, q_T) &
  = \frac{16 C_F}{3} \frac{1}{Q^2} \frac{(1-z_a)^2}{z_a} \delta(1-z_b)\,,
\nn\\
  C_{f_q f'_\bq}^{(2,1)}(z_a, z_b, q_T) &
  = \frac{16 C_F}{3} \frac{1}{Q^2} \delta(1-z_a) \frac{(1-z_b)^2}{z_b}
\,,\nn\\
  C_{f'_q f'_\bq}^{(2,1)}(z_a, z_b, q_T) &= 0
\,.\end{align}
Substituting these results into \eq{sigma_NLP} yields the NLP cross section for $q\bar{q} \to Hg$ at NLO.

\subsubsection{Drell-Yan Production}
\label{sec:DY_results}

We next consider the Drell-Yan process $p p \to Z/\gamma^* \to \ell^+ \ell^-$, and for brevity denote it as $p p \to V$.
In contrast to on-shell Higgs production, it is important to be able to include off-shell effects.
The LO partonic cross section is given by
\begin{align}
 \hat\sigma^\LO(Q) = \frac{4\pi \alpha_{em}^2}{3 N_c Q^2 \Ecm^2}
 \biggl[ Q_q^2 + \frac{(v_q^2 + a_q^2)(v_\ell^2 + a_\ell^2) - 2 Q_q v_q v_\ell (1 - m_Z^2/Q^2)}{(1-m_Z^2/Q^2)^2 + m_Z^2 \Gamma_Z^2 / Q^4} \biggr]
\,,\end{align}
where $Q$ is the dilepton invariant mass, $v_{\ell,q}$ and $a_{\ell,q}$ are the standard
vector and axial couplings of the leptons and quarks to the $Z$ boson,
and the $\ell^+ \ell^-$ phase space has already been integrated over.
At NLO , there are two distinct partonic channels, $q\bar q\to Vg$ and $qg \to Vq$,
which we consider separately.
Here, we calculate the full LL and NLL kernels for all channels.
The LL results will be summarized in \sec{discuss}.

\paragraph{\boldmath $q\bar q\to Vg$}

We first consider the $q\bar q\to Vg$ channel, for which the spin- and color-averaged squared amplitude is given by \cite{Gonsalves:1989ar}
\begin{align} \label{eq:M2_qqbar_Zg}
 |\cM_{q\bar q\to Vg}|^2 &
 = |\cM_{q\bar q\to V}|^2 \times \frac{8 \pi \as C_F \muMS^{2\eps}}{Q^2}
  \left[ (1-\epsilon) \left(\frac{s_{ak}}{s_{bk}}+\frac{s_{bk}}{s_{ak}}\right)+\frac{2 s_{ab} Q^2}{s_{ak} s_{bk}} -2\epsilon \right]
\,.\end{align}
The full result from combining the soft, $n$-collinear, and $\bn$-collinear contributions is given by
\begin{align} \label{eq:C_qqZg}
  C_{f_q f_{\bar q}}^{(2,1)}(z_a, z_b, q_T) &
  = 2 C_F \frac{1}{Q^2} \biggl[
      -4 \delta(1-z_a)\delta(1-z_b)
      \nn\\*&\hspace{2cm}
      - \delta(1-z_a) \frac{1+z_b^2-4z_b^3}{2z_b}
      - \frac{1+z_a^2-4z_a^3}{2z_a} \delta(1-z_b)
    \biggr]
\,,\nn\\
  C_{f'_q f_{\bar q}}^{(2,1)}(z_a, z_b, q_T) &
  = 2 C_F \frac{1}{Q^2} \biggl\{
      \biggl[- \ln\frac{Q^2}{q_T^2} - 1\biggr] \delta(1-z_a)\delta(1-z_b)
      \nn\\*&\hspace{2cm}
      + \delta(1-z_a) \biggl[ \frac32 + \frac{1}{2z_b} + z_b - \cL_0(1-z_b) \biggr]
      \nn\\*&\hspace{2cm}
      - \biggl[ \frac{1+z_a+2z_a^3}{2z_a} + \cL_0(1-z_a) \biggr] \delta(1-z_b)
    \biggr\}
\,,\nn\\
  C_{f_q f'_{\bar q}}^{(2,1)}(z_a, z_b, q_T) &
  = 2 C_F \frac{1}{Q^2} \biggl\{
      \biggl[- \ln\frac{Q^2}{q_T^2} - 1\biggr] \delta(1-z_a)\delta(1-z_b)
      \nn\\*&\hspace{2cm}
      - \delta(1-z_a) \biggl[ \frac{1+z_b+2z_b^3}{2z_b} + \cL_0(1-z_b) \biggr]
      \nn\\*&\hspace{2cm}
      + \biggl[ \frac32 + \frac{1}{2z_a} + z_a - \cL_0(1-z_a) \biggr] \delta(1-z_b)
    \biggr\}
\,,\nn\\
  C_{f'_q f'_{\bar q}}^{(2,1)}(z_a, z_b, q_T) &
  = 2 C_F \frac{1}{Q^2} \biggl\{
      \biggl[ 2 \ln\frac{Q^2}{q_T^2} + 4\biggr] \delta(1-z_a)\delta(1-z_b)
      \nn\\*&\hspace{2cm}
      + \delta(1-z_a) \biggl[ \frac{1-2z_b-z_b^2}{2z_b} + 2 \cL_0(1-z_b) \biggr]
      \nn\\*&\hspace{2cm}
      + \biggl[ \frac{1-2z_a-z_a^2}{2z_a} + 2 \cL_0(1-z_a) \biggr] \delta(1-z_b)
    \biggr\}
\,.\end{align}
Substituting these results into \eq{sigma_NLP} yields the NLP cross section for $q\bar{q} \to Vg$ at NLO.

\paragraph{\boldmath $qg\to Vq$}

The spin- and color-averaged squared amplitude for the $qg\to Vq$ channel is given by \cite{Gonsalves:1989ar}
\begin{align} \label{eq:M2_qg_Zq}
 \Msquared_{qg \to Vq}(Q,Y,\{k\})
 = - \Msquared^\LO_{q \bar q \to V}(Q) \times \frac{8 \pi \as T_F \muMS^{2\eps}}{Q^2 (1-\eps)}
  \left[ (1-\epsilon) \left(\frac{s_{ab}}{s_{bk}}+\frac{s_{bk}}{s_{ab}}\right)+\frac{2 s_{ak} Q^2}{s_{ab} s_{bk}} -2\epsilon \right]
\,.\end{align}
The full result from combining the soft, $n$-collinear, and $\bn$-collinear contributions is given by
\begin{align} \label{eq:C_qgZq}
  C_{f_q f_g}^{(2,1)}(z_a, z_b, q_T) &
  = 2 T_F \frac{1}{Q^2} \biggl\{
    \delta(1-z_a) \delta(1-z_b)
    \nn\\*&\hspace{2cm}
    + \delta(1-z_a) \frac{-1 + z_b^2 + 24 z_b^3 - 18 z_b^4}{2z_b}
    + 2 z_a \delta(1-z_b)
    \biggr\}
\,,\nn\\
  C_{f'_q f_g}^{(2,1)}(z_a, z_b, q_T) &
  = 2 T_F \frac{1}{Q^2} \biggl\{
    \biggl[ \ln\frac{Q^2}{q_T^2} + 2\biggr] \delta(1-z_a) \delta(1-z_b)
    \nn\\*&\hspace{2cm}
    + \delta(1-z_a) \biggl[ \frac{1-z_b-2z_b^2-6z_b^3}{2z_b} + \cL_0(1-z_b) \biggr]
    \nn\\*&\hspace{2cm}
    + \bigl[ 1 - z_a + \cL_0(1-z_a) \bigr] \delta(1-z_b)
    \biggr\}
\,,\nn\\
  C_{f_q f'_g}^{(2,1)}(z_a, z_b, q_T) &
  = 2 T_F \frac{1}{Q^2} \delta(1-z_a) \frac{-1 + 5 z_b + z_b^2 - 10 z_b^3 + 6 z_b^4}{2z_b}
\,,\nn\\
  C_{f'_q f'_g}^{(2,1)}(z_a, z_b, q_T) &
  = 2 T_F \frac{1}{Q^2} \delta(1-z_a) \frac{1-z_b + 2 z_b^3}{2z_b}
\,.\end{align}
Substituting these results into \eq{sigma_NLP} yields the NLP cross section for $q g \to V q$ at NLO.

\paragraph{\boldmath $g q\to Vq$}

The result for $gq \to Vq$ can be obtained from \eq{M2_qg_Zq}
by exchanging $a \leftrightarrow b$ and $f_q \leftrightarrow f_g$,
\begin{align} \label{eq:C_gqZq}
  C_{f_g f_q}^{(2,1)}(z_a, z_b, q_T) &
  = 2 T_F \frac{1}{Q^2} \biggl\{
    \delta(1-z_a) \delta(1-z_b)
    \nn\\*&\hspace{2cm}
    + 2 \delta(1-z_a) z_b
    + \frac{-1 + z_a^2 + 24 z_a^3 - 18 z_a^4}{2z_a} \delta(1-z_b)
    \biggr\}
\,,\nn\\
  C_{f'_g f_q}^{(2,1)}(z_a, z_b, q_T) &
  = 2 T_F \frac{1}{Q^2} \frac{-1+5z_a+z_a^2-10z_a^3 + 6z_a^4}{2z_a} \delta(1-z_b)
\,,\nn\\
  C_{f_g f'_q}^{(2,1)}(z_a, z_b, q_T) &
  = 2 T_F \frac{1}{Q^2} \biggl\{
    \biggl[ \ln\frac{Q^2}{q_T^2} + 2\biggr] \delta(1-z_a) \delta(1-z_b)
    \nn\\*&\hspace{2cm}
    + \delta(1-z_a) \bigl[ 1 - z_b + \cL_0(1-z_b) \biggr]
    \nn\\*&\hspace{2cm}
    + \biggl[ \frac{1-z_a-2z_a^2-6z_a^3}{2z_a} + \cL_0(1-z_a) \biggr] \delta(1-z_b)
    \biggr\}
\,,\nn\\
  C_{f'_g f'_q}^{(2,1)}(z_a, z_b, q_T) &
  = 2 T_F \frac{1}{Q^2} \frac{1-z_a+2z_a^3}{2z_a} \delta(1-z_b)
\,.\end{align}
Substituting these results into \eq{sigma_NLP} yields the NLP cross section for $g q \to V q$ at NLO.

\subsection{Discussion}
\label{sec:discuss}

Since the full calculation of the power corrections is rather involved, and contains a number of moving pieces, here we highlight several interesting features of the calculation, and compare them to
the perturbative power corrections for beam thrust. For the purposes of this discussion, it is convenient to recall the form of the LL power corrections for the Born partonic configurations
\begin{align} \label{eq:sigma_Higgs_LL_intro}
 \frac{\df\sigma^{(2),\text{LL}}_{g g \to H g}}{\df Q^2 \df Y\df q_T^2} &
 = \hat\sigma^\LO_{gg\to H}(Q) \times \frac{\as C_A}{4\pi} \frac{2}{Q^2} \ln\frac{Q^2}{q_T^2}
   \Bigl[ 8 f_g(x_a) f_g(x_b) +f^{gg}_{\text{uni}}(x_a,x_b) \Bigr]\,,\\
 \frac{\df\sigma^{(2),\text{LL}}_{q \bq \to V g}}{\df Q^2 \df Y\df q_T^2} &
 = \hat\sigma^\LO_{q\bar q\to V}(Q) \times \frac{\as C_F}{4\pi} \frac{2}{Q^2} \ln\frac{Q^2}{q_T^2}
   \Bigl[ f^{q\bar q}_{\text{uni}}(x_a,x_b) \Bigr]
\,,\nn\end{align}
where
\begin{align}
f^{gg}_{\text{uni}}(x_a,x_b)&=- x_a f'_g(x_a) f_g(x_b) - f_g(x_a) \, x_b f'_g(x_b)+ 2 x_a f'_g(x_a) \, x_b f'_g(x_b)\,, \\
f^{q\bq}_{\text{uni}}(x_a,x_b)&=- x_a f'_q(x_a) f_\bq(x_b) - f_q(x_a) \, x_b f'_\bq(x_b)+ 2 x_a f'_q(x_a) \, x_b f'_\bq(x_b)\,,
\end{align}
are identical up to switching of the labels on the PDFs. For the channels with a quark emission, we have
\begin{align}
\frac{\df\sigma^{(2),\text{LL}}_{g q \to H q}}{\df Q^2 \df Y\df q_T^2} &
 = \hat\sigma^\LO_{gg\to H}(Q) \times \frac{\as C_F}{4\pi} \frac{2}{Q^2} \ln\frac{Q^2}{q_T^2}
   \Bigl[ f_g(x_a) f_q(x_b) + f^{gq}_{\text{uni}}(x_a,x_b) \Bigr]
\,, \\
 \frac{\df\sigma^{(2),\text{LL}}_{g q \to V q}}{\df Q^2 \df Y\df q_T^2} &
 = \hat\sigma^\LO_{q\bar q\to Z}(Q) \times \frac{\as T_F}{4\pi} \frac{2}{Q^2} \ln\frac{Q^2}{q_T^2}
   \, \Bigl[ f^{qg}_{\text{uni}}(x_b,x_a) \Bigr]\,,
\end{align}
where
\begin{align}
f^{gq}_{\text{uni}}(x_a,x_b)&=x_a f'_g(x_a) f_q(x_b)\,, \\
f^{qg}_{\text{uni}}(x_b,x_a)&=f_g(x_a) \, x_b f'_q(x_b)\,,
\end{align}
are again identical up to the switching of the labels on the PDFs.

First, we note that these results involve a more complicated structure of derivatives than the power corrections to the \sceti~beam thrust observable, where at most a single derivative appeared in a given term \cite{Moult:2016fqy, Boughezal:2016zws, Moult:2017jsg}. Furthermore, for beam thrust, at LL there are no derivatives for the channels involving quark emission. Interestingly, the explanation for this arises from very different reasons in the soft and collinear sectors. In the soft sector, it is a simple consequence of the modified power counting of the soft modes, which implies that they must be expanded to two orders in the power counting. In the collinear sector, where the power counting is the same for $q_T$ and beam thrust, it arises from the presence of the power law singularities, which must be expanded against the PDFs. The cancellation of rapidity divergences between the soft and collinear sectors therefore exhibits a much more nontrivial relationship.

Another feature of the LL power corrections is the independence from explicit factors of the color-singlet rapidity $Y$, suggesting that the expansion parameter is indeed $q_T^2/Q^2$, as is expected from the fact that $q_T$ is boost invariant. In fact, the rapidity dependence is induced purely by the PDFs and their derivatives. This is particularly interesting for the case of Drell-Yan, where the only terms that contribute arise from derivatives acting on the PDFs, which leads to a more nontrivial rapidity dependence, and in particular, a rapidity dependence that is different from that at leading power. This has potentially interesting implications for power corrections for $q_T$ subtractions, and we will show this rapidity dependence numerically in \sec{numerics}.

It is also interesting to discuss the universality of these results between Higgs and Drell-Yan production. For the case of beam thrust, the LL results are related by a Casimir scaling, $C_A \leftrightarrow C_F$. Here we see explicitly that this is not the case for $q_T$. However, we see that all terms involving the derivatives of the PDFs are universal up to exchanges of the partonic indices, and it is only the coefficients of the $ff$ PDF structure that are non-universal.  One way of understanding this difference in universality between beam thrust, which is an \sceti~observable, and $q_T$, which is an \scetii~observable, is the different power counting of the soft sector. Since soft momenta in \scetii~scale as ${\cal O}(\lambda)$ rather than ${\cal O}(\lambda^2)$ this requires that for $q_T$ the soft matrix element must be expanded to one higher power, at which point there is a breaking of their universality.
However, the terms involving derivatives of the PDF get part of their power suppression from expanding the momenta entering the PDFs, and therefore are effectively expanded to the same power as for an \sceti~observable such as beam thrust. It would be interesting to understand this universality structure in more detail, in particular how it extends to other processes, and to higher orders.

\subsection{Numerical Results}
\label{sec:numerics}

In this section, we validate our results by numerically comparing the NLP spectrum to the full $q_T$ spectrum,
which we obtain by numerically integrating \eq{sigmaNLO_1}.
For Drell-Yan production, we fix $Q = m_Z = 91.1876~{\rm GeV}$ and use $\as(m_Z) = 0.118$.
For Higgs production, we work in the on-shell limit with $Q = m_H = 125~{\rm GeV}$
and $\as(m_H) = 0.1126428$ corresponding to a three-loop running from $\as(m_Z)$.
In both cases, we use $\Ecm=13~{\rm TeV}$ and the NNPDF31 NNLO PDFs~\cite{Ball:2017nwa} with
fixed factorization and renormalization scales $\mu_f = \mu_r = Q$.
We also fix the rapidity to $Y=2$ to have a nontrivial test of the rapidity dependence of our results
and to break the degeneracy between the $qg$ and $gq$ channels.

We compare the nonsingular cross section at NLO$_0$,%
\footnote{From the point of view of the $q_T$ factorization theorem, the leading-order Born process is $p p \to X$, and hence $\sigma^\LO \sim \delta(q_T^2)$. A nonvanishing transverse momentum is first obtained for $p p \to X+j$, which is the real part of the NLO correction to $p p \to X$, but the LO contribution for $q_T>0$. For clarity, we denote this order as NLO$_0$ to stress that it is counted with respect to the Born process $p p \to X$.}
which is obtained by subtracting all singular terms which diverge as $1/q_T^2$ from the full $q_T$ spectrum,
against our predictions for the NLP cross section.
The dependence of the nonsingular cross section on $q_T$ is given by
\begin{align} \label{eq:xs_nonsing}
 \frac{\df\sigma^{\rm nonsing}_{{\rm NLO}_0}}{\df Q^2 \df Y \df q_T^2} =
 c_1(Q,Y) \ln\frac{Q^2}{q_T^2} + c_0(Q,Y) + \cO\biggl(\frac{q_T^2}{Q^2}\biggr)
\,,\end{align}
where $c_1$ is predicted by the LL term at NLP and $c_0$ is predicted by the NLL term at NLP.
Note that $c_0$ is independent of $q_T$, but has a nontrivial dependence on $Q$ and $Y$.
The $\cO(q_T^2)$ corrections arise at subsubleading power.

\begin{figure*}[t!]
 \centering
 \includegraphics[height=5.6cm]{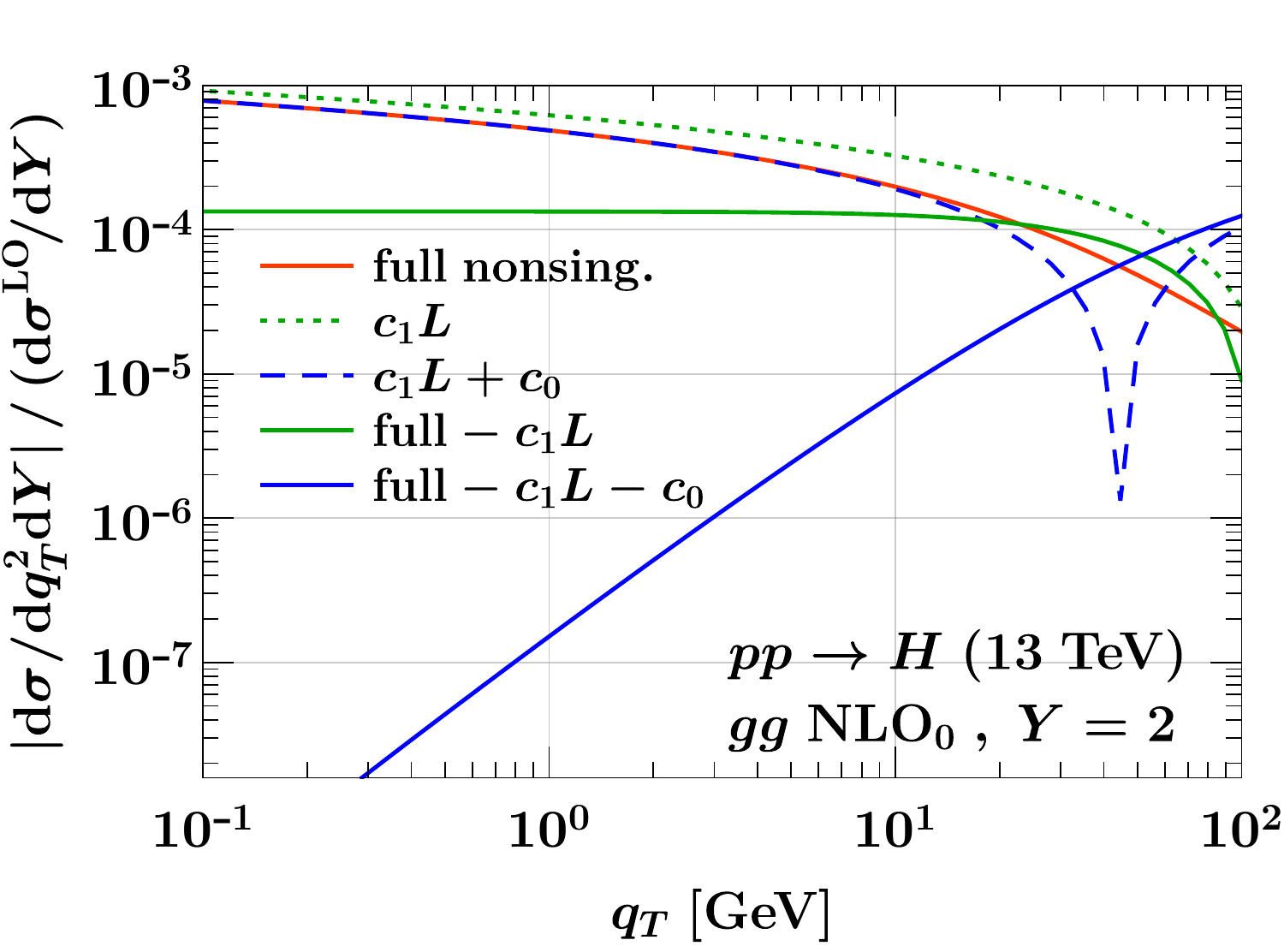}
   \hfill
 \includegraphics[height=5.6cm]{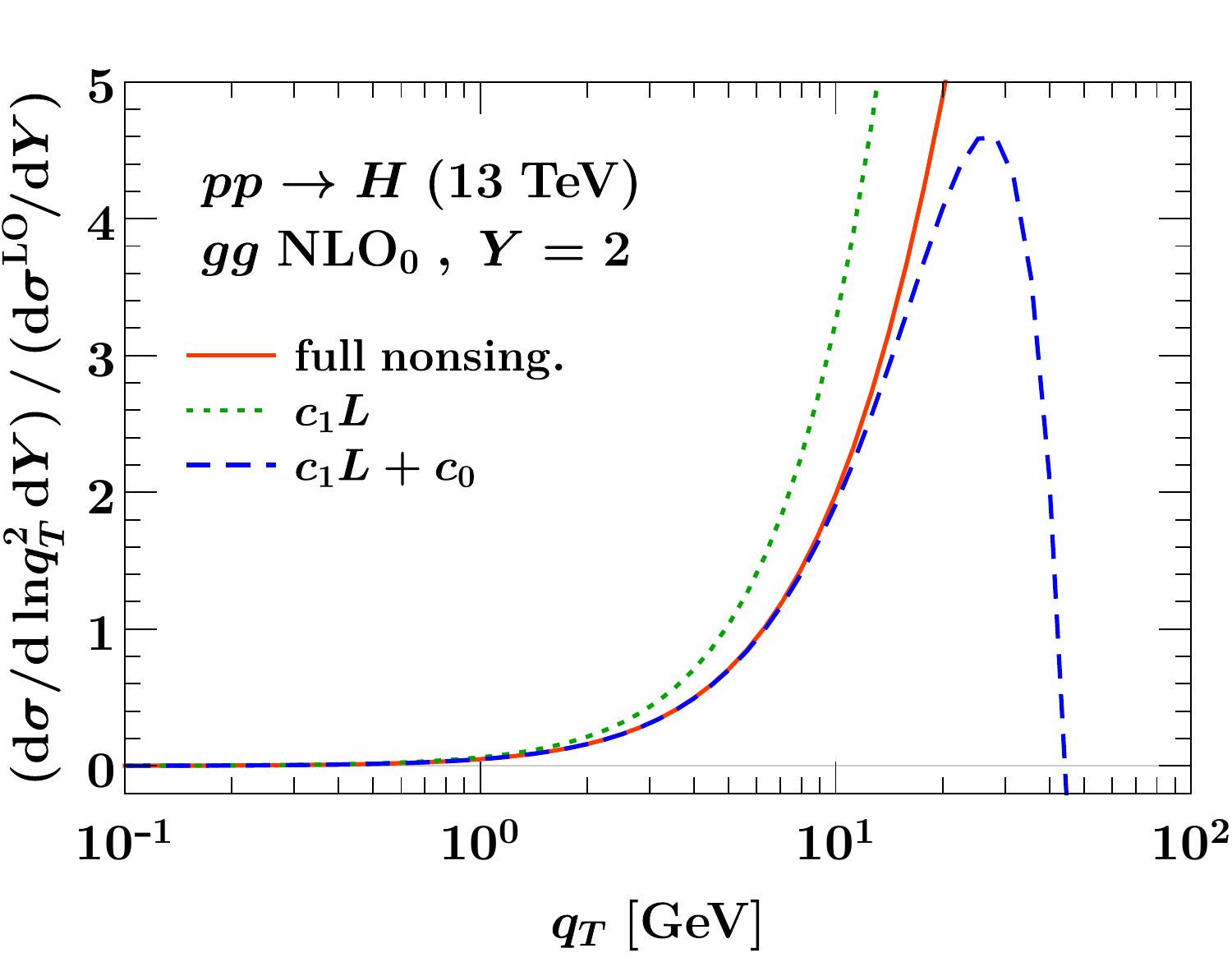}
 \includegraphics[height=5.6cm]{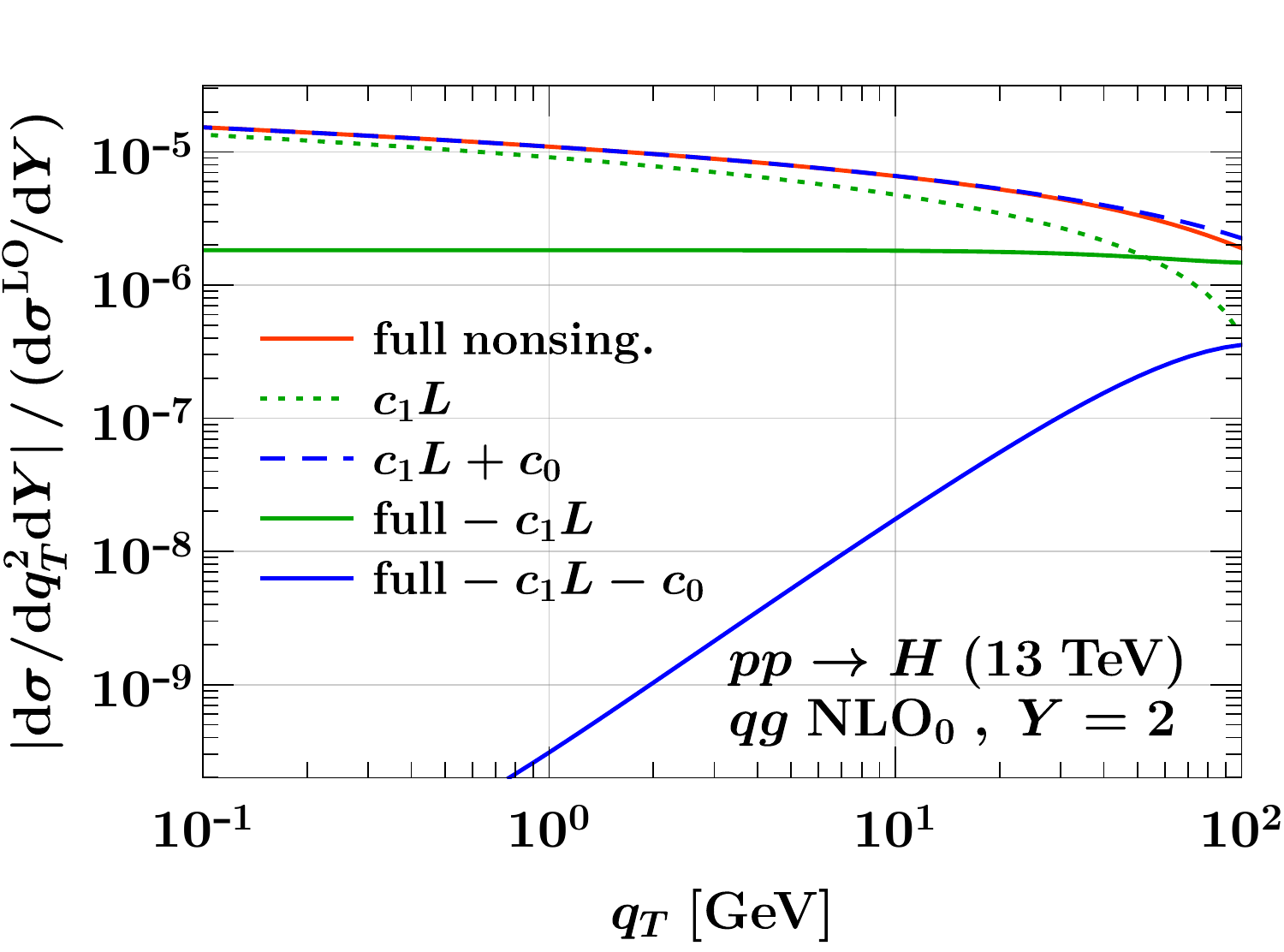}%
   \hfill
 \includegraphics[height=5.6cm]{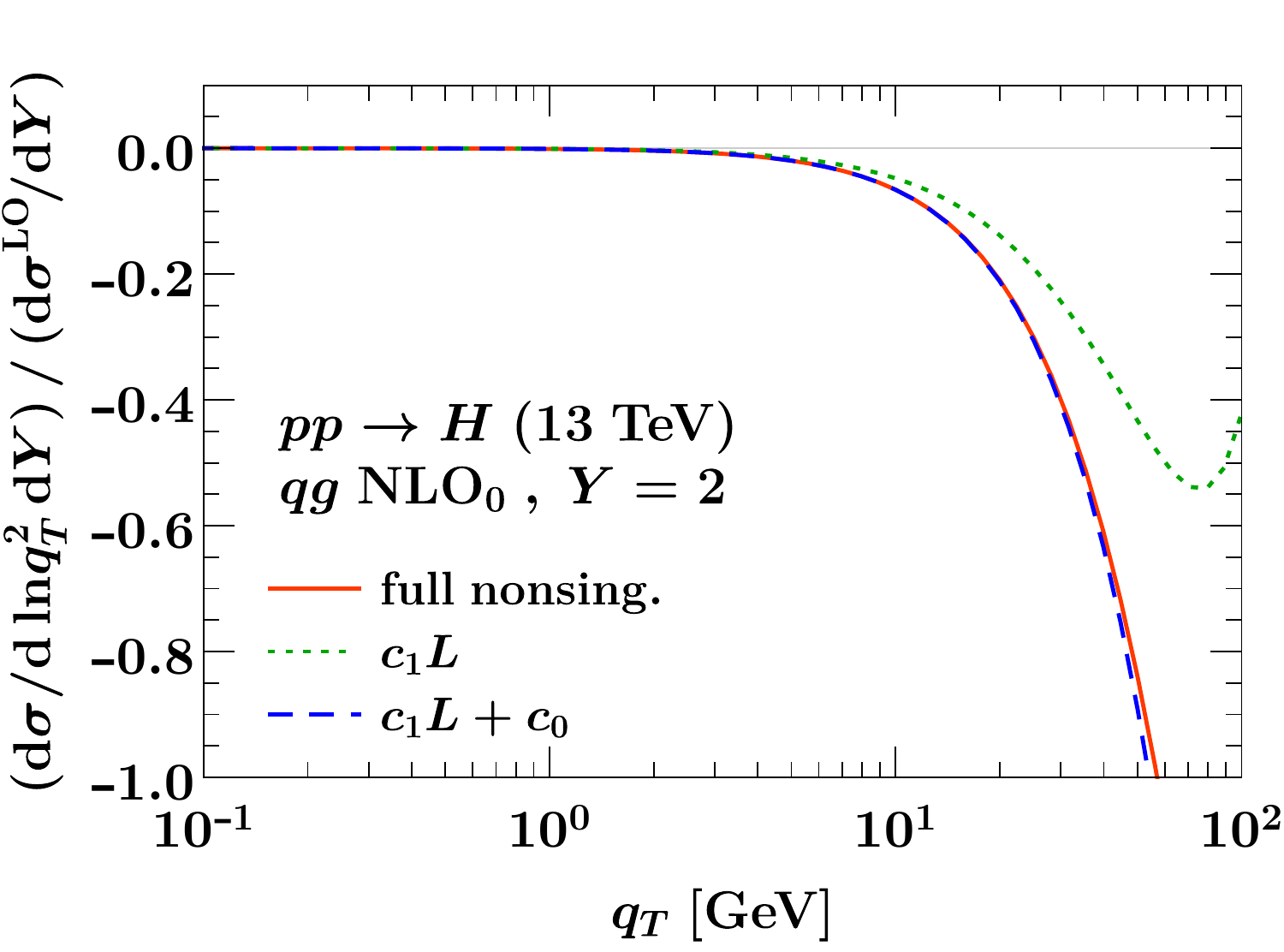}
 \includegraphics[height=5.6cm]{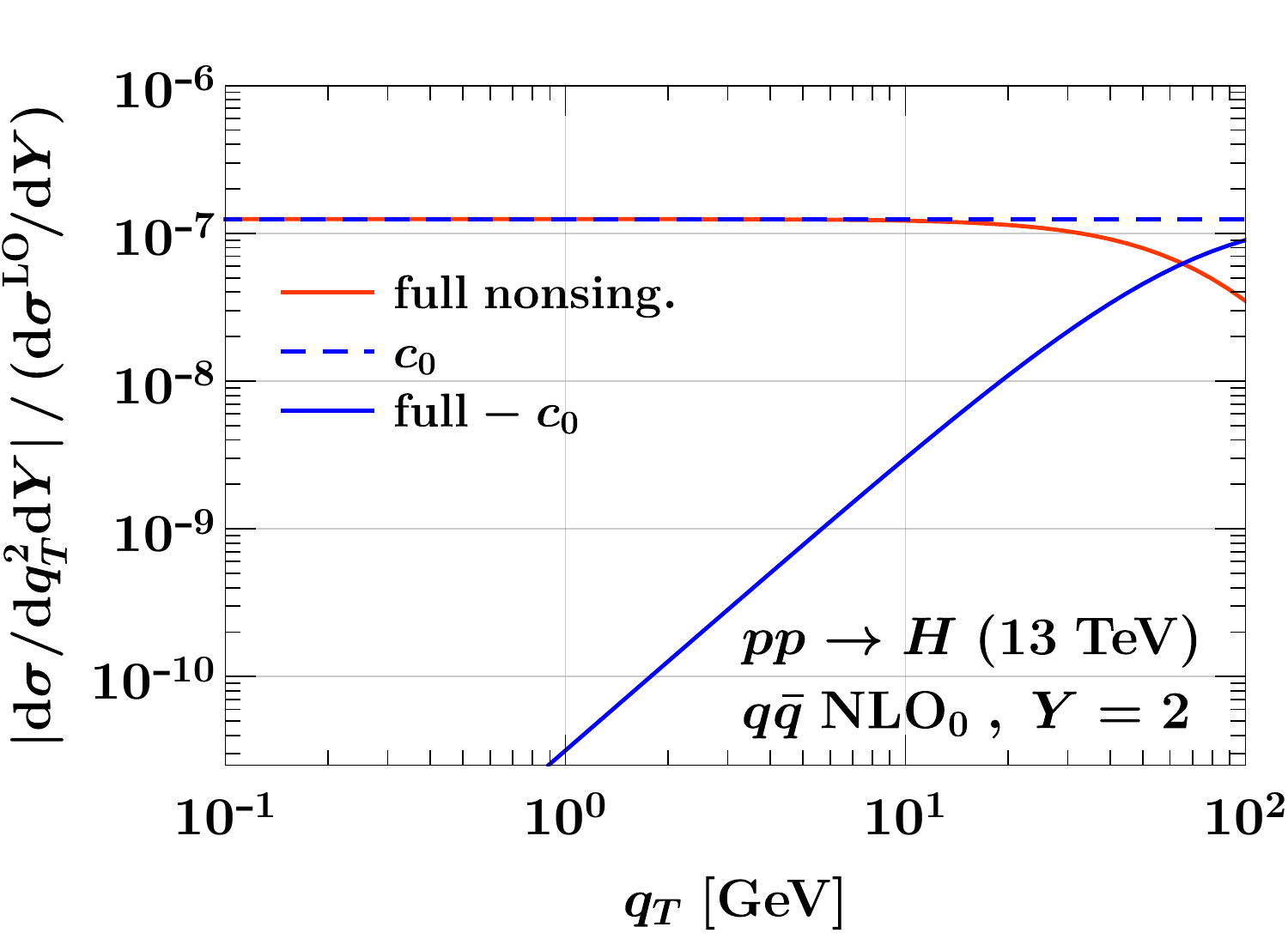}%
   \hfill
 \includegraphics[height=5.6cm]{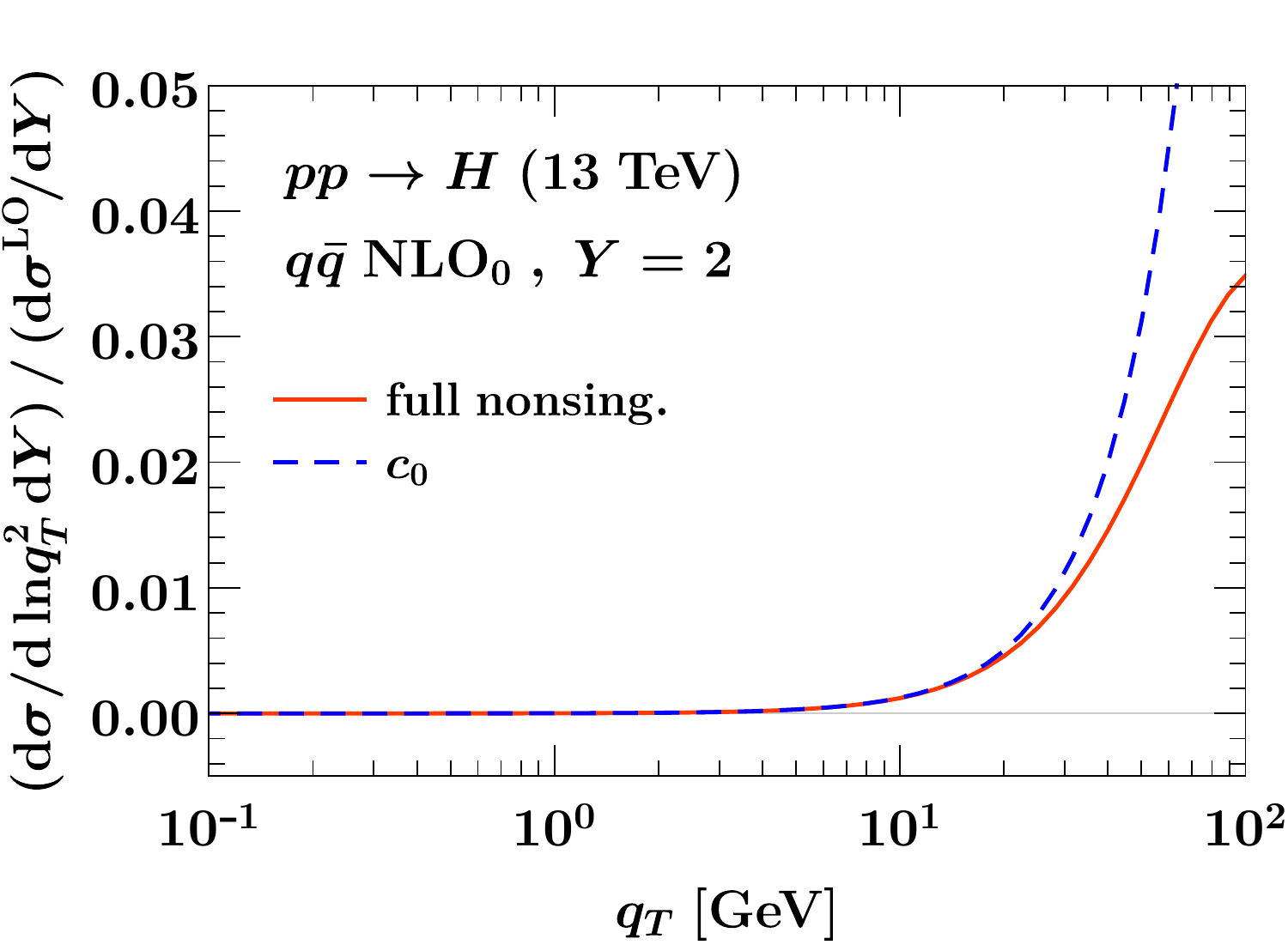}
\caption{Comparison of the LL and NLL corrections at subleading power with the full
nonsingular $q_T$ spectrum for all partonic channels contributing to Higgs production at NLO$_0$.}
 \label{fig:Higgs}
\end{figure*}

\begin{figure*}[t!]
 \centering
 \includegraphics[height=5.6cm]{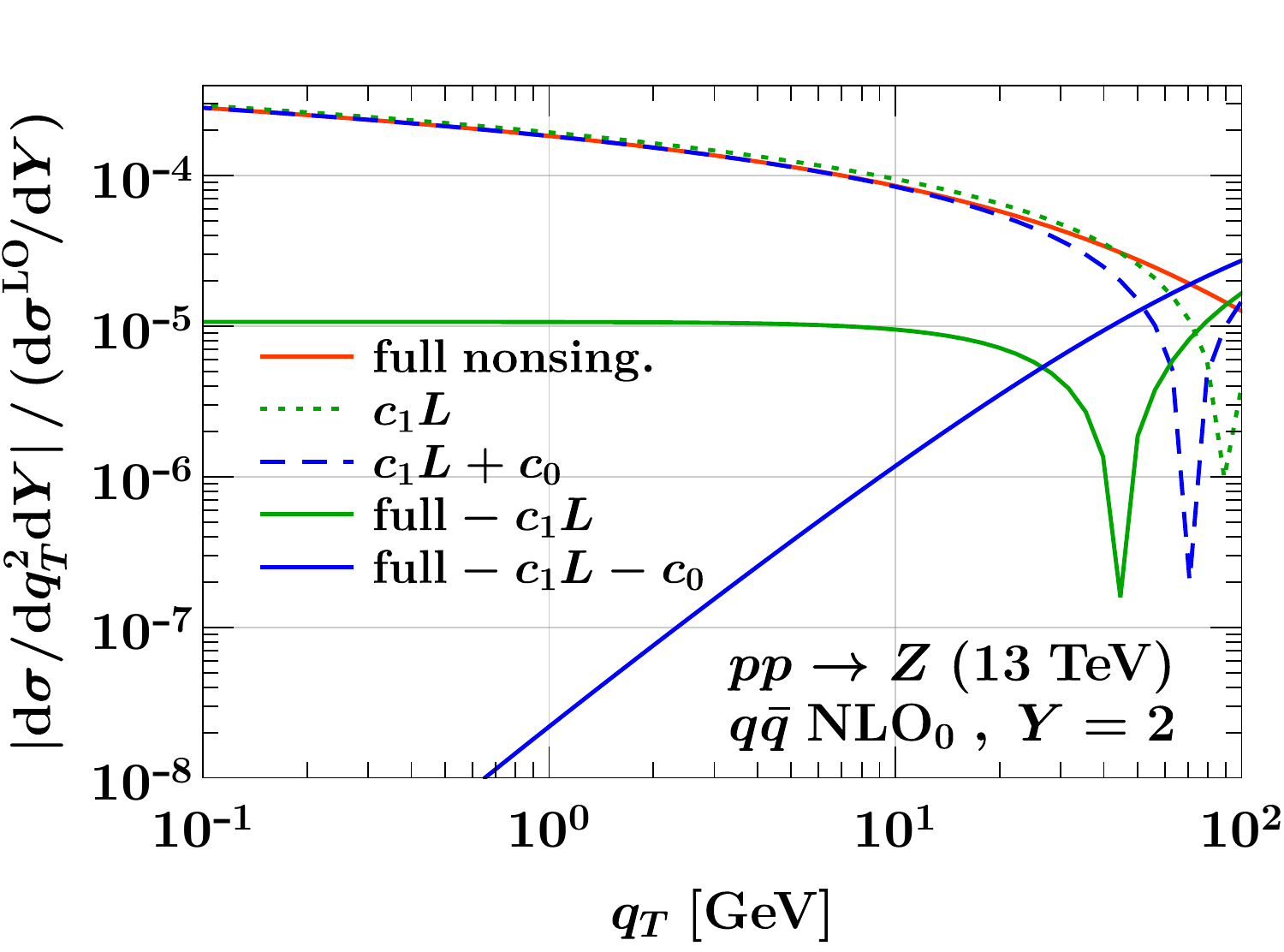}%
   \hfill
 \includegraphics[height=5.6cm]{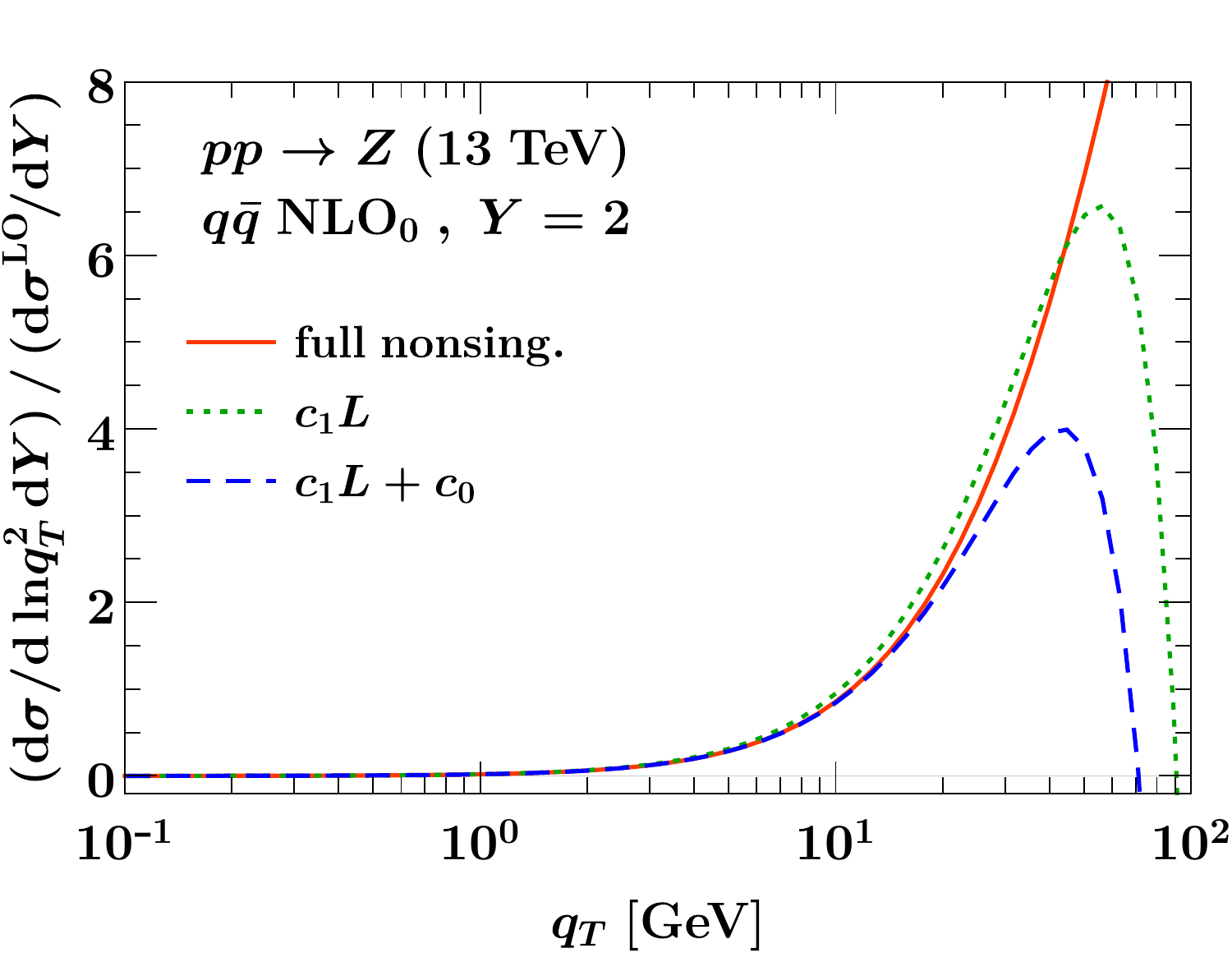}
 \includegraphics[height=5.6cm]{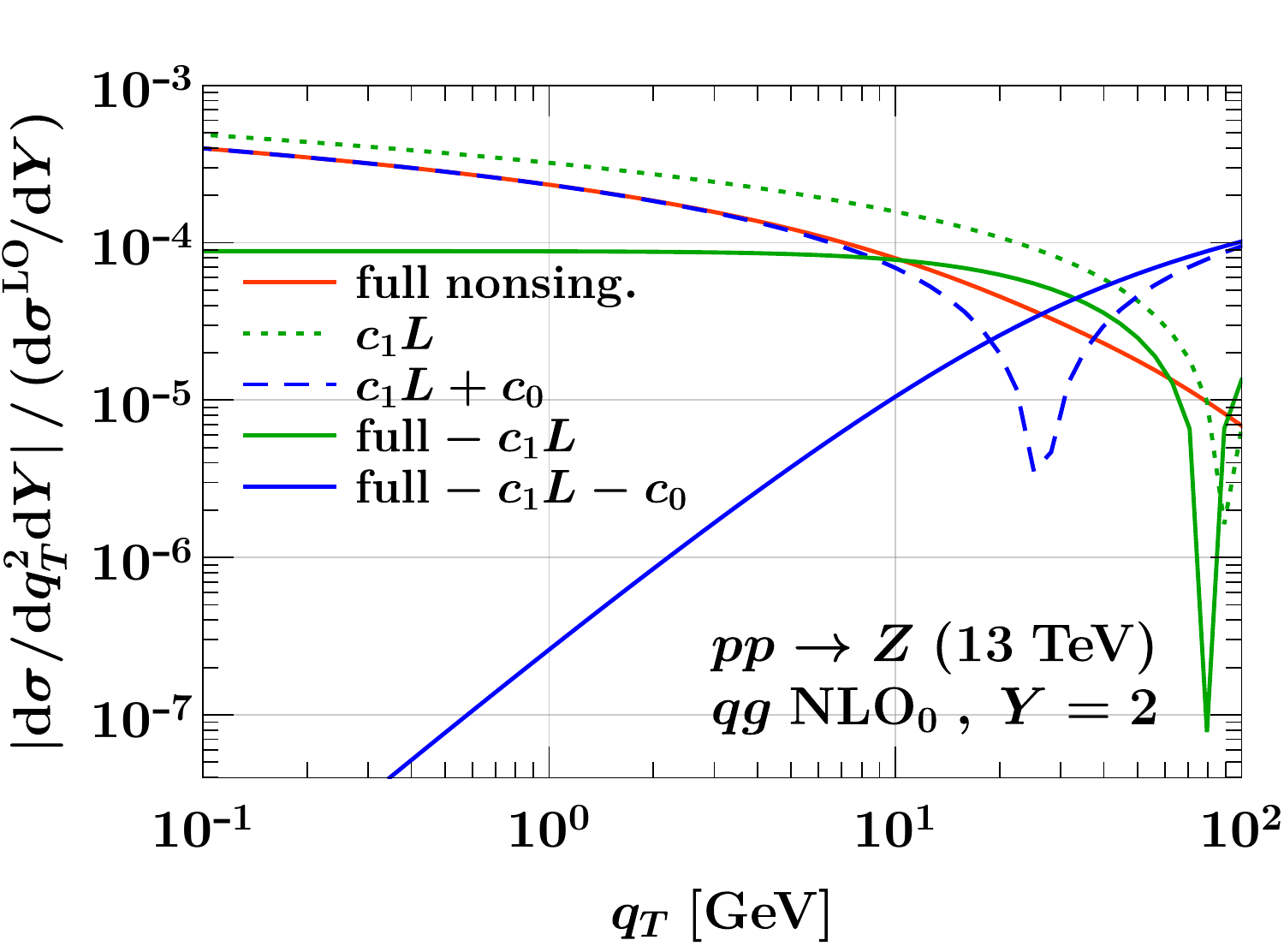}%
   \hfill
 \includegraphics[height=5.6cm]{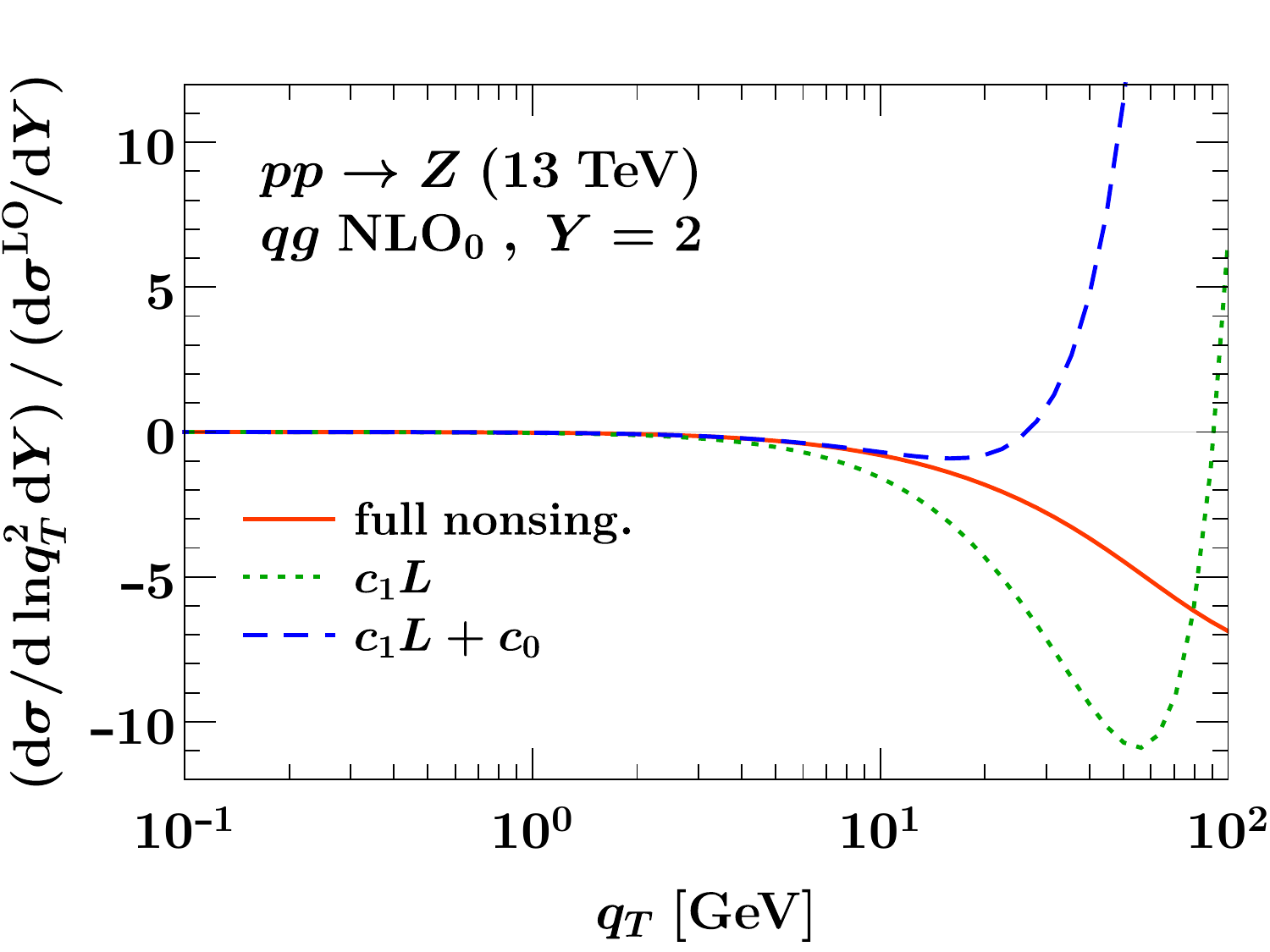}
\caption{Comparison of the LL and NLL corrections at subleading power with the full
nonsingular $q_T$ spectrum for all partonic channels contributing to Drell-Yan production at NLO$_0$.}
 \label{fig:DY}
\end{figure*}

\begin{figure*}[t!]
 \centering
 \includegraphics[height=5.6cm]{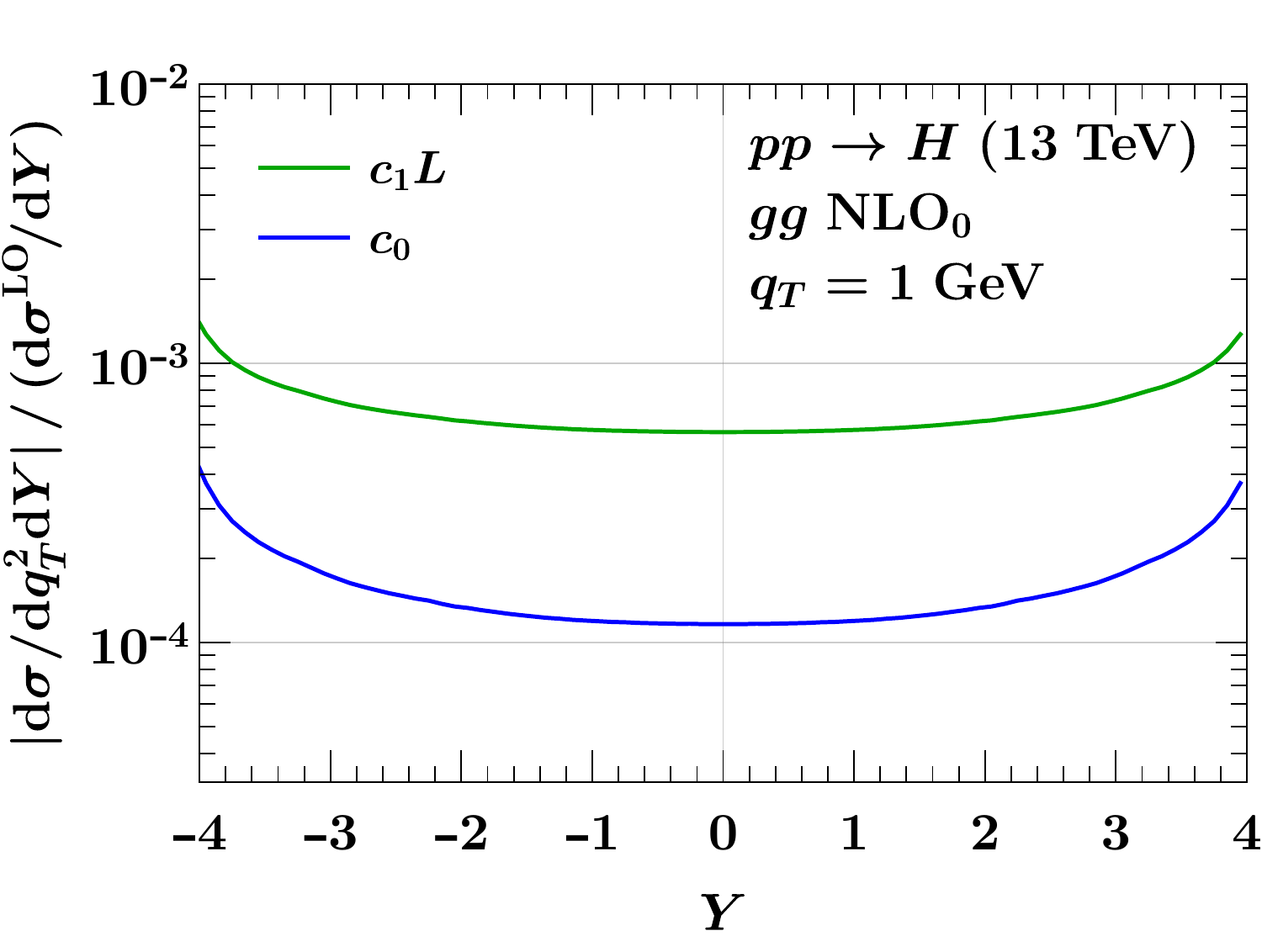}%
   \hfill
 \includegraphics[height=5.6cm]{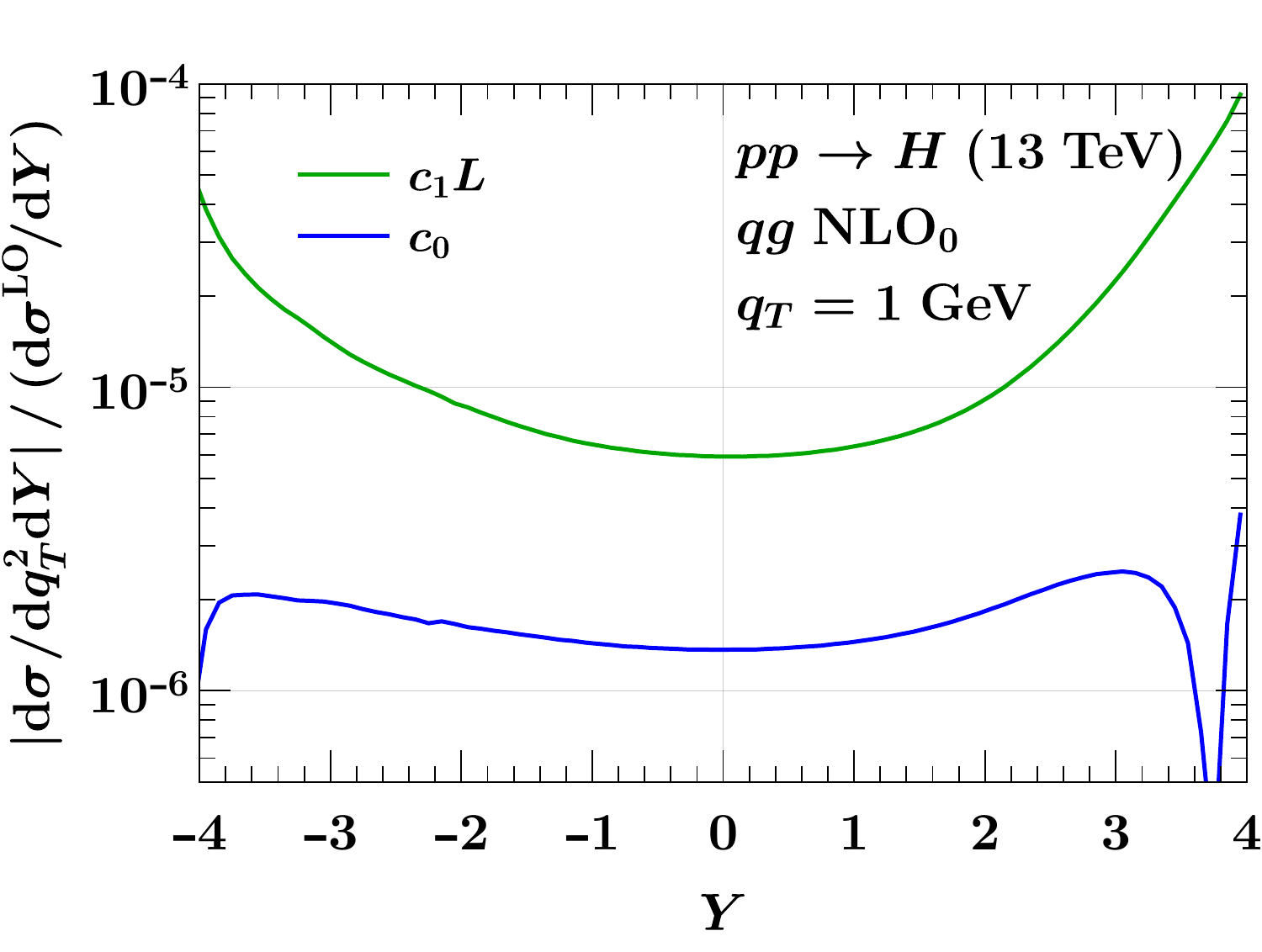}\\
 \includegraphics[height=5.6cm]{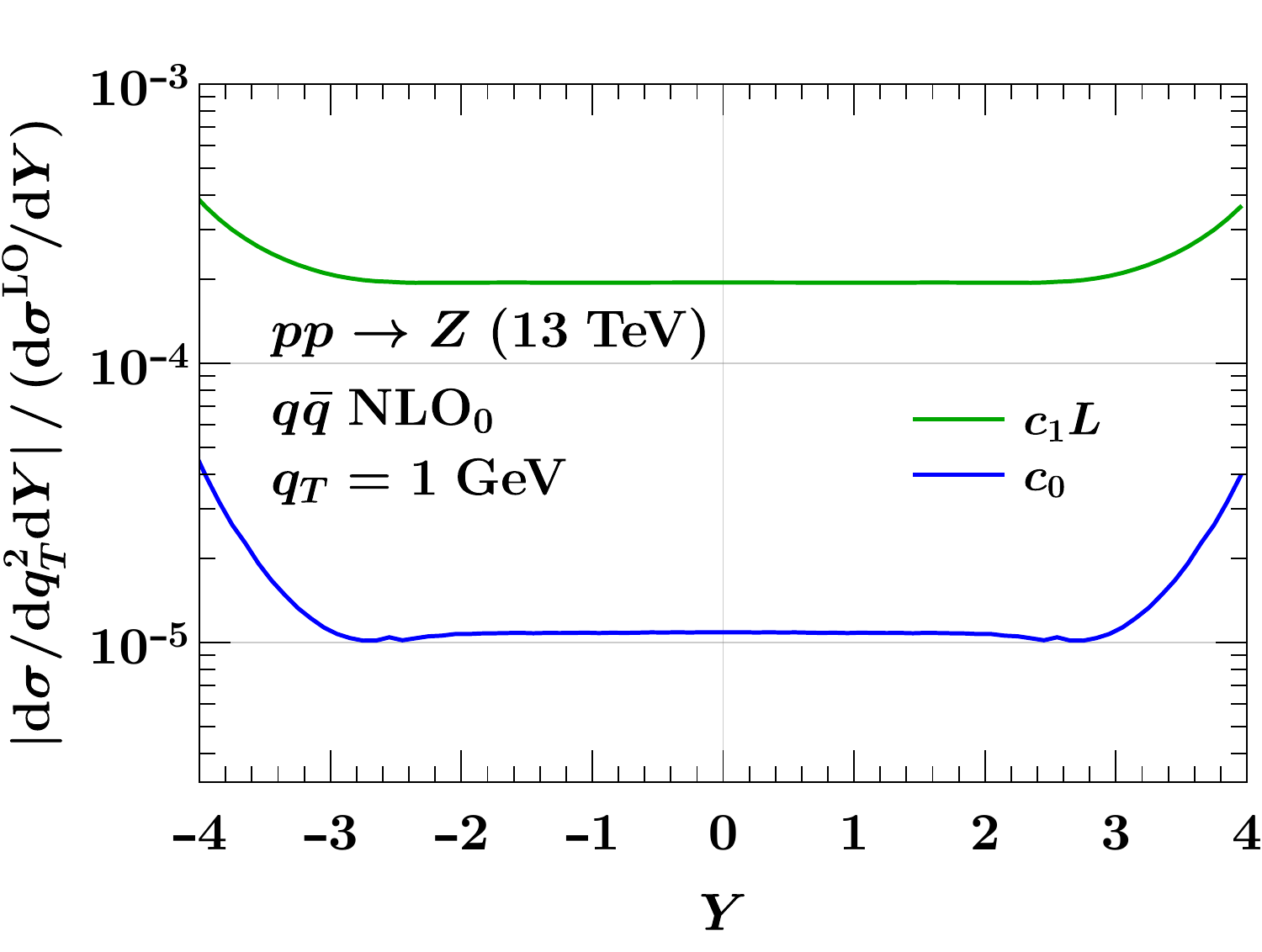}%
  \hfill
 \includegraphics[height=5.6cm]{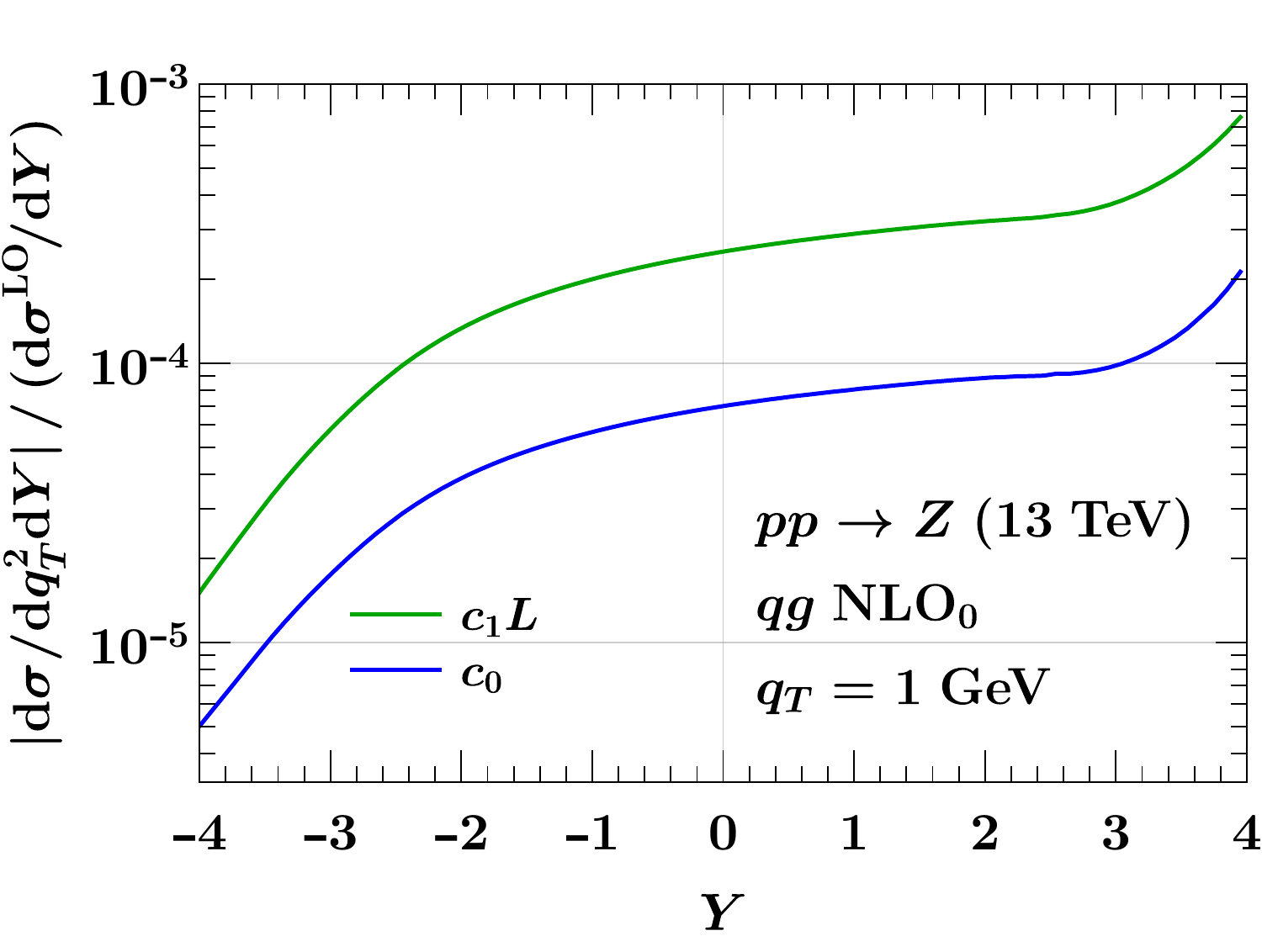}%
 \caption{Rapidity dependence of the LL (green) and NLL (blue) power corrections for Higgs and Drell-Yan production at NLO, relative to the LO rapidity dependence.
  The $q\bar{q}$ channel for Higgs production is not shown, as its LL power corrections vanish.
 }
 \label{fig:rapidity}
\end{figure*}

In \fig{Higgs}, we show the $q_T$ spectrum for all channels contributing to Higgs production.
The corresponding results for Drell-Yan production are shown in \fig{DY}.
In the left panel, we compare the nonsingular $q_T$ spectrum (solid red) against the NLP LL (green dashed) and full NLP (blue dashed) predictions.
For all channels, the NLP NLL result is an excellent approximation of the nonsingular spectrum up to $q_T \sim 10~\GeV$.
The solid green line shows the nonsingular spectrum minus the NLP LL correction, which in all cases is almost perfectly constant up to $q_T \sim 10~\GeV$, as expected from the structure of \eq{xs_nonsing}.
The solid blue line shows the nonsingular spectrum minus the full NLP correction, which vanishes as $q_T^2$ for small $q_T$ as expected from \eq{xs_nonsing}. This provides a strong numerical check
of our analytic results of the NLP contributions.
The right panels of \figs{Higgs}{DY} compare the nonsingular spectrum $q_T^2 \, \df\sigma/\df q_T^2$ with the NLP LL and NLP NLL approximations.
Again, we find excellent agreement up to $q_T \sim 10~\GeV$.

In \fig{rapidity}, we show the rapidity dependence of the power corrections for the $gg$ and $qg$ channels for Higgs production and for the $q\bar{q}$ and $qg$ channels for Drell-Yan production.
We show the individual NLP terms as given in \eq{xs_nonsing}, with the LL term proportional to $c_1$ shown in green and the NLL term proportional to $c_0$ shown in blue.
Since their $q_T$ dependence is trivial, we fix $q_T = 1~\GeV$, which only affects the overall size of the LL term, and we normalize the results to the LO rapidity spectrum.
Despite the fact that the kernels have no explicit rapidity dependence, we observe a nontrivial rapidity dependence due to the PDF derivatives, and in the case of the $qg$ channels also because they involve different PDFs than the Born process.
This is different than the case of beam thrust, which for certain definitions has an explicit rapidity dependence through factors of $e^{\pm Y}$ in both the LL and the NLL kernels \cite{Moult:2016fqy, Moult:2017jsg, Ebert:2018lzn}.
The rapidity dependence is particularly interesting for Drell-Yan production, where the term proportional to the PDFs themselves vanishes, see \eq{sigma_Higgs_LL_intro}, and so the power corrections are determined solely by the structure of the PDF derivatives.
At large values of $|Y|$, this leads to a relatively large dependence of the power corrections on the rapidity.
For Higgs production this effect is more moderate due to the appearance of a term proportional to PDFs as present at LO, which dominates the rapidity dependence.
This observation, which we believe is likely to persist at higher perturbative orders, could have important implications in the context of $q_T$ subtractions \cite{Catani:2007vq},
where it is important to understand the rapidity dependence of the power corrections. Our results suggest that the rapidity dependence may be well behaved for the case of Higgs production
but could be more problematic for Drell-Yan production. We leave the investigation of the structure at higher perturbative orders to future work.

\section{Conclusions}
\label{sec:conc}

In this paper, we have studied in detail the structure and consistent regularization of rapidity divergences at subleading order in the power expansion. We have discussed several
new features appearing at subleading power that put additional requirements on the rapidity regulator.
As a result, most of the rapidity regulators that have been used in the literature at leading power
become either unsuitable or inconvenient at subleading power.
In particular, we have shown that the $\eta$ regulator, which in principle can be applied at subleading power, is not homogeneous in the power expansion, which leads to undesirable complications at subleading power.
We have introduced a new pure rapidity regulator, which is homogeneous in the power counting. It allows us to regulate rapidity divergences appearing in $q_T$ distributions at any order in the power expansion, while respecting the power counting of the EFT. This significantly simplified the analysis of rapidity divergences and the associated logarithms at subleading power. It would be interesting to study its application to other physical problems of interest and to further study its properties.

We have also found a rich structure of power-law divergences at subleading power, which can
have a nontrivial effect on the final NLP result. Furthermore, at subleading power, rapidity divergences arise not only from gluons, but also from quarks. It would be interesting to further understand their formal properties.

As an explicit application of our formalism to a physical observable, we considered the $q_T$ spectrum for color-singlet production, for which we computed the complete NLP corrections, i.e., including both the logarithmic and nonlogarithmic contributions, at fixed $\ord{\alpha_s}$.
This provides a highly nontrivial test of our regulator.
In this case, the power-law rapidity divergences have the effect of inducing derivatives of the PDFs in the final NLP result for the $q_T$ spectrum. We also find that unlike for the case of beam thrust, where the LL power corrections for Higgs and Drell-Yan production are related by $C_A\leftrightarrow C_F$, this is not the case for the LL power corrections for $q_T$, which have a different structure for these two processes.

Our results represent a first important step in systematically studying subleading power corrections for observables with rapidity divergences. It opens the door for addressing a number of interesting questions. It will be important to extend our results and to better understand the structure of subleading-power rapidity divergences at higher perturbative orders. As a particularly interesting application, the power corrections for the $q_T$ spectrum can be used to improve the numerical performance and to better understand the systematic uncertainties of $q_T$ subtractions, whose feasibility at next-to-next-to-next-to-leading order has recently been demonstrated in \refcite{Cieri:2018oms} for Higgs production. We also hope that recent advances in the renormalization at subleading power, which has enabled the all-orders resummation of subleading-power logarithms, can also be extended to enable the resummation of subleading-power rapidity logarithms, with possible applications in a variety of contexts.

\begin{acknowledgments}
This work was supported in part by the Office of Nuclear Physics of the U.S.
Department of Energy under Contract No.~DE-SC0011090, by the Office of High Energy Physics of the U.S.~Department of Energy under Contract No.~DE-AC02-05CH11231, by the Simons Foundation Investigator Grant No.~327942, by the Alexander von Humboldt Foundation through a Feodor Lynen Research Fellowship, by the LDRD Program of LBNL, and within the framework of the TMD Topical Collaboration. IM thanks Zhejiang University and the MIT Center for Theoretical Physics for hospitality while portions of this work were performed. IM, GV, and ME also thank DESY for hospitality.
\end{acknowledgments}

\appendix

\section{NLO Results for \texorpdfstring{\boldmath $q_T$}{qT} at Leading Power}
\label{app:LP}

In this section we derive the LP beam and soft functions using the $\eta$ regulator
as a validation of our general setup.

\subsection{Soft Function}
\label{sec:LP_soft}

The bare soft function at LP can be calculated using the known LP soft limit of a matrix element given in \eq{Msquared_soft_LP},
\begin{align}
 \Msquared^{(0)}_s(Q,Y; \{k\}) &
 = \frac{16 \pi \as \muMS^{2\eps} \mathbf{C}}{k_T^2} \times \Msquared^\LO(Q,Y)
\,.\end{align}
We suppress that this limit only exists if either $ab=gg$ or $ab=q\bar q$.
Inserting into \eq{sigma_soft_LP} and using \eq{soft_integrals_a}, we have
\begin{align} \label{eq:sigma_soft_LP_1}
 \frac{\df\sigma^{(0)}_s}{\df Q^2 \df Y \df q_T^2} &
 = \frac{f_a(x_a) f_b(x_b)}{2 x_a x_b \Ecm^4} \Msquared^\LO(Q,Y)
   \times \frac{\as \mathbf{C}}{\pi} \frac{(4\pi \muMS^2)^\eps}{\Gamma(1-\eps)} q_T^{-2-2\eps}
   \, w^2 I_s^{(0)}
\nn\\*&
 = \frac{\df\sigma^\LO}{\df Q^2 \df Y}
   \times \frac{\as \mathbf{C}}{\pi} \frac{\mu^{2\eps} e^{\eps \gamma_E}}{\Gamma(1-\eps)} q_T^{-2-2\eps - \eta}
   \, w^2 \nu^\eta
   \, \sin\Bigl(\frac{\eta \pi}{2}\Bigr) \,
   \frac{1}{\pi} \Gamma(1-\eta) \Gamma^2\Bigl(\frac{\eta}{2}\Bigr)
\,.\end{align}
Here, we also replaced the MS scale $\muMS$ in terms of the $\MS$ scale $\mu$ using
\begin{align}
 \mu^2 \equiv \mu^2_{\overline{\rm MS}} = \frac{4\pi}{e^{\gamma_E}} \muMS^2
\,.\end{align}
Choosing instead $\mu^{2\eps} = \frac{(4\pi)^\eps}{\Gamma(1-\eps)} \muMS^{2\eps}$ would modify the $\cO(\eps^0)$ piece by $\pi^2/3$.

The divergence as $q_T\to0$ is regulated using the distributional identity
\begin{align}
 \mu^{2\eps} q_T^{-2 - 2\eps - \eta}
 = - \frac{2 \mu^{-\eta}}{2 \eps + \eta} \delta(q_T^2)
   + \mu^{-2-\eta} \biggl[ \biggl(\frac{\mu^2}{q_T^2}\biggr)^{1+\eps+\eta/2} \, \biggr]_+
\,.\end{align}
Inserting this into \eq{sigma_soft_LP_1}, we have to first expand in $\eta\to0$ and then in $\eps\to0$, which gives
\begin{align} \label{eq:sigma_soft_LP_}
 \frac{\df\sigma^{(0)}_s}{\df Q^2 \df Y \df q_T^2}
 = \frac{\df\sigma^\LO}{\df Q^2 \df Y} \times
   \frac{\as \mathbf{C}}{4\pi}
   \biggl[& \biggl(\frac{4}{\eps^2} - \frac{8}{\eps \eta}  + \frac{8}{\eps} \ln\frac{\mu}{\nu}\biggr) + \frac{8}{\eta} \frac{1}{\mu^2}\cL_0(q_T^2/\mu^2)
   \nn\\*&
   - 4 \frac{1}{\mu^2} \cL_1(q_T^2/\mu^2) - 8 \ln\frac{\mu}{\nu} \frac{1}{\mu^2} \cL_0(q_T^2/\mu^2) - \frac{\pi^2}{3} \delta(q_T^2)
   \biggr]
\,.\end{align}
The terms in brackets yield the one-loop soft function integrated over the azimuthal angle of $\qt$.
The fully differential result can be read of as
\begin{align} \label{eq:soft_NLO_LP}
 S^{(1)}_b(\qt, \mu,\nu) = \frac{\as \mathbf{C}}{4\pi} \biggl[ &
   \delta(\qt) \biggl( \frac{4}{\eps^2} - \frac{8}{\eps\eta}  + \frac{8}{\eps} \ln\frac{\mu}{\nu}\biggr)
   + \frac{8}{\eta} \cL_0(\qt,\mu)
   \nn\\&
   - 4\cL_1(\qt,\mu) - 8 \ln\frac{\mu}{\nu} \cL_0(\qt,\mu) - \frac{\pi^2}{3} \delta(\qt) \biggr]
\,,\end{align}
where the two-dimensional plus distributions are defined as in \refcite{Ebert:2016gcn},
\begin{align}
 \cL_n(\qt,\mu) = \frac{1}{\pi \mu^2} \biggl[ \frac{\mu^2}{q_T^2} \ln^n \frac{q_T^2}{\mu^2} \biggr]_+^\mu
 = \frac{1}{\pi \mu^2} \cL_n(q_T^2/\mu^2)
\,.\end{align}
This result agrees exactly with the result in \refcite{Luebbert:2016itl}.

\subsection{Beam Function}
\label{sec:LP_beam}

For illustration, we calculate the $gg$ contribution to the $n$-collinear gluon beam function.
The LP limit of the matrix element is given by
\begin{align} \label{eq:M2_coll_LP_app}
 \Msquared_n^{(0)}(Q, Y, \{k\}) &
 = \frac{8\pi \as \muMS^{2\eps}}{Q e^Y k^+} P_{gg}(z_a, \eps) \Msquared^\LO(Q, Y)
\nn\\&
 = \frac{8\pi \as \muMS^{2\eps}}{k_T^2} \frac{1-z_a}{z_a} \, 2 C_A\biggl[ \frac{z_a}{1-z_a} + \frac{1-z_a}{z_a} + z_a(1-z_a) \biggr] \Msquared^\LO(Q, Y)
\nn\\&
 = \frac{16\pi \as C_A\muMS^{2\eps}}{k_T^2} \biggl[ 1 + \frac{(1-z_a)^2}{z_a^2} + (1-z_a)^2 \biggr] \Msquared^\LO(Q, Y)
\,,\end{align}
where in the second step we used that $k^+ = k_T^2/k^-$ and $k^- = Q e^Y (1-z_a)/z_a$.
Inserting \eq{M2_coll_LP_app} into \eq{sigma_coll_LP}, we obtain
\begin{align} \label{eq:sigma_coll_LP_1}
 \frac{\df\sigma_n^{(0)}}{\df Q^2 \df Y \df q_T^2} &
 = \frac{\Msquared^\LO(Q, Y)}{2 x_a x_b \Ecm^4} \times
   \int_{x_a}^1 \frac{\df z_a}{z_a} f_a\biggl(\frac{x_a}{z_a}\biggr) f_b(x_b)
   \times \frac{\as C_A}{\pi} \frac{\mu^{2\eps} e^{\eps\gamma_E}}{\Gamma(1-\eps)}
          w^2 \biggl|\frac{Q e^Y}{\nu}\biggr|^{-\eta} q_T^{-2-2\eps}
   \nn\\&\hspace{3cm}\times
   \biggl[ \frac{z_a^{1+\eta}}{(1-z_a)^{1+\eta}} + \frac{(1-z_a)^{1-\eta}}{z_a^{1-\eta}} + z_a^{1+\eta} (1-z_a)^{1-\eta} \biggr]
\,.\end{align}
Note that only the first term in the square brackets is singular, and we can regularize it by
\begin{align}
 (1-z_a)^{-1-\eta} = -\frac{\delta(1-z_a)}{\eta} + \cL_0(1-z_a) + \cO(\eta)
\,.\end{align}
The singularity as $q_T \to 0$ is regulated by
\begin{align}
 \mu^{2\eps} q_T^{-2 - 2\eps}
 = - \frac{1}{\eps} \delta(q_T^2)
   + \frac{1}{\mu^2} \biggl[ \biggl(\frac{\mu^2}{q_T^2}\biggr)^{1+\eps} \, \biggr]_+
\,.\end{align}
The LP result becomes
\begin{align} \label{eq:sigma_coll_LP_2}
 \frac{\df\sigma_n^{(0)}}{\df Q^2 \df Y \df q_T^2} &
 = \hat\sigma^\LO(Q,Y) \times
   \int_{x_a}^1 \frac{\df z_a}{z_a} f_a\biggl(\frac{x_a}{z_a}\biggr) f_b(x_b)
   \times \frac{\as C_A}{\pi} w^2 \biggl[- \frac{1}{\eps} \delta(q_T^2) + \frac{1}{\mu^2} \cL_0(q_T^2/\mu^2) \biggr]
   \nn\\*&\hspace{3cm}\times
   \biggl[ - \frac{1}{\eta} \delta(1-z_a) + \frac{P_{gg}(z_a)}{2} + \delta(1-z_a) \ln\frac{Q e^Y}{\nu} \biggr]
\,,\end{align}
where
\begin{align}
 P_{gg}(z) = 2 \frac{(1 - z + z^2)^2}{z} \cL_0(1-z)
\end{align}
is the regularized gluon-gluon splitting function.
The two square brackets give the one-loop matching kernel integrated over the azimuthal angle of $\qt$.
Restoring the full $\qt$ dependence, the bare NLO matching kernel relating the beam function to the PDF, see \eq{q_match}, is given by
\begin{align} \label{eq:Igg_NLO_LP}
 I_{gg}(z, \qt, \omega, \nu) &
 = w^2 \frac{\as C_A}{\pi} \biggl[
   \delta(\qt) \biggl( \frac{\delta(1-z)}{\eta \eps} - \frac{P_{gg}(z)}{2 \eps} - \frac{\delta(1-z)}{\eps} \ln\frac{\omega}{\nu} \biggr)
   - \frac{\delta(1-z)}{\eta} \cL_0(\qt,\mu)
   \nn\\*&\hspace{2cm}
   + \cL_0(\qt,\mu) \biggl(\frac{P_{gg}(z)}{2} + \delta(1-z) \ln\frac{\omega}{\nu} \biggr) \biggr]
\,,\end{align}
where $\omega = Q e^Y$.
The finite part agrees with \refcite{Chiu:2012ir}, and thus after renormalization will give the same renormalized beam function kernel.
Also note that the $\eta$ poles cancel with the soft function \eq{soft_NLO_LP} after adding the $\bn$-collinear beam function.
The $P_{gg}(z)/\eps$ pole cancels with the UV divergence from the bare gluon PDF.
The remaining $\eps$ pole and the $\eps^2$ pole in the soft function, \eq{soft_NLO_LP}, only cancel after taking virtual corrections into account.

\section{Higher-Order Plus Distributions}
\label{app:plus_distr}

Subleading power corrections often involve divergences of the form
\begin{equation} \label{eq:power_law_n}
  \frac{1}{(1-z)^{a + \eta}} \,,\qquad a \in \mathbb{N}
\,.\end{equation}
In \sec{distribution} we encountered the two cases $a=2$ and $a=3$, which were treated using integration by parts to relate them to the case $a=1$, where one can use the relation
\begin{align} \label{eq:power_law_1_reg}
 \frac{1}{(1-z)^{1+\eta}} &
 = -\frac{\delta(1-z)}{\eta} + \biggl[ \frac{1}{(1-z)^{1+\eta}} \biggr]_+^1
\nn\\*&
 = -\frac{\delta(1-z)}{\eta} + \cL_0(1-z) - \eta \cL_1(1-z) + \cO(\eta^2)
\,.\end{align}
Here $\cL_n(x) = \bigl[\ln^n x/x\bigr]_+^1$ is defined in terms of standard plus distributions, which regulate functions $g(x)$ with support $x\ge0$ diverging less than $1/x^2$ as $x\to0$.
The defining properties of such plus distributions are
\begin{align} \label{eq:plus_dist_n1}
 \bigl[ g(x) \bigr]_+^1 &= g(x) \,,\qquad x \ne 0
\,,\nn\\*
 \int_0^1 \df x \, \bigl[ g(x) \bigr]_+^1 &= 0
\,.\end{align}
One can also treat the power-law divergences in \eq{power_law_n} similar to \eq{power_law_1_reg} using higher-order plus distributions.
We therefore generalize \eq{plus_dist_n1} as
\begin{align} \label{eq:plus_dist_n}
 \bigl[ g(x) \bigr]_{+(a)}^1 &= g(x) \,,\qquad x \ne 0
\,,\nn\\
 \int_0^1 \df x \, x^k \, \bigl[ g(x) \bigr]_{+(a)}^1 &= 0 \,,\hspace{1.2cm} \forall\, k < a
\,,\end{align}
where $g(x)$ has support $x\ge0$ and diverges less than $1/x^{1+a}$ as $x\to0$.
For $a=1$, this naturally reduces to \eq{plus_dist_n1}.
For $a=2$, one obtains the $++$ distributions used e.g.\ in \refcite{Mateu:2012nk}.

The distributions defined in \eq{plus_dist_n} can be integrated against any test function $f(x)$ that is at least $a{-}1$-times differentiable at $x=0$.
To be specific, consider the example integral
\begin{align} \label{eq:plus_dist_n_example}
 &\int_0^{x_0} \df x \, f(x) \bigl[ g(x) \bigr]_{+(a)}^1
\nn\\*&
 = \int_0^{x_0} \df x \, \biggl[ f(x) - \sum_{k=0}^{a-1} \frac{f^{(k)}(0)}{k!} x^k \biggr] \bigl[ g(x) \bigr]_{+(a)}^1
 + \sum_{k=0}^{a-1} \frac{f^{(k)}(0)}{k!} \int_0^{x_0} \df x \, x^k \bigl[ g(x) \bigr]_{+(a)}^1
\nn\\*&
 = \int_0^{x_0} \df x \, \biggl[ f(x) - \sum_{k=0}^{a-1} \frac{f^{(k)}(0)}{k!} x^k \biggr] g(x)
 - \sum_{k=0}^{a-1} \frac{f^{(k)}(0)}{k!} \int_{x_0}^1 \df x \, x^k g(x)
\,,\end{align}
where we assume $x_0 > 0$ and $f^{(k)}(0)$ is the $k$-th derivative of $f(x)$ at $x=0$.
In \eq{plus_dist_n_example}, we used that the term in square brackets in the first integral behaves as $\cO(x^a)$ and thus cancels the divergent behavior of $g(x)$ as $x\to0$, which allows us to drop the plus prescription in the first integral in the last line.
In the second integral, we used \eq{plus_dist_n} to change the integration bounds from $[0,x_0]$ to $[x_0,1]$. In the latter interval, $g(x)$ is regular and the plus prescription can be dropped.

The power-law divergence in \eq{power_law_n} can be regularized in terms of the higher-order plus distributions in \eq{plus_dist_n} as
\begin{align} \label{eq:power_law_n_reg}
 \frac{1}{(1-z)^{a+\eta}} &
 = \biggl[ \frac{1}{(1-z)^{a+\eta}} \biggr]_{+(a)}^1 + \sum_{k=0}^{a-1} \frac{(-1)^k}{k!} \frac{\delta^{(k)}(1-z)}{1+k-a-\eta}
\,, \qquad a \in \mathbb{N}
\,.\end{align}
This result can be verified by integrating both sides against a test function $(1-z)^m$ with $m<a$, and treating $\eta$ as in dimensional regularization to render all integrals finite.
In \eq{power_law_n_reg}, $\delta^{(k)}(1-z)$ is the $k$-th derivative on $\delta(1-z)$, which thus induces a sign $(-1)^k$ in an integral over $z$ and picks out the $k$-th derivative of any test function it acts on.
Note that only the $k=a-1$ term in \eq{power_law_n_reg} diverges for $\eta\to0$,
\begin{align}
 \frac{1}{(1-z)^{a+\eta}} &
 = -\frac{1}{\eta} \frac{(-1)^{a-1}}{(a-1)!} \delta^{(a-1)}(1-z)
 + \biggl[ \frac{1}{(1-z)^{a}} \biggr]_{+(a)}^1 + \sum_{k=0}^{a-2} \frac{(-1)^k}{k!} \frac{\delta^{(k)}(1-z)}{1+k-a}
 + \cO(\eta)
\,,\end{align}
so irrespective of the power $a$, any power law divergence $(1-z)^{-a-\eta}$ has exactly one single pole.

\section{Derivation of the Master Formula for Generic \texorpdfstring{\boldmath $c$}{c}}
\label{app:master_formula_c}

In \secs{master_formula}{master_formula_2}, we derived master formulas for the NLP correction to the $q_T$ spectrum using the $\eta$ regulator and the pure rapidity regulator, respectively.
In \sec{upsilonreg}, we also introduced a class of homogeneous rapidity regulators spanned by a parameter $c \neq 1$.
Here, we give the master formulas for this regulator for generic $c\ne1$.
In this regulator, the soft contribution is scaleless and vanishes, similar to the pure rapidity regulator.
Thus, one only needs to consider the $n$-collinear and $\bn$-collinear limits.

The derivation of the $n$-collinear expansion proceeds similar to the calculation shown in \sec{collinear_master}.
One can also obtain it from the result for the pure rapidity regulator, \eq{sigma_ncoll_NLP_LL_vita}, using the replacement
\begin{align}
 &\upsilon^\eta \biggl|\frac{k^-}{k^+}\biggr|^{-\eta/2} = \upsilon^\eta q_T^\eta |k^-|^{-\eta}
 \nn\\*
\quad\to\quad
 &\upsilon^{(1-c)\eta/2} \biggl|\frac{k^-}{\nu}\biggr|^{-\eta/2} \biggl|\frac{k^+}{\nu}\biggr|^{-c \eta/2}
 = \biggl[\upsilon \Bigl(\frac{\nu}{q_T} \Bigr)^{\frac{1+c}{1-c}}\biggr]^{(1-c)\eta/2} q_T^{(1-c)\eta/2} |k^-|^{-(1-c)\eta/2}
\,.\end{align}
Thus, in \eq{sigma_ncoll_NLP_LL_vita} one has to shift $\eta \to (1-c)\eta/2$ and $\upsilon \to \upsilon (\nu/q_T)^{\frac{1+c}{1-c}}$, giving
\begin{align} \label{eq:sigma_ncoll_NLP_LL_vita_c}
 \frac{\df\sigma^{(2),\text{LL}}_n}{\df Q^2 \df Y \df q_T^2} &
 = \frac{1}{(4\pi)^2} \frac{q_T^2}{2Q^2}
   \frac{1}{2 x_a x_b \Ecm^4}\,w^2\biggl(\frac{2}{(1-c)\smallupsilon} - \ln\frac{Q e^Y}{q_T} + \frac{1+c}{1-c} \ln\frac{\nu}{q_T} + \ln(\upsilon) \biggr)
   \nn\\&\quad \times  \biggl\{
   f_a(x_a) f_b(x_b) \Bigl[ \altMp{2}_n(1) - 2 \altMp{0}_n(1) \Bigr]
   + f_a(x_a) \, x_b f'_b(x_b) \Bigl[ \altM{0}_n(1) + 2 {\altMp{0}_n}(1) \Bigr]
   \nn\\*&\qquad
   + x_a f'_a(x_a) f_b(x_b) \Bigl[ \altM{0}_n(1) - \altM{2}_n(1) \Bigr]
   - 2 x_a f'_a(x_a) \, x_b f'_b(x_b)  \altM{0}_n(1)
   \biggr\}
\,.\end{align}
This result is well-defined for all $c\ne1$, whereas one encounters two explicit poles as $c\to1$.
This behavior is expected because for $c=1$ the regulator depends on the boost-invariant product $k^+ k^- = q_T^2$ and therefore does not regulate rapidity divergences, as explained at the end of \sec{upsilonreg}.
For $c=-1$ we recover the result of pure rapidity regularization of \eq{sigma_ncoll_NLP_LL_vita}.
In this case, the $\nu$ dependence in the regulator \eq{vita_regulator} cancels,
which is reflected by the vanishing of the coefficient of $\ln(\nu/q_T)$ in \eq{sigma_ncoll_NLP_LL_vita_c}.

In the $\bn$-collinear limit, the regulator for arbitrary $c\ne1$ is obtained from the pure rapidity regulator through
\begin{align}
 &\upsilon^\eta \biggl|\frac{k^-}{k^+}\biggr|^{-\eta/2} = \upsilon^\eta q_T^{-\eta} |k^+|^{\eta}
 \nn\\*
\quad\to\quad
 &\upsilon^{(1-c)\eta/2} \biggl|\frac{k^-}{\nu}\biggr|^{-\eta/2} \biggl|\frac{k^+}{\nu}\biggr|^{-c \eta/2}
 = \biggl[\upsilon \Bigl(\frac{\nu}{q_T} \Bigr)^{\frac{1+c}{1-c}}\biggr]^{(1-c)\eta/2} q_T^{-(1-c)\eta/2} |k^+|^{(1-c)\eta/2}
\,.\end{align}
Thus, in \eq{sigma_nbarcoll_NLP_LL_vita} one has to shift $\eta \to (c-1)\eta/2$ and $\upsilon \to \upsilon (\nu/q_T)^{\frac{1+c}{1-c}}$, giving
\begin{align} \label{eq:sigma_nbarcoll_NLP_LL_vita_c}
 \frac{\df\sigma^{(2),\text{LL}}_{\bn}}{\df Q^2 \df Y \df q_T^2} &
 = \frac{1}{(4\pi)^2} \frac{q_T^2}{2Q^2}
   \frac{1}{2 x_a x_b \Ecm^4}\,w^2\biggl(\frac{2}{(c-1)\smallupsilon} - \ln\frac{Q e^{-Y}}{q_T} - \frac{1+c}{1-c} \ln\frac{\nu}{q_T}  - \ln(\upsilon) \biggr)
   \\&\quad \times  \biggl\{
   f_a(x_a) f_b(x_b) \Bigl[ \altMp{2}_{\bn}(1) - 2 \altMp{0}_{\bn}(1) \Bigr]
   + f_a(x_a) \, x_b f'_b(x_b) \Bigl[ \altM{0}_{\bn}(1) - \altM{2}_{\bn}(1) \Bigr]
   \nn\\*&\qquad
   + x_a f'_a(x_a) f_b(x_b)  \Bigl[ \altM{0}_{\bn}(1) + 2 {\altMp{0}_{\bn}}(1) \Bigr]
   - 2 x_a f'_a(x_a) \, x_b f'_b(x_b)  \altM{0}_{\bn}(1)
   \biggr\}
\nn\,.\end{align}
Summing \eqs{sigma_ncoll_NLP_LL_vita_c}{sigma_nbarcoll_NLP_LL_vita_c}, the poles in $\smallupsilon$ precisely cancel,
and the dependence on $c$, $\bigupsilon$ and $e^Y$ cancels as well to yield a pure logarithm in $\ln(Q/q_T)$.
As for the pure rapidity regulator, this cancellation has to occur between the two collinear sectors,
since the soft sector does not give a contribution.

The NLP NLL result is identical to that in pure rapidity regularization, which is given by \eq{sigma_coll_NLP_NLL} upon dropping all regulator-dependent pieces, as explained in \sec{master_formula_2}. This provides another check of our regularization procedure.

\bibliography{../subleading}
\bibliographystyle{jhep}

\end{document}